\documentclass[useAMS,usenatbib]{mn2e}
\usepackage{amssymb}
\usepackage{graphicx}
\usepackage{multirow}
\usepackage{amsmath}
\usepackage{epstopdf}
\usepackage{color}
\definecolor{red}{rgb}{1.0,0.0,0.0}
\definecolor{blue}{rgb}{0.0,0.0,1.0}
\definecolor{green}{rgb}{0.0,1.0,0.0}

\title[Cluster simulations with AGN feedback]{Cosmological simulations of galaxy clusters with feedback from active galactic nuclei: profiles and scaling relations}

\author[Pike et al.]{Simon R. Pike,$^{1}$
Scott T. Kay,$^{1}$\thanks{E-mail: scott.kay@manchester.ac.uk}
Richard D. A. Newton,$^{1,2}$ 
Peter A. Thomas$^{3}$ 
and \newauthor Adrian Jenkins$^{4}$\\
$^{1}$Jodrell Bank Centre for Astrophysics, School of Physics and Astronomy, The University of Manchester, Manchester M13 9PL\\
$^{2}$International Centre for Radio Astronomy Research, University of Western Australia, 35 Stirling Highway, Crawley, \\Western Australia 6009, Australia\\
$^{3}$Astronomy Centre, University of Sussex, Falmer, Brighton BN1 9QH\\
$^{4}$Institute for Computational Cosmology, Department of Physics, University of Durham, Durham DH1 3LE\\
}
\begin{document}

\date{Accepted 1988 December 15. Received 1988 December 14; in original form 1988 October 11}

\pagerange{\pageref{firstpage}--\pageref{lastpage}} \pubyear{2002}

\maketitle

\label{firstpage}

\begin{abstract}
We present results from a new set of 30 cosmological simulations of galaxy clusters, including
the effects of radiative cooling, star formation, supernova feedback, black hole growth and AGN
feedback. We first demonstrate that our AGN model is capable of reproducing the observed cluster
pressure profile at redshift, $z \simeq 0$, once the AGN heating temperature of the 
targeted particles is made to scale with the final virial temperature of the halo. This allows the ejected 
gas to reach larger radii in higher-mass clusters than would be possible had a fixed heating 
temperature been used. Such a model also successfully reduces the star formation rate in 
brightest cluster galaxies and broadly reproduces a number of other observational properties at low redshift, 
including baryon, gas and star fractions; entropy profiles outside the core; and the X-ray luminosity-mass relation. 
Our results are consistent with the notion that the excess entropy is generated via selective removal of the densest 
material through radiative cooling; supernova and AGN feedback largely serve as regulation mechanisms, moving heated 
gas out of galaxies and away from cluster cores. However, our simulations fail to address a number of
serious issues; for example, they are incapable of reproducing the shape and diversity of the observed entropy 
profiles within the core region. We also show that the stellar and black hole masses are sensitive to numerical resolution, 
particularly the gravitational softening length; a smaller value leads to more efficient black hole growth at early times 
and a smaller central galaxy. 
\end{abstract}

\section{Introduction}

It has long been known that the observable properties of the intracluster medium (ICM hereafter), especially X-ray luminosity,
do not scale with mass as expected if gravitational heating is the only important physical process at work
(e.g. \citealt{Voit2005}). \cite{Ponman1999} confirmed that the reason for this {\it similarity breaking} is
due to low-mass groups and clusters having excess entropy in their cores. A large body of work has since been 
accumulating using X-ray data, measuring the detailed thermal structure of the ICM and how it depends on cluster mass, 
redshift and dynamical state 
(e.g. \citealt{Vikhlinin2006,Pratt2009,Sun2012,Eckert2013}). 

Complementary to the X-ray work, observations of the Sunyaev-Zel'dovich (hereafter SZ) effect \citep{Sunyaev1972} 
are now providing independent measurements of the ICM pressure distribution and scaling relations 
(e.g. \citealt{Planck2011e,Planck2013,Andersson2011,Marrone2012,Sifon2013}). 
Furthermore, optical-infrared studies are measuring the stellar mass component, both in galaxies and the intracluster light  
(e.g. \citealt{Stott2011,Lidman2012,Budzynski2014}).
It is clear that the majority of the baryons are in the ICM, with only a few per cent of the total cluster mass locked in stars.

The physical origin of the excess entropy (and why star formation is so inefficient)
continues to be a subject of debate. Early work suggested that the ICM 
was pre-heated at high redshift, prior to cluster formation \citep{Evrard1991,Kaiser1991}. However, cluster models 
with pre-heating have shown that it produces isentropic cores in low-mass systems (e.g. \citealt{Borgani2001,Babul2002}),
at odds with the observational data (e.g. \citealt{Ponman2003}). Pre-heating simulations that include radiative cooling
also tend to produce too little star formation, but this somewhat depends on numerical resolution  (e.g. \citealt{Muanwong2002}).

An alternative model is to exploit the radiative cooling of gas directly. As the lowest entropy gas cools and forms stars, it allows 
the remaining, higher entropy material to flow towards the centre of the cluster, creating an overall excess in the core. Since this 
effect is more prominent in lower mass systems where the cooling time is shorter, it leads to the desired outcome \citep{Bryan2000}. 
The {\it radiative} model was confirmed with fully-cosmological simulations (e.g. \citealt{Pearce2000,Muanwong2001,Dave2002}) 
but it is ultimately flawed as it requires an unrealistic amount of gas to cool and form stars (the so-called {\it overcooling} problem; 
see \citealt{Balogh2001,Borgani2002}). 

The most promising solution to both entropy and overcooling problems is negative {\it feedback}, i.e. energetic galactic 
outflows that remove the densest gas and reduce the star formation efficiency in galaxies. The first models
focused on supernova feedback but these fail to produce enough entropy to remove material from cluster
cores (e.g. \citealt{Borgani2004}) unless the energy is targeted at a small amount of mass (e.g. \citealt{Kay2003,Kay2004}). 
A more appealing solution, on energetic grounds, is feedback from active galactic nuclei (AGN; e.g. \citealt{Wu2000}),
where around 10 per cent of the mass accreted on to a super-massive black hole (BH) is potentially available as feedback energy.
High resolution X-ray observations have now firmly established that AGN are interacting with the ICM in low-redshift
clusters through the production of jet-induced cavities and weak shocks (e.g. \citealt{Fabian2012,McNamara2012}). It is also likely
that BH's are even more active in high redshift clusters, given that the space density of quasars peaks at $z \simeq 2$ 
\citep{Shaver1996}.

Including AGN feedback in cosmological simulations is a highly non-trivial task, given the disparity in scales between the
accreting BH ($< 1$ pc) and the host galaxy ($\sim 10$ kpc). As a result, a range of models for both the accretion and feedback 
processes, have been developed and applied to simulations of galaxies 
(e.g. \citealt{SpringelDMH2005,Booth2009,Power2011,Newton2013}). Due to the infancy of these models,
much of the simulation work on cluster scales has been done using idealised, or cosmologically-influenced, initial conditions 
(e.g. \citealt{Morsony2010,Gaspari2011,Hardcastle2013}). However, a growing number of groups
are now starting to incorporate AGN feedback in fully-cosmological simulations of groups and clusters, with some success. 
We summarise a few of their results below.

\cite{Sijacki2007} included models for both a {\it quasar} mode (heating the gas local to the BH) and a {\it radio} mode (injecting 
bubbles into the ICM when the accretion rate is low), showing that such feedback could produce a realistic entropy profile in clusters 
while suppressing their cooling flows. \cite{Puchwein2008,Puchwein2010} applied this model to a larger cosmological sample 
of clusters and showed that the AGN feedback reduced the overcooling on to brightest cluster galaxies (BCGs), resulting in X-ray and
optical properties that are more realistic, but producing a large fraction of intracluster stars. \cite{Dubois2011} ran a cosmological 
re-simulation of a cluster and were also able to prevent overcooling with AGN feedback, producing gas profiles that were consistent with 
cool-core clusters when metallicity effects were neglected. \cite{Fabjan2010} ran re-simulations for 16 clusters and found that BCG growth 
was sufficiently quenched at redshifts, $z<4$, and their runs produced reasonable temperature profiles of galaxy groups. However in 
massive clusters, the AGN model is unable to create cool cores, producing an excess of entropy within $r_{2500}$. \cite{Planelles2013} 
further showed that AGN feedback in their simulations is capable of reproducing observed cluster baryon, gas and star fractions. 
\cite{Short2010} included AGN feedback into cosmological simulations using a semi-analytic galaxy formation model to infer the heating
rates from the full galaxy population and showed that such a model could reproduce a range of X-ray cluster properties, although neglected
the effects of radiative cooling (however, see also \citealt{Short2012}). \cite{McCarthy2010,McCarthy2011} simulated the effects of AGN feedback in
galaxy groups and showed that they could reproduce a number of their observed properties. Their feedback model, based on \cite{Booth2009}, 
works by ejecting high entropy gas out of the cores of proto-group haloes at high redshift and thus generates the excess entropy in a similar
way to the {\it radiative} model described above, while also regulating the amount of star formation.\footnote{This mechanism was 
originally described in a model by \cite{Voit2001}, who phrased it in terms of feedback from supernovae.} 

In this paper, we introduce a new set of cosmological simulations of clusters and use them to further our understanding of 
how non-gravitational processes (especially AGN feedback) affect such systems, comparing to observational data where appropriate.
Our study has the following particular strengths. 
Firstly, we have selected a representative sample of clusters to assess their properties across the full cluster mass range.  
Secondly, all objects have around the same number of particles within their virial radius at $z=0$,
removing potential bias due to low-mass systems being less well resolved. Thirdly, we have
run our simulations several times, incrementally adding radiative cooling and star formation; supernova feedback and AGN feedback.
This allows us to assess the relative effects of these individual components. Finally, we use the AGN feedback model from 
\cite{Booth2009}; since we apply it to cluster scales our results complement those on group scales by \cite{McCarthy2010}. In particular, 
we show that our simulations can reproduce observed ICM pressure profiles at $z\simeq 0$ particularly well, once the AGN heating
temperature is adjusted to scale with the final virial temperature of the cluster. 

The remainder of the paper is laid out as follows. Section~\ref{sec:method} provides details of the sample selection, our implementation of 
the sub-grid physics and the method by which radial profiles and scaling relations are estimated. Our main results are then
presented in Sections~\ref{sec:baryons} (global baryonic properties), 
\ref{sec:profiles} (radial profiles) and \ref{ss:SR} (scaling relations). In Section~\ref{sec:res}, 
we present a resolution study before drawing conclusions and discussing our results in the context 
of recent work by others (\citealt{LeBrun2013,Planelles2013,Planelles2014}) in Section~\ref{sec:discuss}.

\section{Simulation details}
\label{sec:method}

Our main results are based on a sample of 30 clusters, re-simulated from a large cosmological simulation of structure formation within the $\Lambda$CDM cosmology.
The sample size was chosen as it was deemed to be large enough to produce reasonable statistical estimates of cluster properties over the appropriate range of masses 
and dynamical states, while small enough to allow a competitive resolution to be used. We outline how the clusters were selected below, before summarising details
of the baryonic physics in our simulations.

\subsection{Cluster sample}

The clusters were selected from the Virgo Consortium's MR7 dark matter-only simulation, available online via the 
Millennium database.\footnote{\tt{http://gavo.mpa-garching.mpg.de/Millennium/}}
The simulation also features in \cite{Guo2013} with the name MS-W7. It is similar to the original Millennium simulation 
\citep{Springeletal2005}, with $2160^{3}$ particles within a $500\,h^{-1}{\rm Mpc}$ comoving volume, but uses different 
cosmological parameters and phases. The cosmological parameters are consistent with the {\it WMAP} 7-year data 
\citep{Komatsu2011}, with $\Omega_{\rm m} =0.272, \Omega_{\Lambda}=0.728, \Omega_{\rm b}=0.0455, h=0.704$ and 
$\sigma_{8}=0.81$. The phases for the MR7 volume were taken from the public multi-scale Gaussian white noise field
Panphasia (\citealt{Jenkins2013}; referred to as MW7 in their Table~6).

Clusters were identified in the parent simulation at $z=0$ using the Friends-of-Friends algorithm \citep{Davis1985} with dimensionless linking
length, $b=0.2$. The SUBFIND \citep{Springel2001} routine was also run on-the-fly and we used the position of the particle with the
minimum energy (from the most massive sub-halo within each Friends-of-Friends group) to define the cluster centre. We sub-divided the clusters with masses 
$10^{14} < \log_{10}(M_{200}/h^{-1}{\rm M}_{\odot})<10^{15}$
into five mass bins, equally spaced in $\log_{10}(M_{\rm 200})$.\footnote{The mass, $M_{200}$, is that contained within a sphere of 
radius $r_{200}$, enclosing a mean density of 200 times the critical density of the Universe.}
Six objects were then chosen at random from within each bin, yielding a sample
of 30 objects. Particle IDs within $3r_{\rm{200}}$ (centred on the most bound particle) were recorded and their coordinates 
at the initial redshift ($z=128$) used to define a Lagrangian region to be re-simulated at higher resolution. 
Finally, initial conditions were generated for each object with a particle mass chosen to produce a fixed number of 
particles within $r_{200}$, $N_{200} \simeq 10^{6}$. The advantage of this choice is that the same dynamic range in 
internal substructure is resolved within each object, regardless of its mass. The particle mass varies from 
$m=1\times 10^{8}\,h^{-1}{\rm M}_{\odot}$ for the lowest-mass clusters, to 
$m=8\times 10^{8}\,h^{-1}{\rm M}_{\odot}$ for the highest-mass clusters.

The method used to make the initial conditions for the re-simulations was essentially that described in 
\cite{Springel2008} for the Aquarius project. The large-scale power, from Panphasia, was reproduced and uncorrelated small scale power added to the high resolution region down to the particle Nyquist frequency of that region. These initial conditions were created before the re-simulation method described in \cite{Jenkins2013} was developed. This
means that the added small-scale power was an independent realisation and distinct from that given by the Panphasia field itself.

Each cluster was run several times using a modified version of the Gadget-2 $N$-body/SPH
code \citep{Springel2005}, first with dark matter (DM) only, then with gas and 
varying assumptions for the baryonic physics (discussed below). The gas initial conditions were identical to the DM-only
case, except that we split each particle within the Lagrangian region into a gas particle with mass, $m_{\rm gas}=(\Omega_{\rm b}/
\Omega_{\rm m})m$,  and a DM particle with mass, $m_{\rm DM}=m-m_{\rm gas}$. 
The gravitational softening length was fixed in physical co-ordinates for $z<3$, 
setting the equivalent Plummer value to $\epsilon = 4r_{200}/\sqrt{N_{200}}$ following \cite{Power2003}. Thus, in our lowest-mass clusters 
$\epsilon \simeq 3 h^{-1}{\rm kpc}$, increasing by a factor of two for our highest-mass clusters. The softening was fixed in comoving 
co-ordinates at 
$z>3$. For the gas, the SPH smoothing length was never allowed to become smaller than the softening length, given that gravitational
forces become inaccurate below this value.

\subsection{Baryonic physics}

\begin{table}
\centering
\caption{Summary of the models used for our cluster simulations with
baryonic physics.}
\begin{tabular}{llll}
\hline   
Model & Cooling \& SF & Supernovae & AGN\\
\hline   								
NR & No & No & No\\
CSF & Yes & No & No\\
SFB  & Yes & Yes & No\\
AGN & Yes & Yes & Yes\\
\hline   
\end{tabular}
\label{tab:model}
\end{table}

For our main results, we performed four sets of runs with gas and additional, non-gravitational physics. The first model
(labelled NR) used non-radiative gas dynamics only. 
For the second set of runs, we included radiative cooling and star formation (CSF); in the third, we 
added supernova feedback (SFB); and in the fourth we additionally modelled feedback from active galactic
nuclei (AGN). Table~\ref{tab:model} summarises these choices. We discuss the details of each process
below and refer to \cite{Newton2013} for further information.

\subsubsection{Radiative cooling and star formation}

Gas particles with temperatures, $T>10^{4}$K are allowed to cool radiatively. We assume collisional ionisation 
equilibrium and the gas is isochoric when calculating the energy radiated across each timestep, following 
\cite{Thomas1992}. Cooling rates are calculated using the tables given by \cite{Sutherland1993} for a zero 
metallicity gas. (We note the lack of metal enrichment is a limitation of the simulations and its effect on the
cooling rate will likely be important at high redshift in particular.)

For redshifts, $z<10$ and densities, $n_{\rm H}<0.1\, {\rm cm}^{-3}$, a temperature floor of $10^4$K is imposed,
approximating the effect of heating from a UV background (although this has no effect on our cluster simulations). 
Above this density and at all redshifts, the gas is assumed to be a multi-phase mixture of cold molecular clouds, 
warm atomic gas and hot ionized bubbles, all approximately in pressure equilibrium. Following \cite{Schaye2008},
we model this using a polytropic equation of state
\begin{equation}
\label{eqnstate1_eqn}
P=A \, n_{\rm H}^{\gamma \rm eff},
\end{equation}
where  $P$ is the gas pressure, $A$ is a constant (set to ensure that $T=10^{4}$K at $n_{\rm H}=0.1\,{\rm cm}^{-3}$)
and $\gamma_{\rm eff}=4/3$, causing the Jeans mass to be independent of density \citep{Schaye2008}. 
Gas is allowed to leave the equation of state if its thermal energy increases by at least 0.5 dex, or if it is heated by 
a nearby supernova or AGN.

Each gas particle found on the equation of state is given a probability to form a star particle following the method
of \cite{Schaye2008}. This is designed to match the observed Kennicutt-Schmidt law for a disc whose thickness is approximately 
equal to the Jeans length (i.e. the gas is hydrostatically supported perpendicular to the disc plane). We assume a disc
gas mass fraction, $f_{\rm g}=1$,\footnote{While this is not true in practice, the star formation rate depends weakly on the
gas fraction, $\dot{\rho}_{*} \propto f_{\rm g}^{0.2}$, as discussed in \protect\cite{Schaye2008}} 
and a Salpeter IMF when calculating the star formation rate, which can be expressed as
\begin{equation}
\label{sfr1_eqn}
\dot{m}_{*}= 5.99 \times 10^{-10} \, {\rm M_{\odot} yr^{-1}} \, 
\left( \frac{m_{\rm gas}}{1 \, {\rm M}_{\odot}} \right)  \, 
\left( \frac{P/k}{10^3 \, {\rm cm}^{-3} {\rm K}} \right)^{0.2}.
\end{equation}
It thus follows that the estimated probability of a given gas particle forming a star, $p_{*}$, is given by
{\begin{equation}
\label{sfr2_eqn}
p_{*} = {\rm min}\left( \frac{\dot{m}_{*} \, \Delta t}{m_{{\rm gas}}},1\right),
\end{equation}
where $\Delta t$ is the current time-step of the particle.

\subsubsection{Supernova feedback}

Supernova feedback is an important mechanism for re-heating interstellar gas following star formation. In addition to this
effect (which is already accounted for in our equation of state, above), we also assume that supernovae produce galactic winds.
The method used here follows the prescription outlined in \cite{DallaVecchia2012}. The dominant contribution comes from the 
Type~II (core collapse) supernovae, which occur shortly after formation (up to $\sim$ 10 million years); for simplicity we neglect this 
short delay. The temperature to which a supernova event (associated with a newly formed star particle) can heat the surrounding gas particles, 
$T_{\rm SN}$, is calculated as 
\begin{equation}
\label{sfr2_eqn}
T_{\rm SN} = 2.65 \times 10^7 {\rm K}  \, \left( \frac{\epsilon _{\rm SN}}{N_{\rm SN}} \frac{m_{\rm star}}{m_{\rm gas}} \right), 
\end{equation}
where $\epsilon_{\rm SN}$ is the fraction of supernova energy available for heating, 
$N_{\rm SN}$ is the number of particles to be heated, and $m_{\rm star}$ is
the star particle mass (we set $m_{\rm star}=m_{\rm gas}$). When calculating this temperature
we have assumed that the total energy released per supernovae, $E_{\rm SN}=10^{51}{\rm erg}$.
For our main results (see below), we fix $T_{\rm SN}=10^{7}$K and $N_{\rm SN}=3$, implying an efficiency, $\epsilon_{\rm SN} \simeq 1.1$
for a Salpeter IMF (or $\epsilon_{\rm SN} \simeq 0.7$ for a Chabrier IMF, which predicts relatively more high-mass stars). We discuss
variations in the heating parameters below.

\subsubsection{Black hole growth and AGN feedback}

Black holes are usually included as collisionless sink particles within cosmological simulations, with an initial {\rm seed} placed in every 
Friends-of-Friends group that is newly resolved by the simulation. This requires the group finder to be run on-the-fly; our code is
currently unable to perform this task, instead we place our seed black holes at a fixed (high) redshift. Specifically, we take
the snapshot at redshift, $z_{\rm ini}$ from our SFB model and find all sub-haloes with mass, $M>M_{\rm sub}$, replacing
the most bound (gas or star) particle with a black hole particle (leaving the particle mass, position and velocity unchanged). 
For our default AGN model, we assume $z_{\rm ini}=5.2$ and set $M_{\rm sub}$ to a value that is approximately equal to the 
mass of 50 DM particles. Tests revealed the final hot gas and stellar distributions to be insensitive to the choice of these parameters. 
This is because most of the AGN feedback originates from the central black hole, which gets most of its mass from accretion in the 
cluster at much lower redshift ($z<2$; see Fig.~\ref{plot:res_bh} in Section~\ref{sec:res}).

Black hole accretion and AGN feedback rates are modelled via the \cite{Booth2009} method, based on the original approach
by \cite{SpringelDMH2005}. Black holes grow both via accretion of the surrounding gas and mergers with other black holes. Since 
discreteness effects are severe for all but the most massive black holes, a second {\it internal}
mass variable is tracked to ensure the accretion of the gas onto the central black hole can be modelled smoothly. We give each
black hole an initial internal mass of $10^{5} \, h^{-1}{\rm M}_{\odot}$. All local properties are then estimated by adopting the SPH method
for each black hole particle. A smoothing length is determined adaptively by enclosing a fixed number of neighbours, but it cannot
go lower than the gravitational softening scale. In practice, smoothing lengths for central black holes are nearly always set to this
minimum value which limits the estimate of the local gas density.

Accretion occurs at a rate set by the minimum of the Bondi-Hoyle-Lyttleton \citep{Hoyle1939} and Eddington values
\begin{equation}
\label{bhac_eqn}
\dot{M}_{\rm BH}= {\rm min} \, \left[ \alpha \frac{4 \pi G^2 M^2_{\rm BH} \rho_{\rm gas}}{ (c^2_{\rm s} + v^2)^{3/2}},
{4 \pi G m_{\rm H} M_{\rm BH} \over \epsilon_{\rm r} \sigma_{\rm T} c} \right],
\end{equation}
where $M_{\rm BH}$ is the internal black hole mass, $\epsilon_{\rm r}$ the efficiency of mass-energy conversion, 
$\rho_{\rm gas}$ the local gas density, $c_{\rm s}$  the sound-speed 
and $v$ the relative velocity of the black hole with respect to the gas it inhabits. The value of $\alpha$ is calculated following
\cite{Booth2009}, as 
\begin{equation}
\alpha = {\rm max} \left[ \left( {n_{\rm H} \over 0.1 {\rm cm}^{-3}}\right)^{2},1 \right],
\label{eqn:BSalpha}
\end{equation}
which attempts to correct for the mismatch in scales between where the gas properties are estimated and where the accretion would
actually be going on.\footnote{Note in this method, the accretion rate is a strong function of gas density when sub-Eddington, 
$\dot{M}_{\rm BH} \propto \rho_{\rm gas}^3$.}
If the internal mass exceeds the particle mass (set initially to $m_{\rm gas}$), 
neighbouring gas particles are removed from the simulation at the appropriate rate. Black holes may also grow via mergers with 
other black holes, when the least massive object comes within the smoothing radius of the more massive object and the two are
gravitationally bound.\footnote{We note that black hole particle mass is conserved in our simulations, thus when many mergers 
occur at high redshift, the
mass of a black hole particle can significantly exceed its internal mass.}
The latter is irrelevant in practice as we force the position of a black hole to be at the local potential minimum. This leads to
some over-merging of black holes but avoids spurious scattering, causing the accretion (and therefore feedback) rate to be
severely underestimated.

For the AGN feedback, the heating rate is assumed to scale with the accretion rate as
\begin{equation}
  \label{bhac3_eqn}
  \dot{E}_{\rm AGN} = \epsilon_{\rm f} \epsilon_{\rm r} \dot{M}_{\rm BH} c^2,
\end{equation}
where $\epsilon_{\rm f}$ is the efficiency with which the energy couples to the gas. 
For our default models we set both efficiency parameters to the values used in 
\cite{Booth2009}, namely $\epsilon_{\rm f}=0.15$ and $\epsilon_{\rm r}=0.1$.

Due to the limitations in resolution it is unclear what the best method for distributing the energy is. In order to create outflows 
the surrounding gas needs to be given enough energy to rise out of the potential well, before it is able to radiate it away. In 
order to achieve this, an amount of feedback energy, $E_{\rm crit}$ is stored until there is enough to heat at least $N_{\rm AGN}$ 
neighbouring gas particles to a temperature $T_{ \rm AGN}$, i.e.
\begin{equation}
E_{\rm crit} = N_{\rm AGN} \, m_{\rm gas} \, {3 \over 2} {kT_{\rm AGN} \over \mu m_{\rm H}}, 
\end{equation}
where $\mu=0.59$ is the mean atomic weight for an ionised gas with primordial ($X=0.76, Z=0$) composition.
In our default AGN model we set $N_{\rm AGN}=1$ (i.e. heat a minimum of one particle at a time) but vary 
$T_{\rm AGN}$ in proportion to the final virial temperature of the cluster (from $T_{\rm AGN}=10^{8}$K
in the lowest mass objects to $10^{8.5}$K at the highest mass (further details are given below).

\subsection{Calculation of cluster properties}

For our main results, we focus on the radial distribution of observable cluster properties (profiles)
and the scaling of integrated properties with mass (cluster scaling relations). 
Unless specified, we measure all properties within a radius $r_{500}$, as this is the most common scale 
used for the observational data. Details of how we calculate these properties are provided 
in Appendix~\ref{app:measure}; we also summarise the observational data that we compare our results with 
in Appendix~\ref{app:obs}.

An issue that we report here is the large discrepancy between spectroscopic-like temperature $T_{\rm sl}$ (a proxy for X-ray
temperature; \citealt{Mazzotta2004}) 
and mass-weighted temperature  $T_{\rm m}$ (more relevant for SZ observations). The former was found to be
significantly lower (and noisier) than the latter in our simulations. For the AGN model, the ratio between the two
temperatures at $z=0$ varies from $T_{\rm sl}/T_{\rm m}=0.6-0.7$ in low-mass clusters, decreasing to $0.3-0.4$ in high-mass clusters. It 
is particularly problematic for the most massive clusters, where the virial temperature is significantly higher than the cut-off 
temperature for calculating $T_{\rm sl}$ (0.5 keV). 

The origin of this discrepancy is two-fold. 
Firstly, large, X-ray bright substructures may contain gas that is sufficiently cold and dense to produce a significant 
bias in the spectroscopic-like temperature. 
Such substructure would normally be masked out of X-ray images (e.g. \citealt{Nagai2007X}). Secondly, even 
when there are no large DM substructures present, the clusters contain a small amount of cool ($\sim 1$ keV), 
dense gas.  It is likely that this material is spurious, caused by the failure of SPH to mix stripped, low entropy
gas with the hot cluster atmosphere. This requires further investigation so we leave this to future work (but comment
on its dependence on resolution, in Section~\ref{sec:res}). In the meantime, we 
remove the spurious gas following the method suggested by \cite{Roncarelli2013}. 
In this method, discussed further in Appendix~\ref{app:measure}, a small amount of gas with the highest density is
excluded from the temperature calculation. In practice, this method also removes the densest, X-ray bright gas
in substructures. The outcome is that the X-ray temperatures are much closer to the mass-weighted temperatures for 
our clusters, so long as the central region is excluded.

\begin{figure*}
\includegraphics[width=80mm]{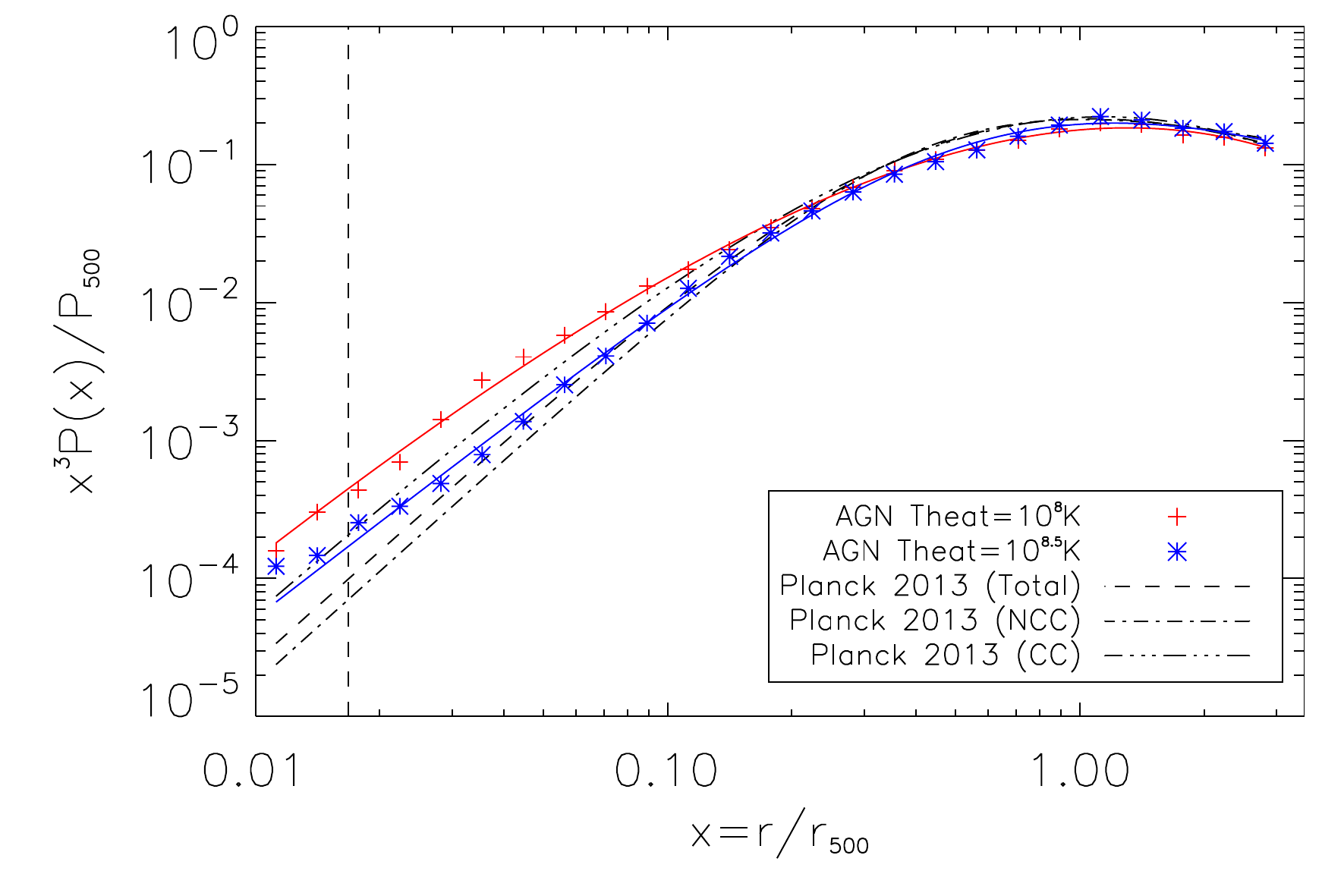}
\includegraphics[width=80mm]{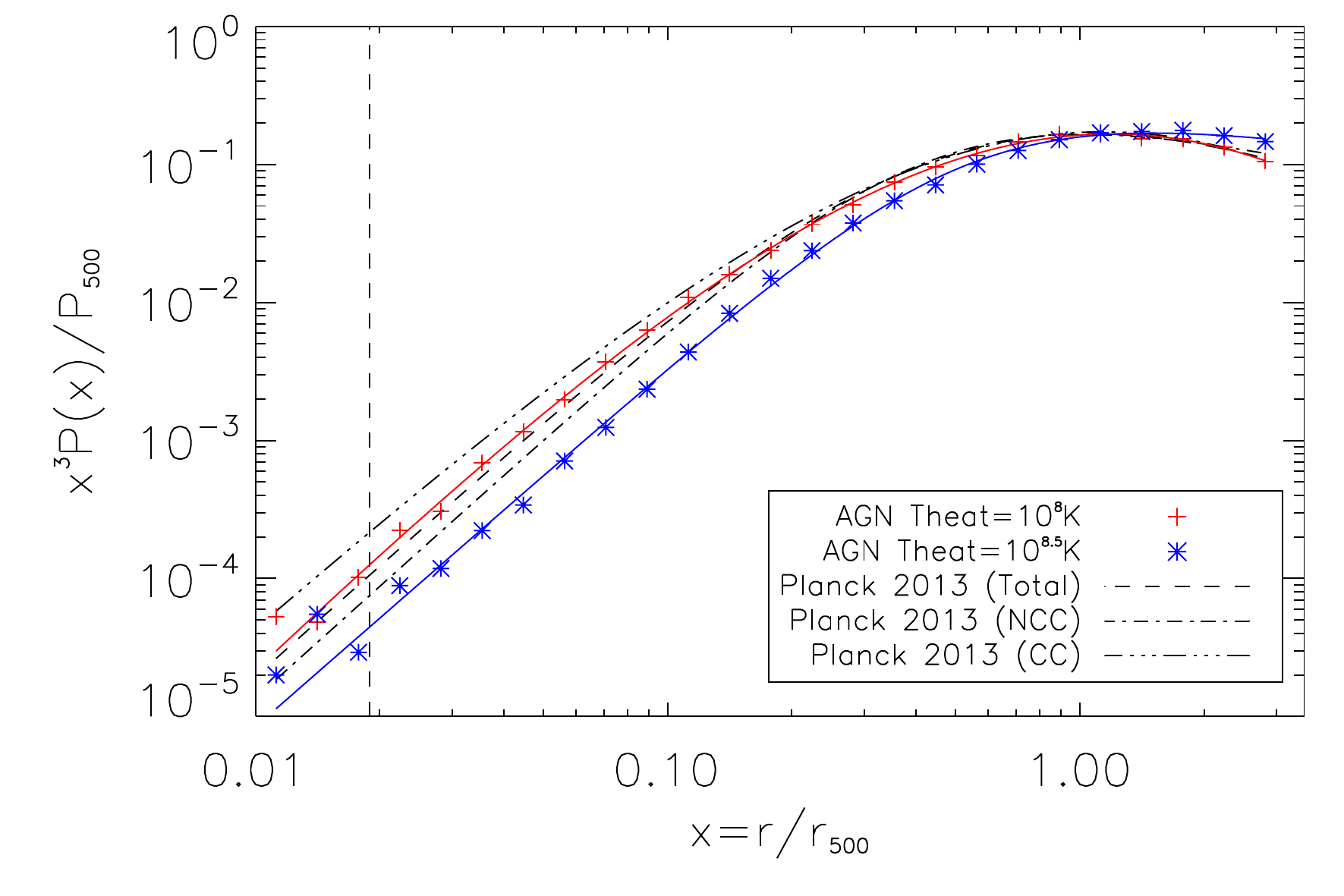}
\includegraphics[width=80mm]{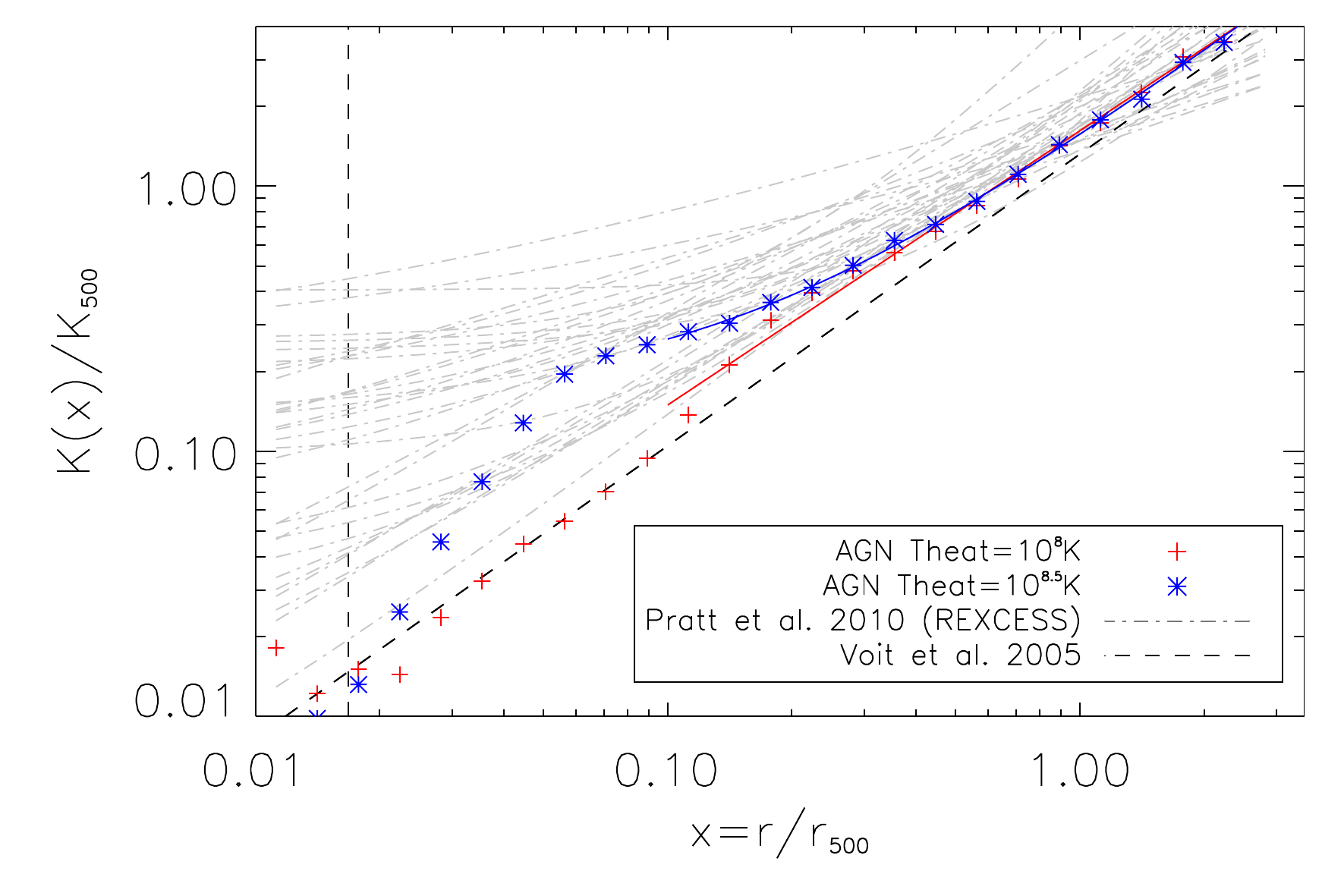}
\includegraphics[width=80mm]{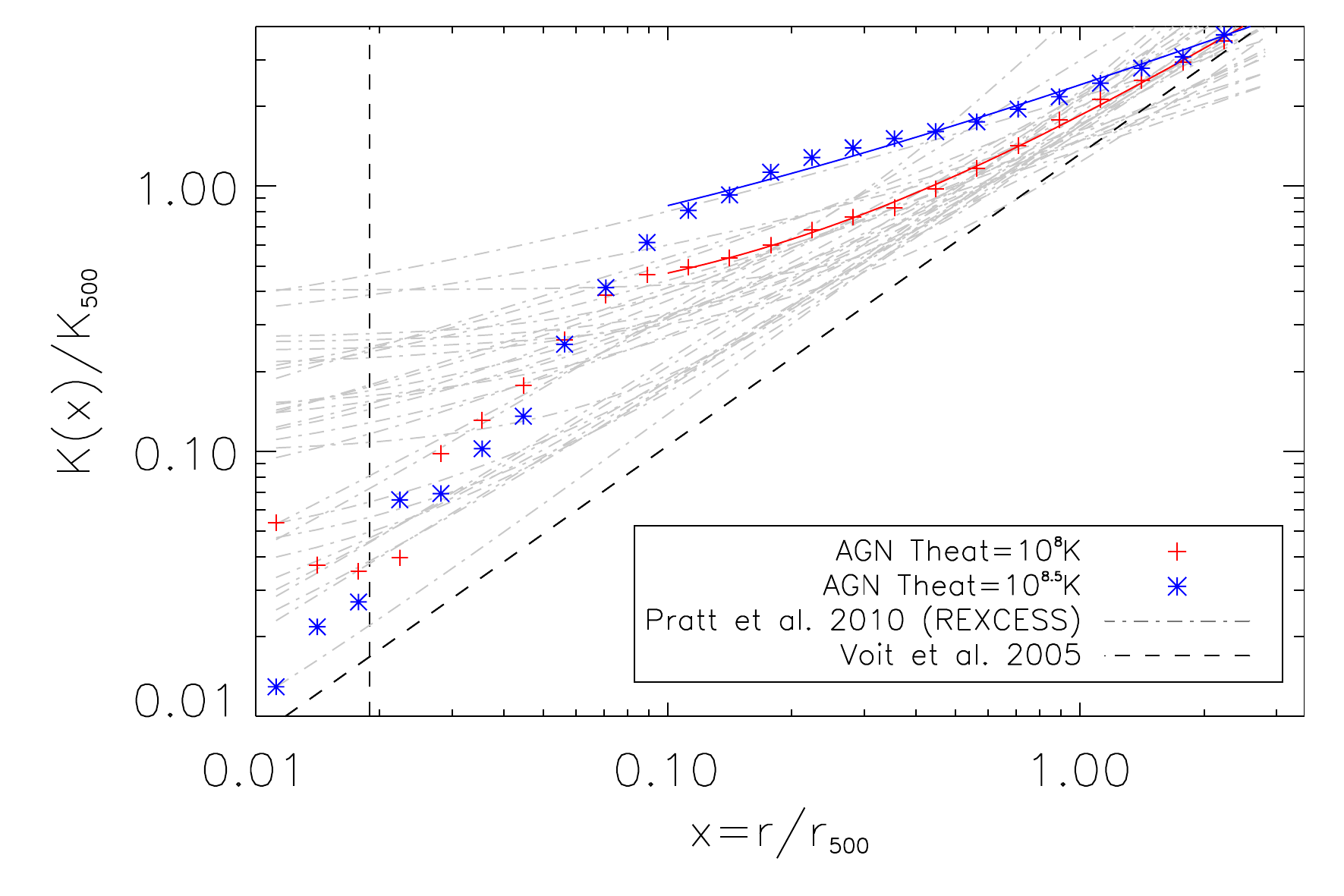}
\caption{Scaled pressure (top) and entropy (bottom) profiles at $z=0$, for the most massive 
($M_{\rm 200} \simeq 10^{15} h^{-1} {\rm M}_{\odot}$; left) and one of the least-massive 
($M_{\rm 200} \simeq 10^{14} h^{-1} {\rm M}_{\odot}$; right) clusters. Results shown in red 
and blue are for runs where $T_{ \rm AGN}$ is set to $10^8$K and $10^{8.5}$K respectively. 
The pressure profiles are compared to the best-fitting observed profiles from \protect\cite{Planck2013}, 
scaled for a cluster with the same mass. In the bottom panels, the grey dot-dashed curves are fits to 
REXCESS entropy profiles from \protect\cite{Pratt2010}, while the black dashed line is the fit to 
non-radiative clusters in \protect\cite{Voit2005}. The vertical dashed line in both panels indicates the 
gravitational softening radius (2.8 times the equivalent Plummer softening length) where the two-body force
 deviates from an inverse square law.}
\label{plot:tagn_prof}
\end{figure*}

\subsection{Choice of feedback parameters}
\label{sec:heattemp}

The physics of supernova and AGN feedback occur on scales much smaller than are resolvable, so it is unclear how the parameters 
which govern the amount and manner in which energy is released in a feedback event should be chosen. In order to make this choice,
the feedback parameters,  $[N_{\rm SN}, T_{ \rm SN}, N_{\rm AGN}, T_{\rm AGN},  \epsilon_{\rm f}]$, were varied over a limited range 
and their effects on the scaling relations and profiles compared. 

The supernova feedback parameters ($T_{ \rm SN}$ and $N_{ \rm SN}$) were varied with the primary intention of matching the cluster 
gas fractions. As we will see in the next section, supernovae play a particularly important role in keeping most of the cluster baryons 
in the gas phase. Our default choice of $T_{ \rm SN}=10^7$K and $N_{ \rm SN}=3$ (also including AGN feedback) produces gas and star 
fractions that are similar to those observed. Lowering the heating temperature (which corresponds to a lower overall amount of 
available energy for constant $N_{\rm SN}$, or equivalently the same amount of energy distributed over more particles) results in 
larger star fractions and lower gas fractions as more gas cools before having a chance to escape from dense regions. 

Regarding the AGN feedback parameters, it was found that varying $N_{ \rm AGN}$  by an order of magnitude 
and $\epsilon_{\rm f}$ by a factor of three had little effect on the cluster properties. We therefore chose to set $N_{\rm AGN}=1$, 
minimising the period over which energy is stored. 
When the efficiency is lowered, the accretion rate increases until the amount of heating is able to shut it off. As a result, the amount of 
energy produced by the black hole is similar but the black hole mass can be very different. For our work, we chose to keep the 
default value of $\epsilon_{\rm f}=0.15$ \citep{Booth2009}, which as we will show, leads to reasonable black hole masses.

The most significant parameter affecting the cluster gas is the AGN heating temperature, $T_{ \rm AGN}$. This is highlighted in 
Fig.~\ref{plot:tagn_prof}, where we show scaled pressure (top panels) and entropy (lower panels) profiles for our most massive
cluster (left panels) and one of our lowest mass clusters (right panels) at $z=0$. Within each panel, we show results from two runs,
one where we set $T_{\rm AGN}=10^{8}$ K (red curve) and one with $T_{\rm AGN}=10^{8.5}$ K (blue curve). 
We also show observational data; in the case of the pressure profiles we show the best-fitting generalised Navarro, Frenk \& White 
(GNFW) models from \citet{Planck2013}, scaled to the appropriate cluster mass (see Section~\ref{subsec:preprof}). For the 
entropy profiles, we compare with fits to the REXCESS X-ray data \citep{Pratt2010}.

In the larger mass halo, it is clear that to match the observed pressure profiles in the central region, 
heating to $T_{ \rm AGN}=10^{8.5}$K is required; a lower temperature leads to the central region 
being over-pressured. It is also apparent from the entropy profiles that this higher heating temperature 
is a better match to the observational data outside the inner core ($r>0.05\,r_{500}$). The importance 
of the heating temperature can be understood by the fact that heating the gas to a higher temperature 
allows it to rise further out of the central potential (because the gas will also have higher entropy) and 
lowers the rate at which its thermal energy is lost to radiative cooling (because the cooling time scales 
as $\sqrt{T}$ for thermal bremsstrahlung). In the case where the gas is heated to $T_{ \rm AGN}=10^{8}$K, 
the pressure is too high in the central region because the heating is less able to expel gas from the central 
region, resulting in a denser core. 

Looking at the results for the lower mass cluster, it is perhaps unsurprising that setting 
$T_{ \rm AGN}=10^{8.5}$K is excessive, creating a pressure profile that is below the observational data 
and an entropy profile that is too high. Instead,  $T_{ \rm AGN}=10^{8}$K gives much better results, more 
similar to the profile for the higher mass halo. Given the order of magnitude range in cluster masses, these 
results suggest that an appropriate heating temperature is that which scales with the virial temperature of the halo 
(since $T_{\rm vir} \propto M^{2/3}$). We therefore choose to scale $T_{\rm AGN}$ in this way, for all clusters 
in our sample. Specifically, we use the central mass within each bin and use the above values for the two extremes. 

It is unclear whether there is any physical basis for this choice of temperature scaling. It may be that the specific 
energy in AGN outflows is somehow intimately connected to the properties of the black hole (i.e. its mass and/or spin),
given that its mass is predicted to be determined by the mass of the dark matter halo \citep{Booth2010}. 
However, the scaling may also be effectively correcting for the effects of limited numerical resolution, and/or 
the heating method itself. In the former case, it may 
be that higher resolution simulations allow the interaction of gas in different phases to be
resolved in more detail, which somehow leads to more effective outflows in higher-mass clusters (where the cooling
time is longer). Alternatively, it may be that if the gas were heated in confined regions (e.g. bubbles), 
this could naturally produce concentrations of higher entropy gas in higher-mass clusters. 
What is clear, is that such fine tuning of the feedback model is still not sufficient to reproduce the entropy profile at all radii
(the inner region in particular) although this does improve at higher resolution, as we will show later. 
Furthermore, the heating temperature could potentially play a role in generating 
scatter in the entropy profile. We will return to this in Section~\ref{sec:profiles}.

\section{Cluster Baryons}
\label{sec:baryons}

We now present results for our full sample of 30 clusters, run with our 4 {\it physics} models (NR, CSF, SFB \& AGN).
In this section, we assess the general validity of our AGN model by investigating the overall distribution of cluster 
baryons. Furthermore, by comparing the different models, we can approximately measure the contribution 
from individual physical processes (cooling and star formation, supernovae and AGN). We start by comparing the baryon, 
gas and star fractions with observational data at $z=0$, before going on to investigate the star formation histories and 
black hole masses.

\subsection{Baryon, gas and star fractions}

\begin{figure}
  \includegraphics[width=80mm]{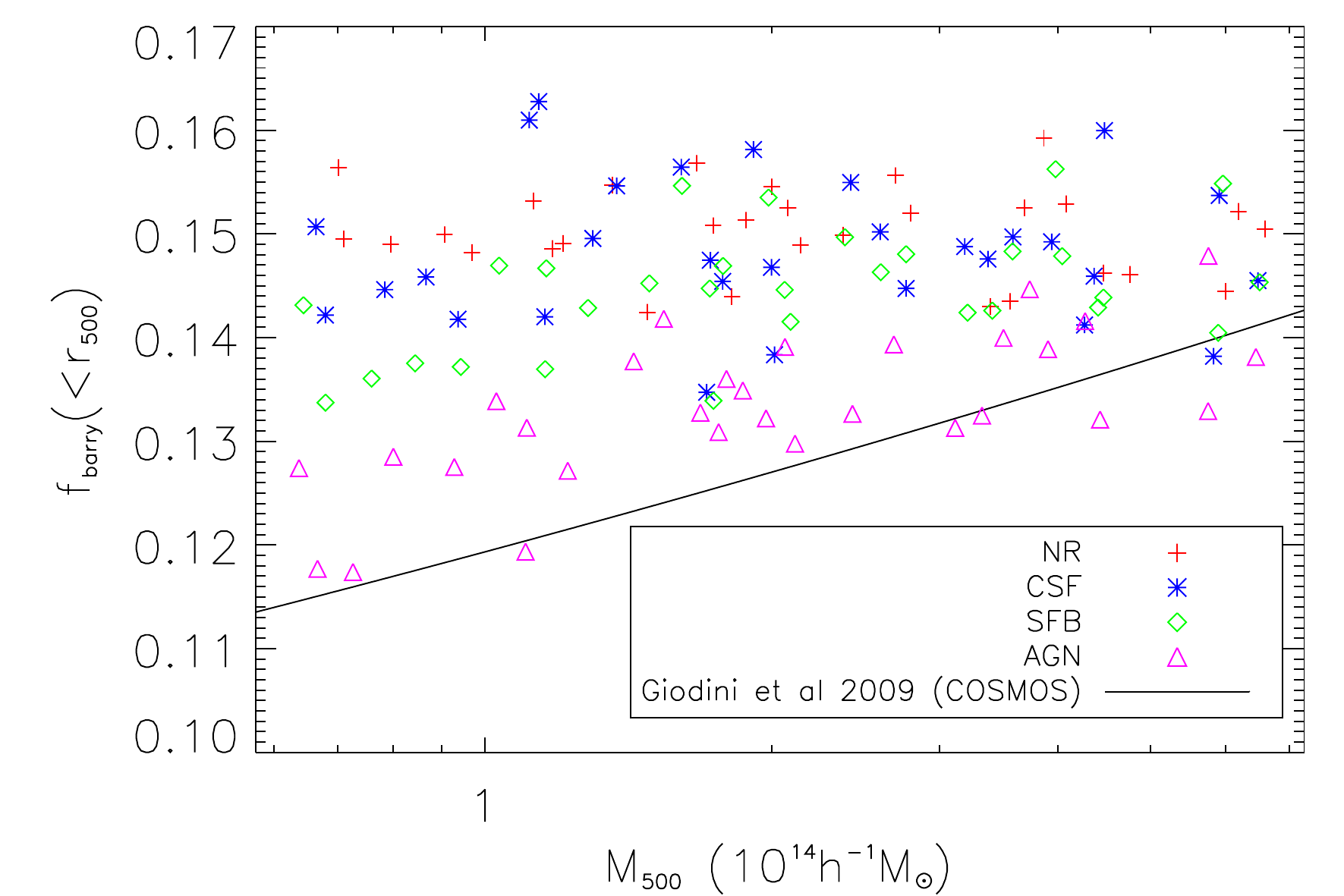}
  \includegraphics[width=80mm]{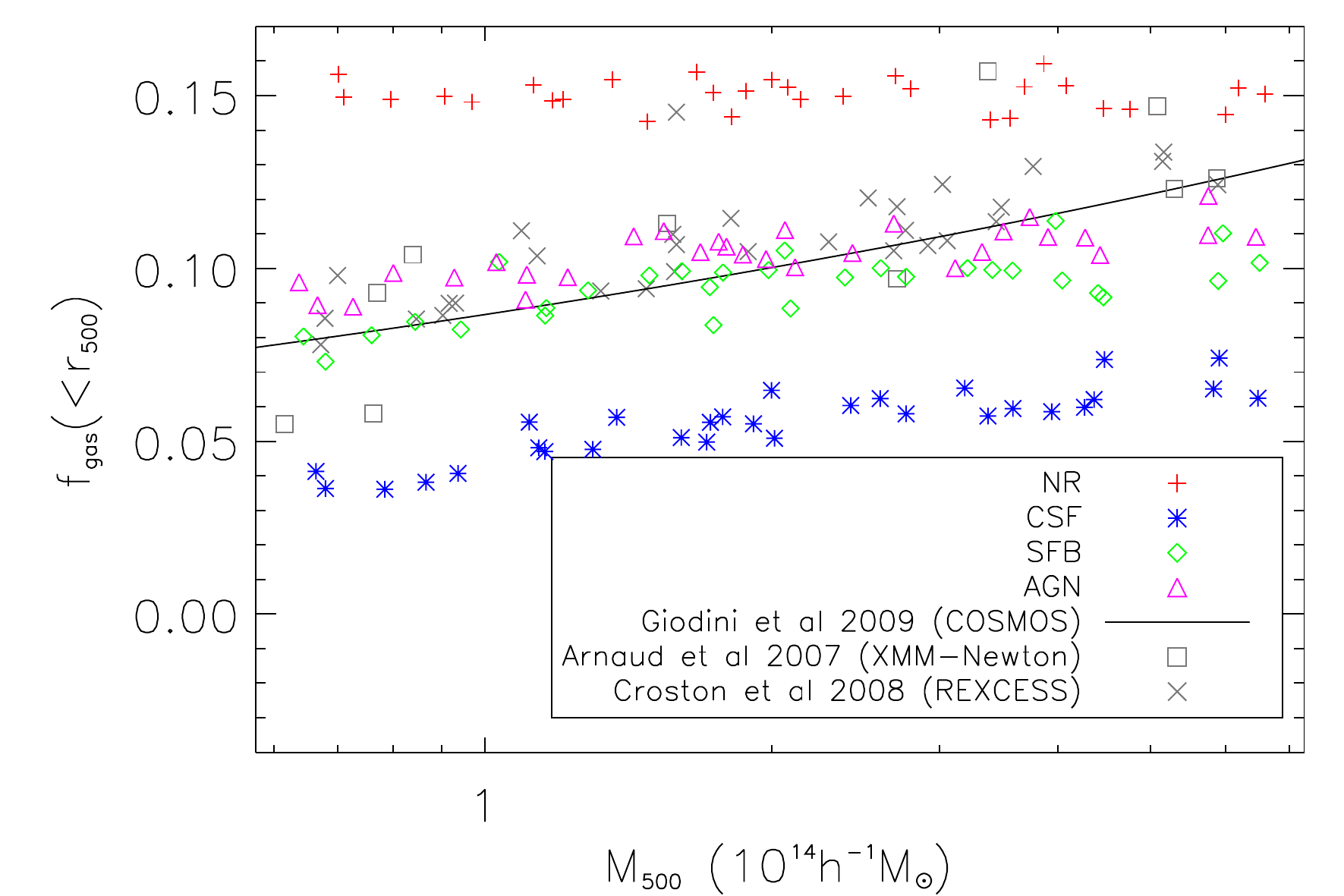}
  \includegraphics[width=80mm]{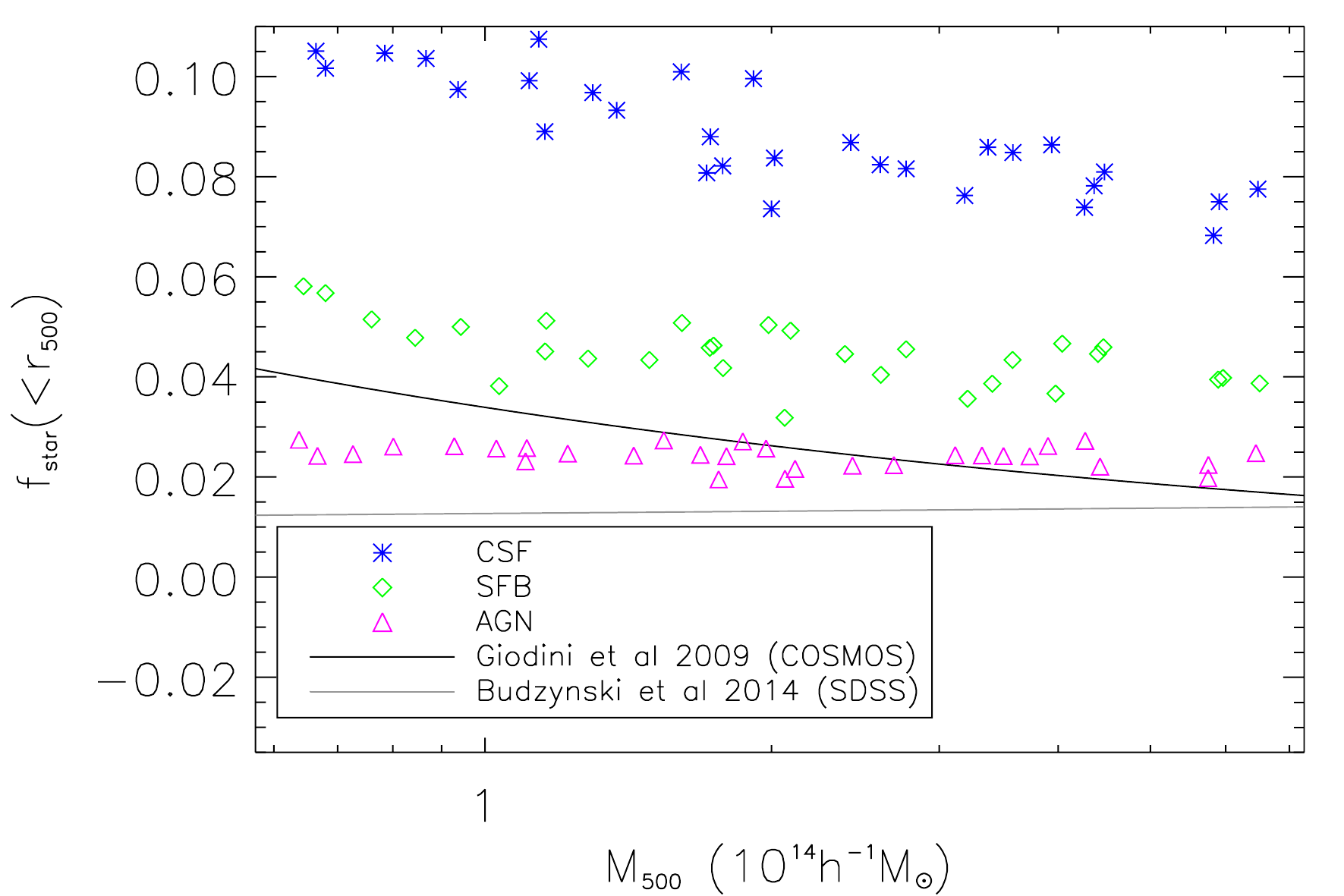}
 \caption{Baryon (top panel), hot gas (middle) and star (bottom) fractions versus halo mass for the four {\it physics} 
 models at $z=0$. In all panels, the solid black curve is the best-fit observed relation from the COSMOS survey 
 \protect\citep{Giodini2009}. We additionally show observed gas fractions from {\it XMM-Newton} data 
 \protect\citep{Arnaud2007,Croston2008} in the middle panel and the best-fit relation between star fraction and
 halo mass from SDSS data \protect\citep{Budzynski2014} in the lower panel.}
\label{plot:bc}
\end{figure}

Baryon, gas and stellar fractions, within $r_{\rm 500}$, are shown versus mass for our four simulation sets at $z=0$, in 
Fig~\ref{plot:bc}. We also compare our results with observational data, as detailed in the legends and caption 
(see also Appendix~\ref{app:obs}).

The baryon fractions (top panel) are similar for the NR and CSF runs and show no dependence on mass. The mean 
baryon fraction is around 90 per cent of the cosmological value ($\Omega_{\rm b}/\Omega_{\rm m}=0.15$), similar to 
previous work (e.g. \citealt{Crain2007}).  Both the SFB and AGN models show more significant (and mass dependent) 
depletion, with the AGN model producing values that are closer to the observations. This is due to the feedback expelling 
some gas from within $r_{\rm 500}$ and being more effective at doing so within smaller clusters, which have shallower 
potential wells.

The middle panel of Fig~\ref{plot:bc} displays the hot gas fractions. As expected, the NR results are too high (because 
radiative cooling is neglected), whilst the CSF values are too low. It is well known that simulations without feedback 
suffer from the {\it over-cooling} problem, where too much gas is converted into stars (e.g. \citealt{Balogh2001}). 
Interestingly, the SFB and AGN runs have similar gas fractions, both which closely match the observations, with 
the AGN result having a slightly higher gas fraction. Clearly, the supernova feedback is strong enough by itself to 
suppress the cooling and star formation in cluster galaxies by about the right amount. As mentioned in the previous section, 
we tuned the feedback parameters to achieve this result; less effective feedback (e.g. by heating fewer gas particles, 
or using a lower heating temperature) would result in lower gas fractions. The AGN feedback additionally affects the gas 
in two competing ways. Firstly, as discussed above, it heats the gas more, making it hotter and ejecting some of it 
beyond $r_{500}$. Secondly, as the gas is less dense and warmer around the black hole particles, star formation is 
reduced. These two effects partly cancel each other out, with the decreased star formation rate being the slightly 
stronger effect, resulting in slightly higher gas fractions in the AGN runs.

Finally, star fractions are presented in the bottom panels of Fig.~\ref{plot:bc} (the NR results are omitted as these runs 
do not produce any stars).  Again, the CSF runs fail due to over-cooling, producing star fractions of order 10 per cent. 
Supernova feedback reduces the fractions by around a factor of two, but still fail to match the observations. (Recall that 
we are already close to maximal heating efficiency for a Salpeter IMF; including metal enrichment would likely make the 
situation even worse). Only when the AGN are included does the star fraction fall to the more reasonable level of 2-3 per cent.
The reason for this will now become evident, when we analyse the cluster star formation histories in more detail.

\subsection{Formation history and distribution of stars}
\label{sec:sfh}

\begin{figure*}
  \includegraphics[width=55mm]{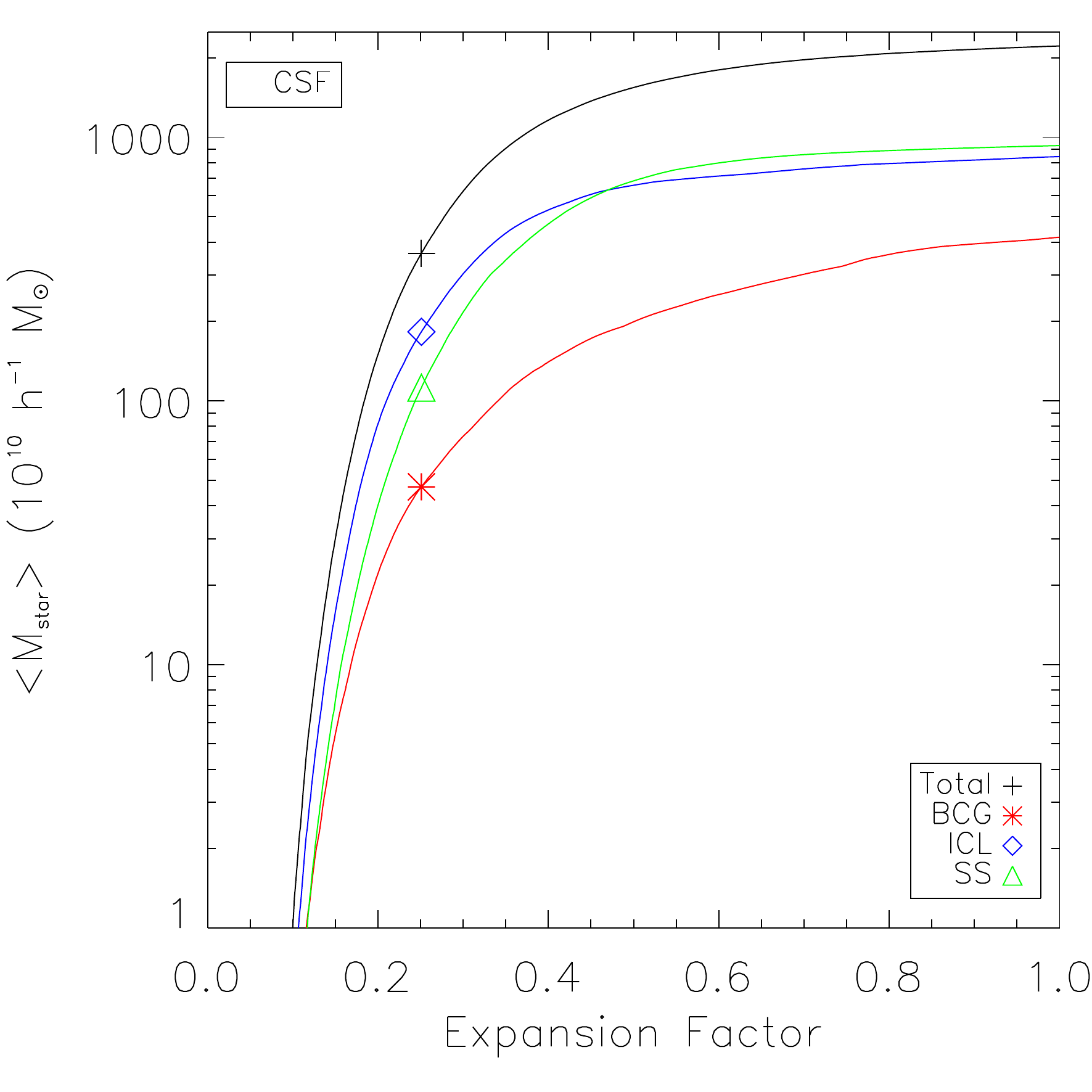}
  \includegraphics[width=55mm]{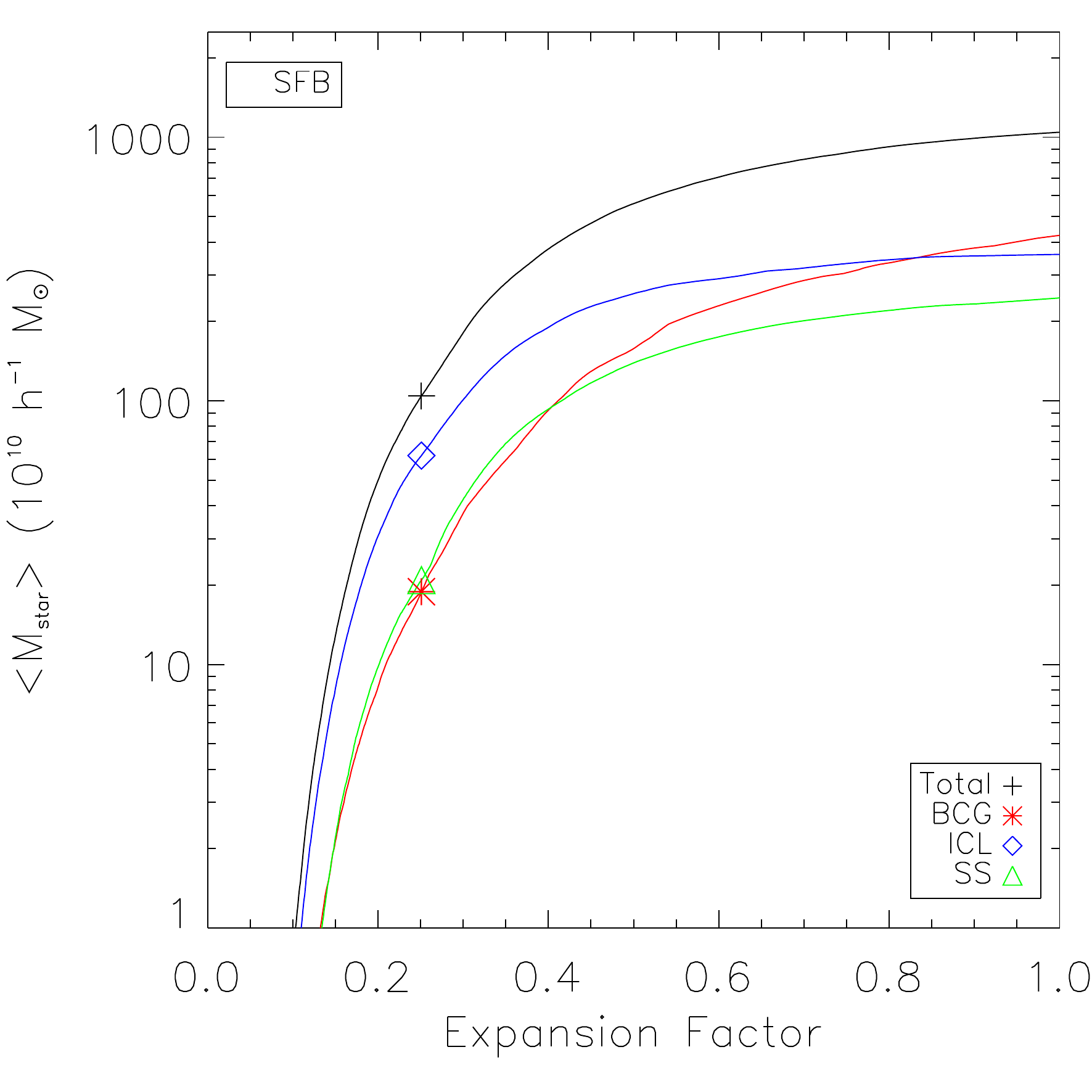}
  \includegraphics[width=55mm]{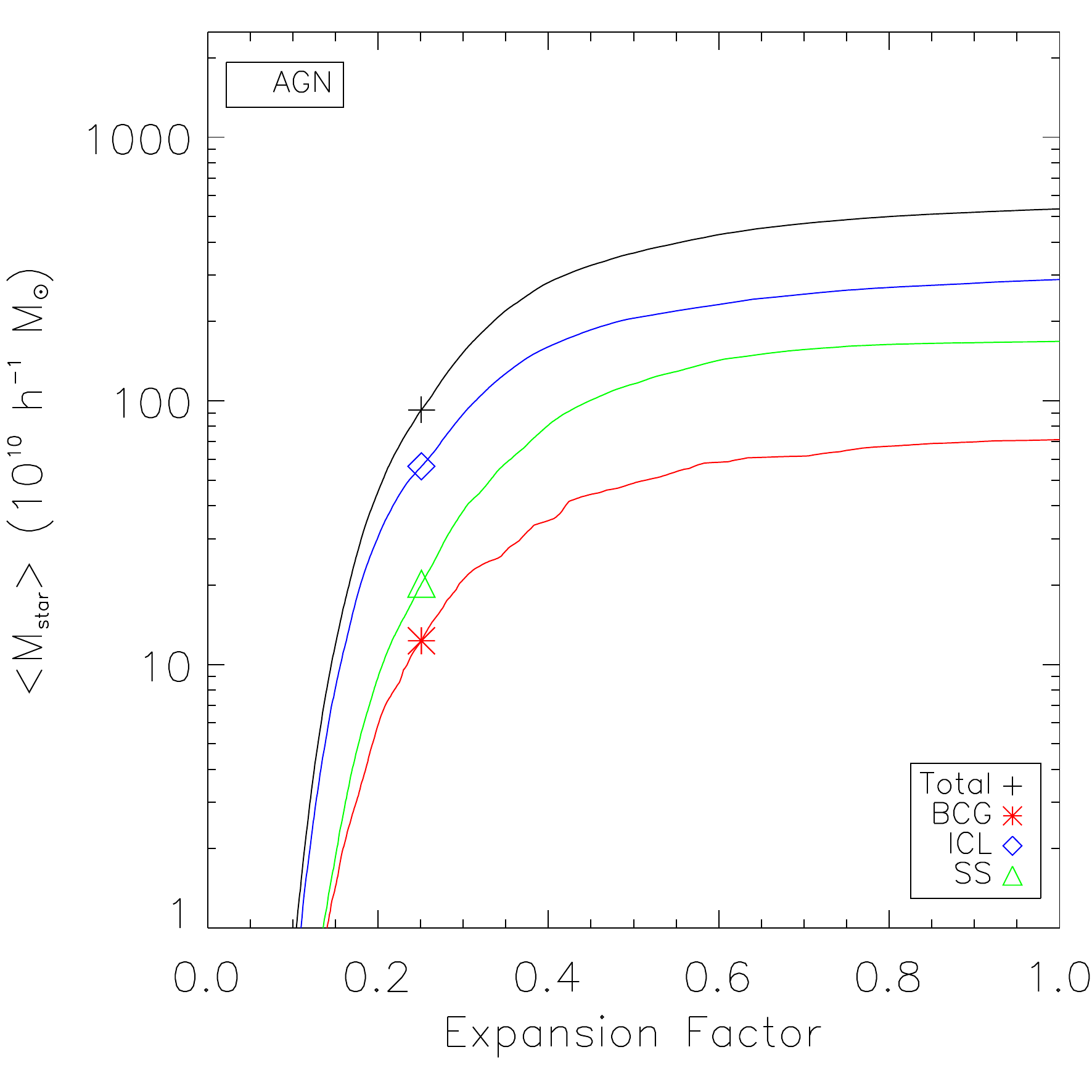}
\caption{Star formation histories of particles that are within $r_{200}$ at $a=1$ (black curves), for the
CSF (left), SFB (middle) and AGN (right panel) runs. Specifically, the cumulative fraction of stellar mass
formed by a given value of $a$ is calculated for each cluster and the median curve is shown, re-scaled to the median
final stellar mass. The stars are also sub-divided into where they end up: red curves represent those 
ending up in the central BCG; blue for those in the ICL; and green for those in substructures (galaxies).}
\label{plot:sfhist}
\end{figure*}

We now study how the cluster star formation rates are affected by supernova and AGN feedback, by considering
the formation times of the stars present in the cluster at $z=0$. For each object, we identify all star particles within
$r_{200}$ and associate them with one of three components: the brightest cluster galaxy (BCG); the intracluster light (ICL)
and cluster substructure (SS). For the latter, which we take as a proxy for the cluster galaxies, we identify all star particles
belonging to subgroups (as found by SUBFIND) other than the most massive one (i.e. the cluster itself). For the BCG and ICL,
we take stars belonging to the most massive subgroup and split them according to their distance from the centre as in
\cite{Puchwein2010}, who set this demarcation distance to be
\begin{equation}
  \label{eqn:rcut}
  r_{\rm cut}=27.3 \, \left( {M_{200} \over 10^{15} h^{-1} {\rm M}_{\odot} } \right)^{0.29} \, h^{-1}{\rm kpc}.
\end{equation}
Thus, all stars with $r<r_{\rm cut}$ are assumed to belong to the BCG. While this is a fairly crude method (e.g. one that is more
consistent with observations would be to use a surface brightness threshold; e.g. \citealt{Burke2012}), it nevertheless allows us to assess the effect
of feedback in the central region versus the rest of the cluster. 

\begin{table}
\centering
  \caption{Comparison of stellar masses formed by $a=1$ and $a=0.5$, between the CSF, SFB and AGN models, 
  for each of the components that they end up in at $a=1$.}
  \begin{tabular}{@{}lcccc@{}}
  \hline   
  & $M_{\rm SFB}/M_{\rm CSF}$ & $M_{\rm AGN}/M_{\rm CSF}$ & $M_{\rm AGN}/M_{\rm SFB}$   \\
\hline  
$a=1$ &&&\\						
Total  & $~0.47$ & $~0.24$ & $~0.50$  \\    
BCG  & $~0.98$ & $~0.16$ & $~0.17$ \\
ICL    & $~0.45$ & $~0.36$ & $~0.80$  \\ 
SS     & $~0.26$ & $~0.18$ & $~0.68$ \\ 
&&&\\
$a=0.5$ &&&\\
Total  & $~0.47$ & $~0.25$ & $~0.53$  \\
BCG    & $~0.79$ & $~0.19$ & $~0.24$  \\
ICL    & $~0.34$ & $~0.32$ & $~0.92$  \\
SS     & $~0.30$ & $~0.26$ & $~0.87$  \\
\hline   
\end{tabular}
\label{tab:sfhist}
\end{table}

In Fig.~\ref{plot:sfhist} we show the stellar mass formed at a given value of $a$, for stars that end up in each of the three components at
$a=1$, as well as the total stellar mass. From left to right, results are shown for the CSF, SFB and AGN models respectively. To
account for cluster-to-cluster variation, we compute the cumulative star fraction for each object individually, then present the 
median curve for the whole sample, multiplied by the median mass at $a=1$. Ratios of median stellar masses between pairs of runs are 
summarised in Table~\ref{tab:sfhist}, for both $a=0.5$ and $a=1$.

In the CSF runs, more than half the stars have already formed by $a=0.4$ ($z=1.5$). Stars in the galaxies (SS) and ICL also tend to have 
earlier formation times than in the BCG, which continues to form stars steadily until the present, due to the continual accretion of cool gas 
on to the centre of the cluster. The stellar mass in the BCG is largely unaffected by the introduction of supernova feedback; most of the 
reduction in stellar mass comes from its effect on the galaxies and ICL. As the ICL is largely stripped material from 
SS \citep{Puchwein2010}, this is not unexpected. When AGN feedback is included, the largest effect is on the stellar mass of the 
BCG. Again, this is not surprising as the central black hole is significantly more massive than the others and thus provides most of the 
heating. This is largely why the AGN clusters have lower star fractions than in the SFB model. 

 \begin{table}
\centering
  \caption{Median fraction of stars within the BCG, ICL and SS at $a=1$, where $M_{\rm T}=M_{\rm BCG}+M_{\rm ICL}+M_{\rm SS}$. 
  The final column lists the fraction of stars in the main subgroup belonging to the ICL, where 
  $M_{\rm SG0}=M_{\rm BCG}+M_{\rm ICL}$.}
\begin{tabular}{@{}lcccc@{}}
\hline   
& $M_{\rm BCG}/M_{\rm T}$ & $M_{\rm ICL}/M_{\rm T}$ & $M_{\rm SS}/M_{\rm T}$ & $M_{\rm ICL }/M_{\rm SG0}$ \\
\hline
CSF & 0.20         & 0.34           & 0.45         &  0.62   \\
SFB & 0.42        & 0.32           & 0.25         &  0.43   \\
AGN & 0.14        & 0.51           & 0.34         &  0.78   \\
\hline   
\end{tabular}
\label{tab:stelfrac}
\end{table}

The median fraction of stars within each component at $a=1$ is shown explicitly in Table~\ref{tab:stelfrac}. When feedback is 
absent (CSF model), nearly half the stars are in satellite galaxies. In the SFB runs, the BCG becomes the largest component as the 
supernova feedback affects the lower-mass haloes. Finally, in the AGN model, the reduction in the BCG mass leads to half 
of the stars now being in the ICL. Our AGN results compare favourably with \cite{Puchwein2010}, who found that $\sim$50 
per cent of stars were in the ICL and $\sim$10 per cent in the BCG. These results appear to be at odds with some 
observations of the ICL in clusters (e.g. \citealt{Gonzalez2005,Gonzalez2007}), which tend to find significantly 
lower fractions. However, more recent work by \cite{Budzynski2014} suggest that the ICL can contribute as much as 40 per cent
to the total stellar mass in clusters.

Another issue of current observational interest is the rate at which the BCG grows, primarily due to the recent availability of 
data for clusters beyond $z=1$. \cite{Lidman2012} find that the mass of BCGs increases by a factor of $1.8\pm 0.3$ between 
$z=0.9$ and $z=0.2$. This is somewhat at odds with the results of \cite{Stott2011}, who found that the BCG stellar masses were 
unchanged at high redshift. By comparing the BCGs in our most massive progenitors at $z=1$ with our results at $z=0$, we find 
that the BCG grows by around a factor of 5, on average, in our AGN model (mainly by dry mergers). This is significantly higher 
than the observations, even when sample selection is accounted for \citep{Lidman2012}, and requires further investigation.
One explanation for these discrepancies is numerical resolution; we will discuss this possibility further in Section~\ref{sec:res}.

\subsection{Black hole masses}

\begin{figure}
\includegraphics[width=85mm]{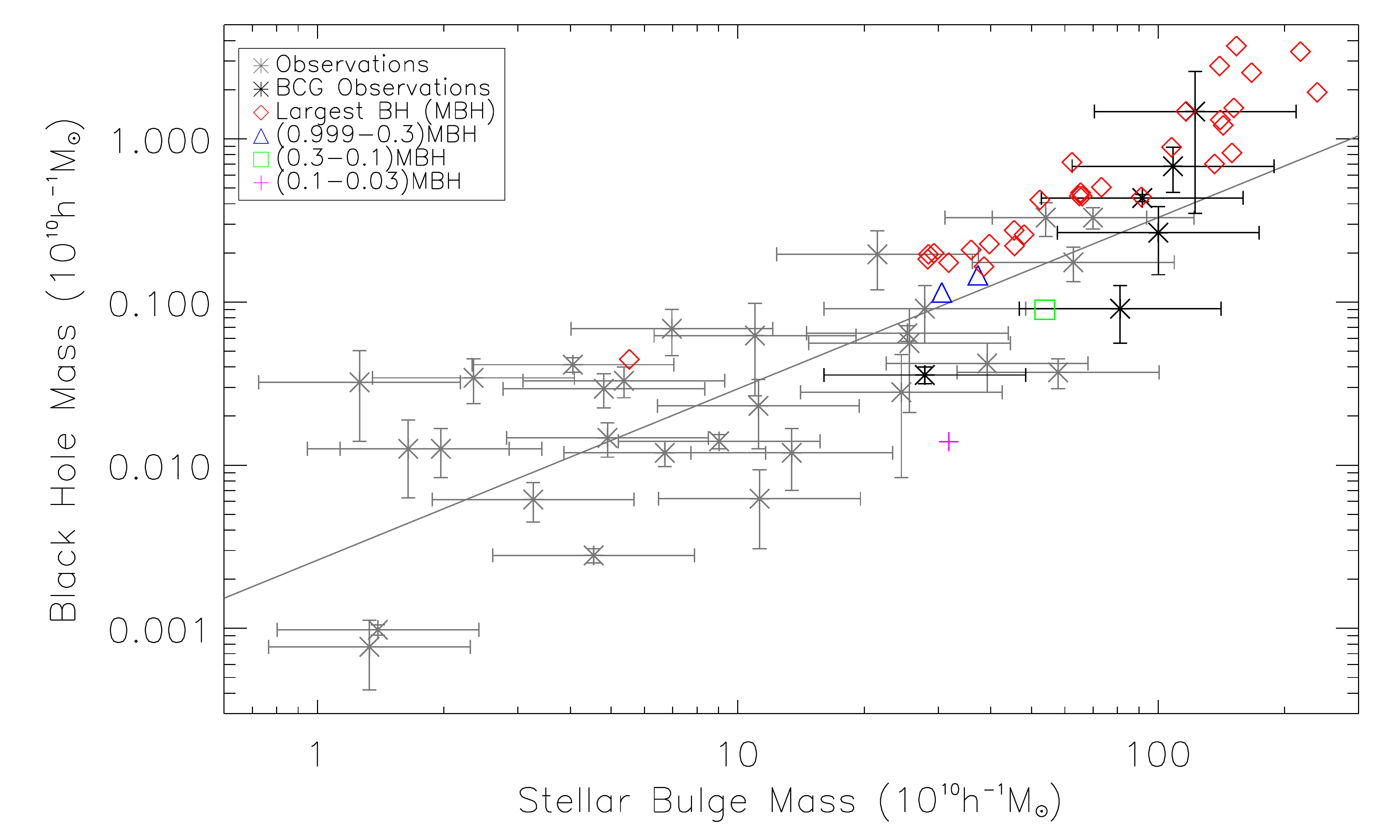}
\caption{Black hole mass against stellar mass for our AGN clusters at $z=0$. Red diamonds are the central 
and most massive) black holes, while the other points correspond to less massive black holes associated with
satellite galaxies. Grey and black asterisks are observational data (\protect\citealt{McConnell2013}), the latter being 
objects associated with BCGs.}
\label{plot:Mag}
\end{figure}

The remaining component in our AGN model clusters are the super-massive black holes. Fig.~\ref{plot:Mag} shows the 
black hole masses within $r_{200}$, plotted against stellar mass at $z=0$. We have sub-divided the black holes into 
those at the cluster centre (red diamonds) and those belonging to cluster galaxies with lower masses. The stellar 
masses are estimated using the crude method outlined above 
for BCGs (i.e. stars with $r<r_{\rm cut}$). 
For satellite galaxies, we use the stellar mass in each sub-halo as found by SUBFIND.
We also show observational data compiled by \cite{McConnell2013}, highlighting BCGs in bold. 

Overall, the simulations are in reasonable agreement with the observational data. As discussed in the previous
section, the black hole mass can be tuned by varying the heating efficiency, $\epsilon_{\rm f}$. Our default value 
($\epsilon_{\rm f}=0.15$) was found by \cite{Booth2009} to reproduce the observed black hole mass-stellar bulge mass 
relation on galaxy scales, so it is somewhat re-assuring that this choice also produces a reasonably good relation for 
our cluster-scale simulations, given that our results are completely independent from theirs. However, we will show in
Section~\ref{sec:res} that the position of an individual cluster on this relation depends on resolution.

\section{Radial profiles}
\label{sec:profiles}

In this section, we are principally concerned with how our AGN feedback model affects the spatial distribution 
of hot gas and stars within our clusters, by considering radial profiles at $z=0$.

\subsection{Baryon, gas and star fraction profiles}

\begin{figure}
\includegraphics[width=85mm]{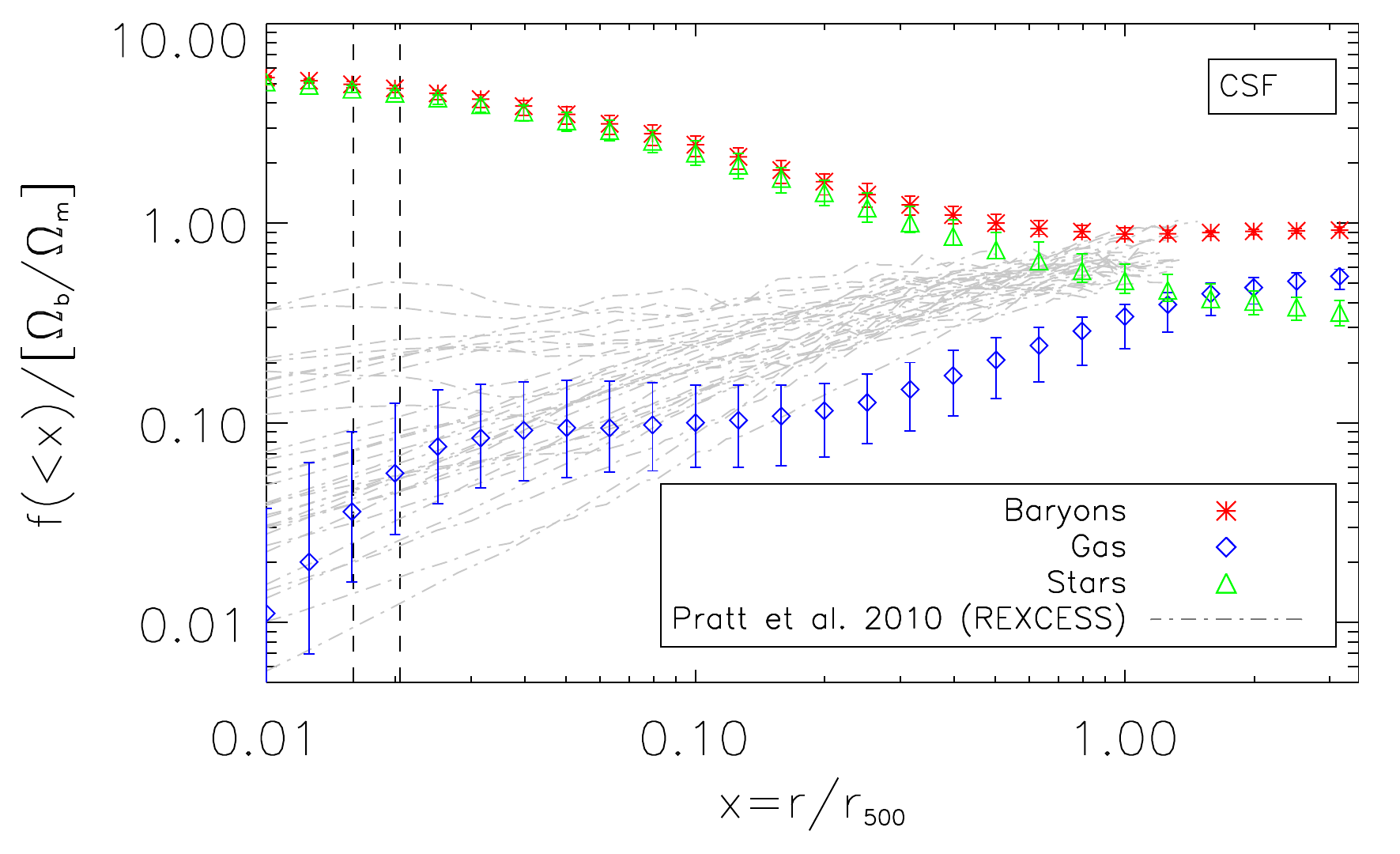}
\includegraphics[width=85mm]{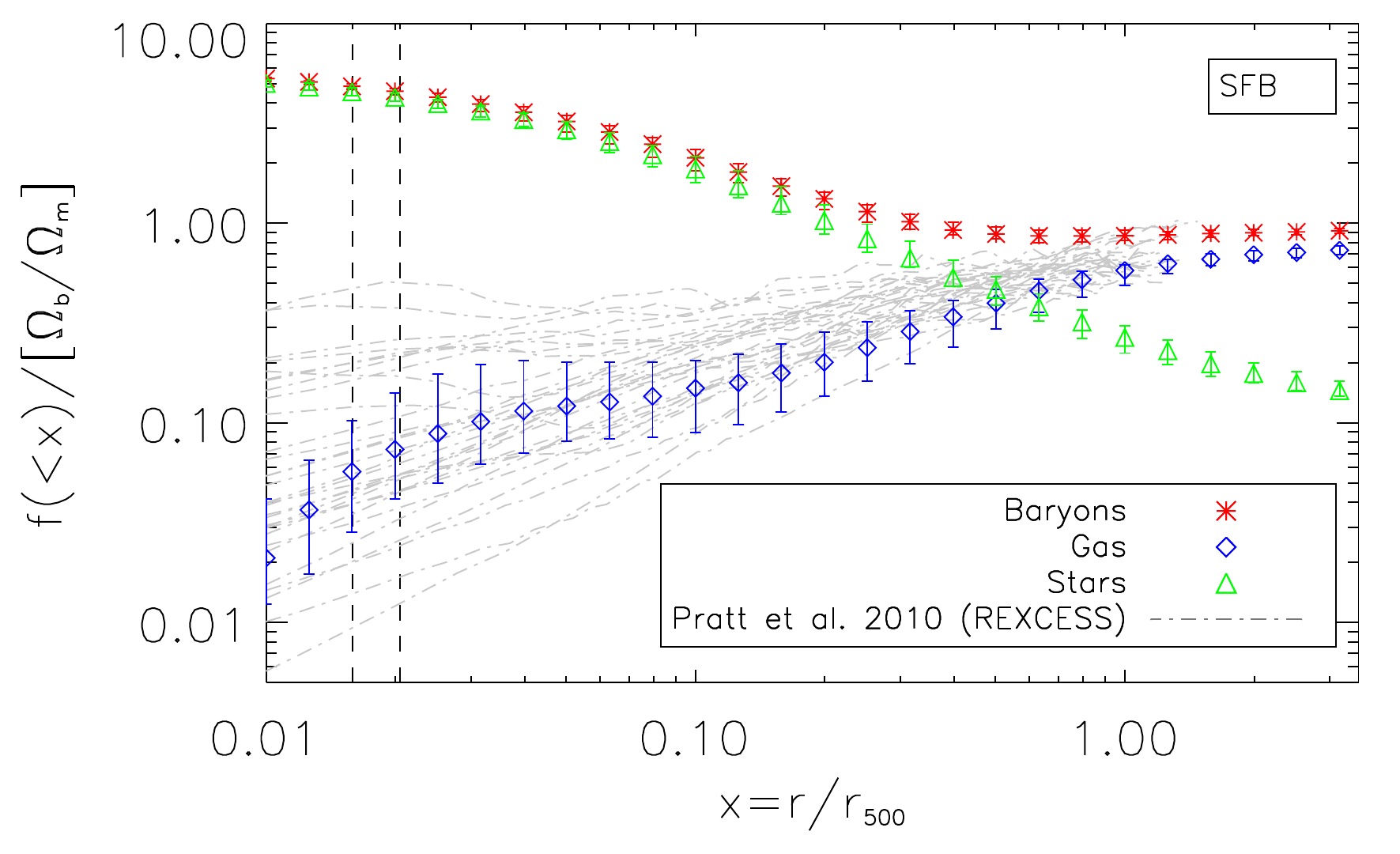}
\includegraphics[width=85mm]{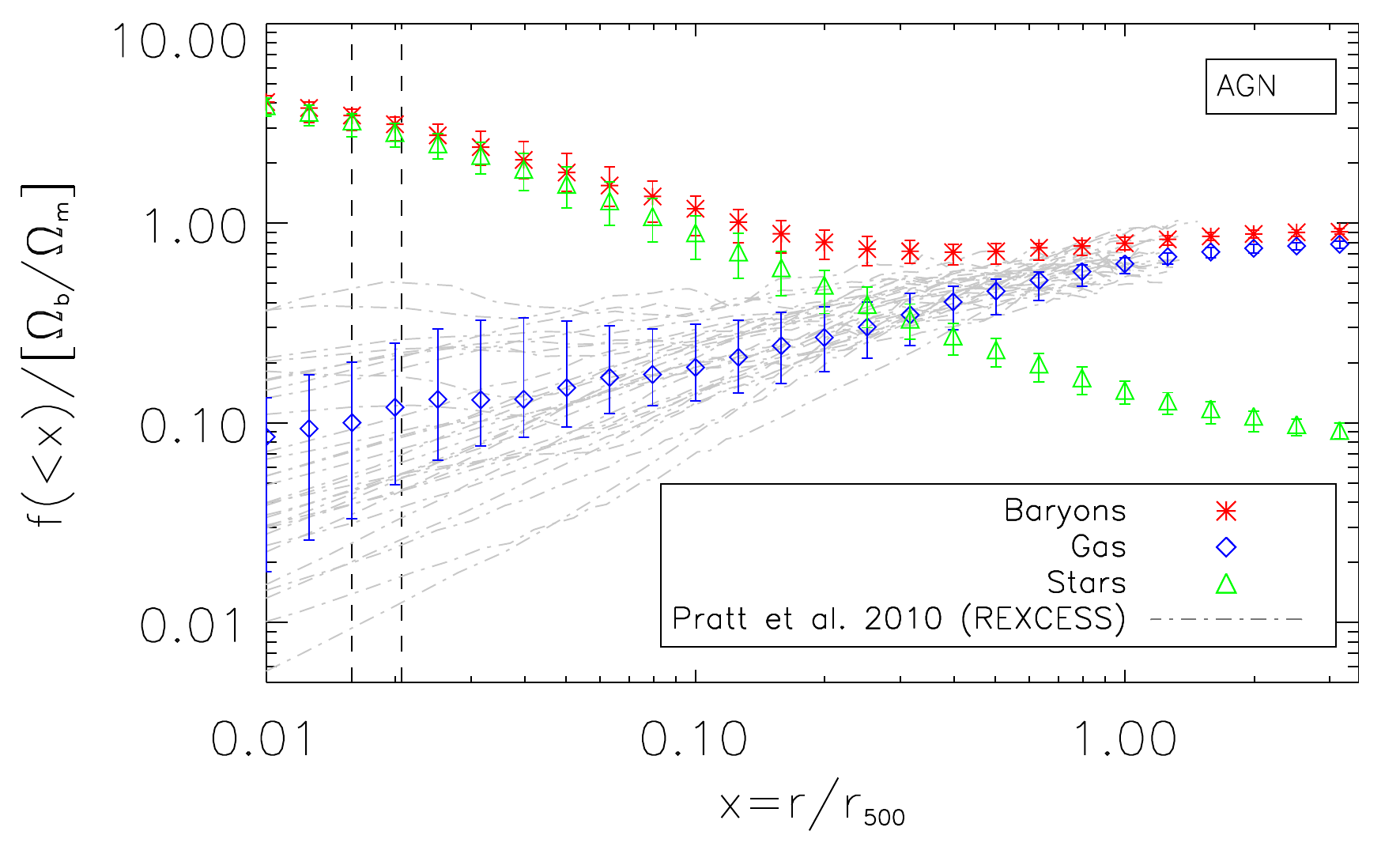}
\caption{Median baryon (red) , hot gas (blue) and star (green) fraction profiles for the 
cooling and star formation (CSF), supernova feedback (SFB) and AGN feedback (AGN) models at $z=0$. 
Error bars indicate the 10th and 90th percentiles, illustrating the cluster-to-cluster scatter within each bin.  
The vertical dashed lines represent the range in force resolution for the sample. Grey curves are observed
gas fraction profiles from REXCESS \protect\citep{Pratt2010}.}
\label{plot:profiles_tot_frac}
\end{figure}

We first consider the radial distribution of baryons, gas and stars. Fig.~\ref{plot:profiles_tot_frac} shows the 
integrated baryon, gas and star fractions within each radius (plotted as a dimensionless quantity, $x=r/r_{500}$), 
for our three radiative models (CSF, SFB $\&$ AGN).  Also plotted are the REXCESS
gas fraction profiles \citep{Pratt2009}. For the NR runs (not shown), baryon fractions reach a constant value 
($\sim 0.9 \Omega_{\rm b}/\Omega_{\rm m}$) by $r\simeq 0.2 r_{500}$. 
In the CSF model, the stars dominate at all radii within $r_{500}$, exceeding the the cosmological baryon fraction by a 
factor of five in the centre due to over-cooling. In the SFB and AGN runs, the dominance of the stellar component is 
reduced; for example, the stellar mass only exceeds the gas mass within $\sim 0.3\, r_{500}$ in the AGN clusters. 
However, the star fraction is still a factor of four higher than the cosmological baryon fraction in the centre. 
Regarding gas fractions, we see that the AGN model best reproduces the observations at all radii, although 
predicts less scatter in the core.

\subsection{Gas density profiles}

\begin{figure}
\includegraphics[width=85mm]{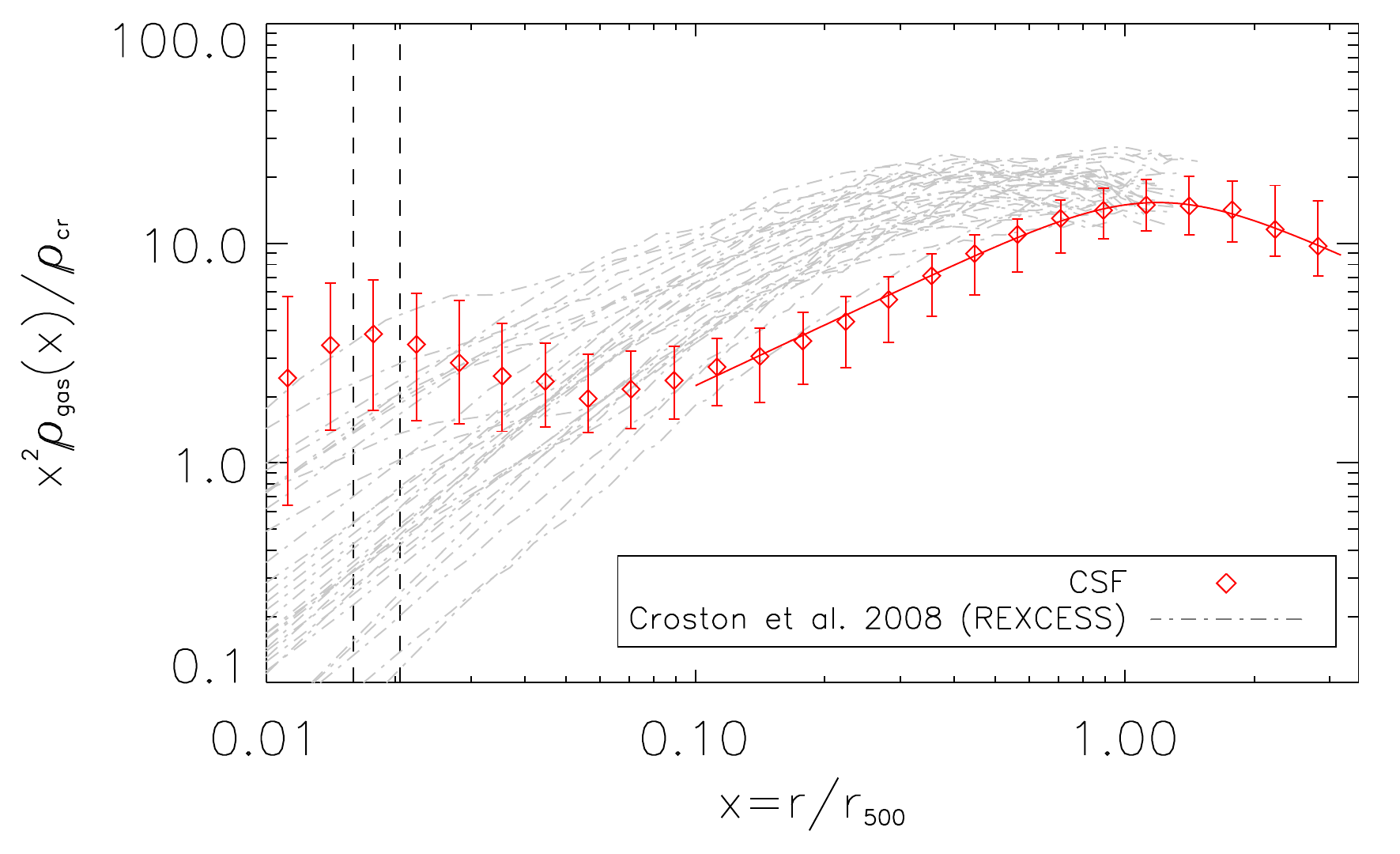}
\includegraphics[width=85mm]{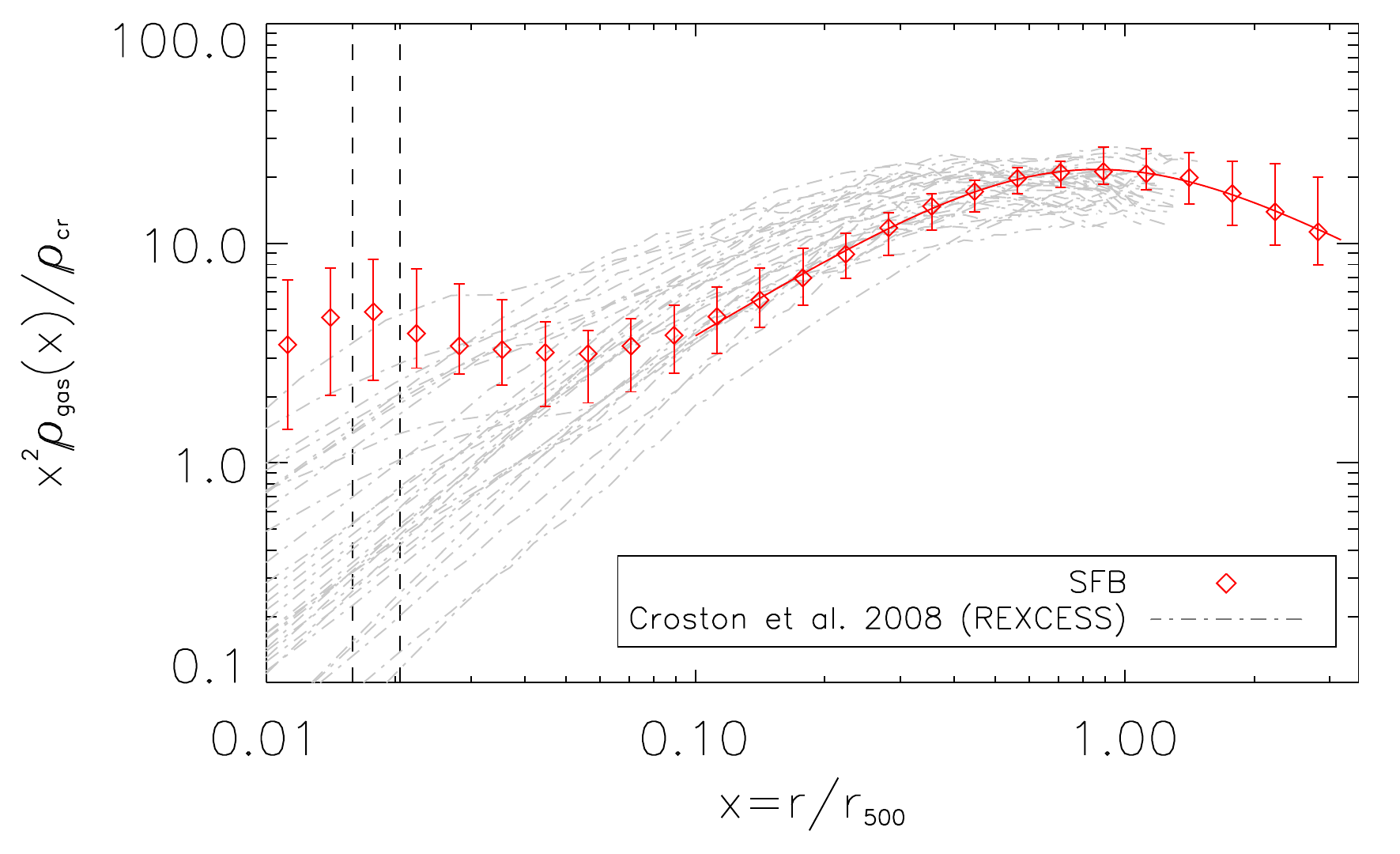}
\includegraphics[width=85mm]{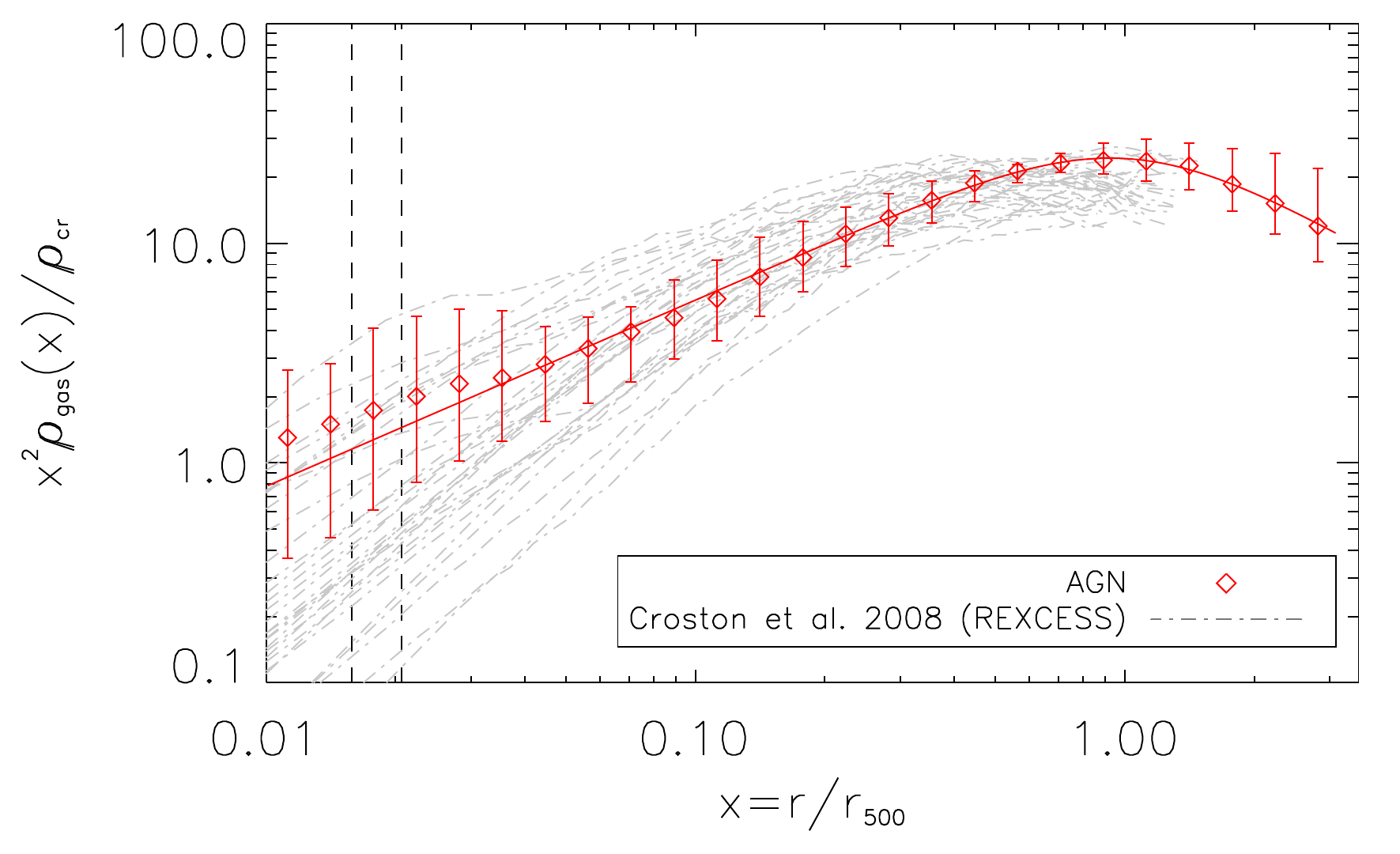}
\caption{Median gas density profiles for the four sets of runs. The red curve is the best fitting GNFW model to the median points. 
The grey lines display observed density profiles from \protect\cite{Croston2008}. Other details are as described in Fig.~\ref{plot:profiles_tot_frac}.}
\label{plot:profiles_gm}
\end{figure}

Gas density profiles are plotted in Fig.~\ref{plot:profiles_gm}, along with observed profiles produced 
by \cite{Croston2008}, from the REXCESS data. We multiply the dimensionless density profile 
($\rho/\rho_{\rm cr}$) by $x^2$ to highlight differences between the models. We also fit the GNFW model 
(see equation~\ref{eqn:GNFW} below) to the median points, restricting the fit to outside the cluster core 
($0.1<x<3$) for the CSF and SFB models.

In accord with the gas fraction profiles, the NR runs (not shown) contain gas that is too dense at all radii 
within $r_{500}$, when compared to observations.  Inclusion of radiative cooling and star formation (CSF; 
top panel) produces a median profile that is too steep in the centre ($x<0.03$) and too low elsewhere. 
This is the classic effect of over-cooling, where the gas loses pressure support and flows into the centre 
before finally being able to cool down sufficiently to form stars. 
(Note the effect is not as clear for the integrated gas fraction due to the large increase in stellar mass
which dominates the total mass in the centre.)
When supernova feedback is included
(SFB; middle panel), the effect of the supernovae is to raise the gas density in the cluster as the feedback 
keeps more of the gas in the hot phase. The result is a gas density profile that matches observations 
reasonably well beyond the core ($x> 0.1$) but the central densities are still too high. This problem is 
largely solved by the inclusion of AGN feedback, which heats the core gas to much higher temperatures 
($T \ge 10^{8}$K), allowing more gas to move out to larger radii. As a result, the agreement between the AGN 
model and the observations is better, although the median profile is a little steep in the centre. This agreement 
is not too surprising, given that our AGN feedback model was tuned to match the observed median pressure 
profile (see below).

We also checked if the density profiles depend on mass. To do this, we first ranked the clusters in mass and 
then divided the ranked list into three bins of 10 objects, before comparing the median profile for each mass bin. 
In all three radiative models we find a small but significant trend such that higher-mass clusters have scaled
density profiles with higher normalisation. As a result, the scaled entropy profiles of the higher-mass objects are 
lower (but no such trend is seen for the pressure profiles). This is expected given the mass-dependent effects of 
cooling and feedback on the gas fraction (as shown in Fig.~\ref{plot:bc}). 

\subsection{Pressure profiles}
\label{subsec:preprof}

\begin{figure}
  \includegraphics[width=85mm]{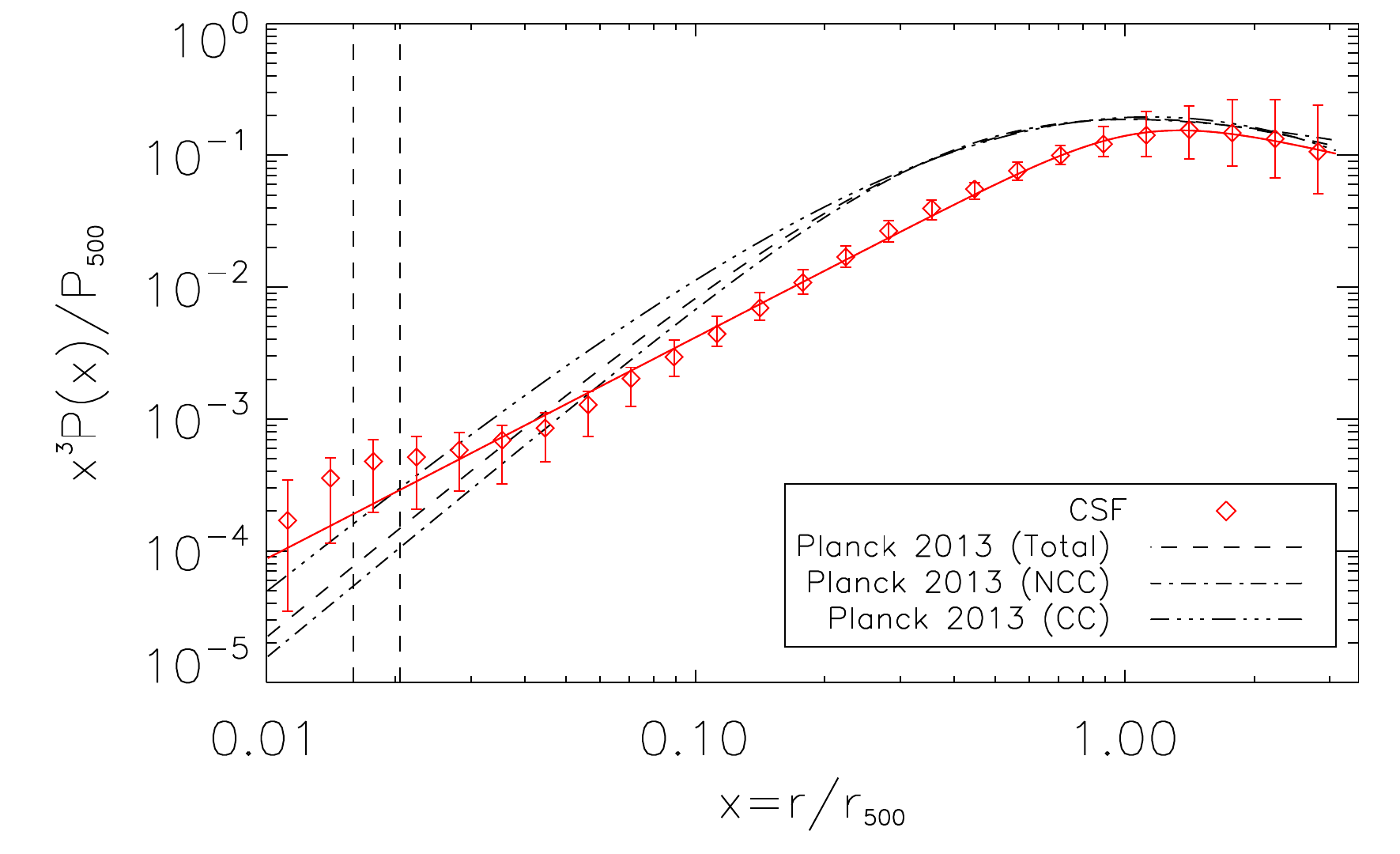}
  \includegraphics[width=85mm]{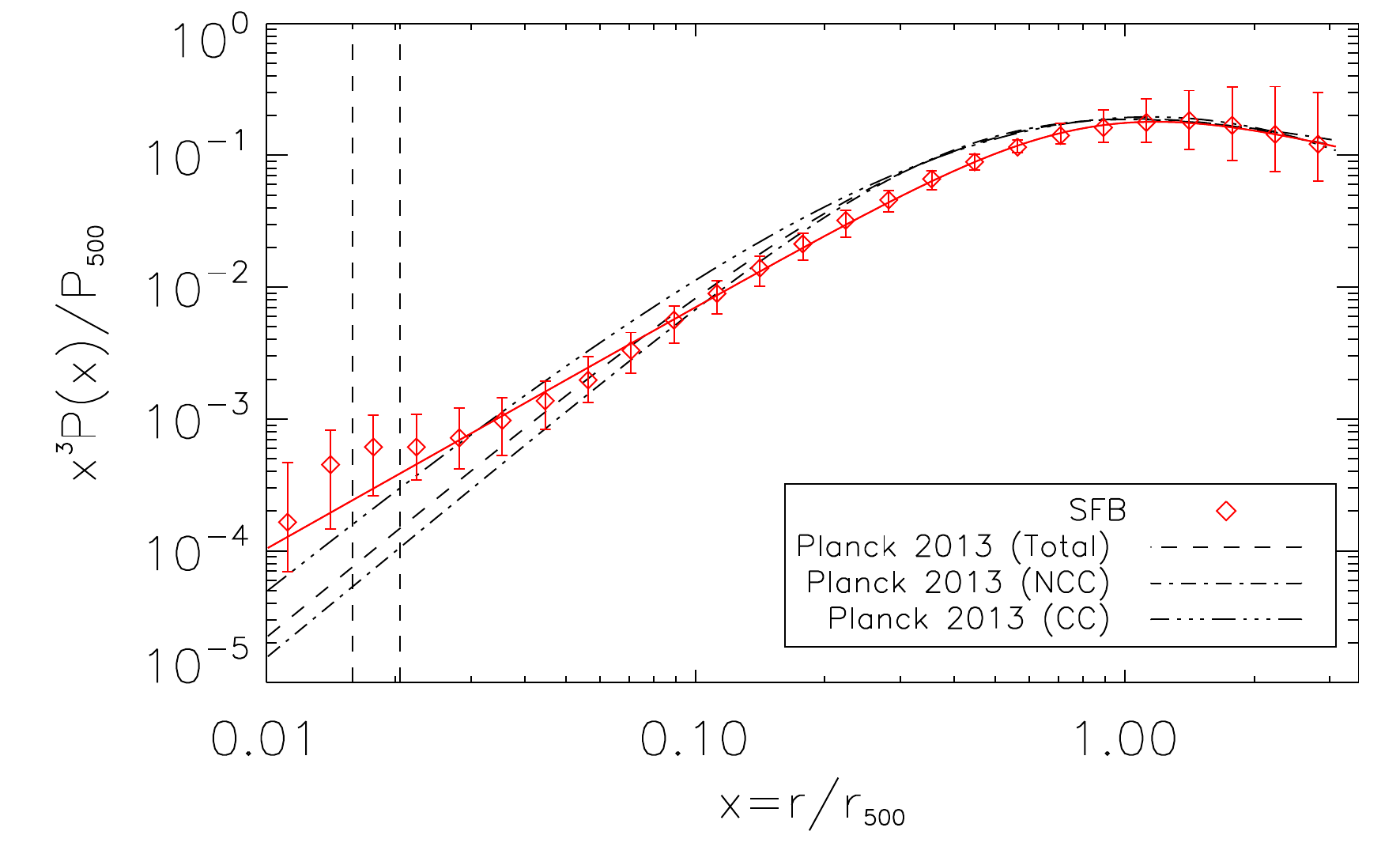}
  \includegraphics[width=85mm]{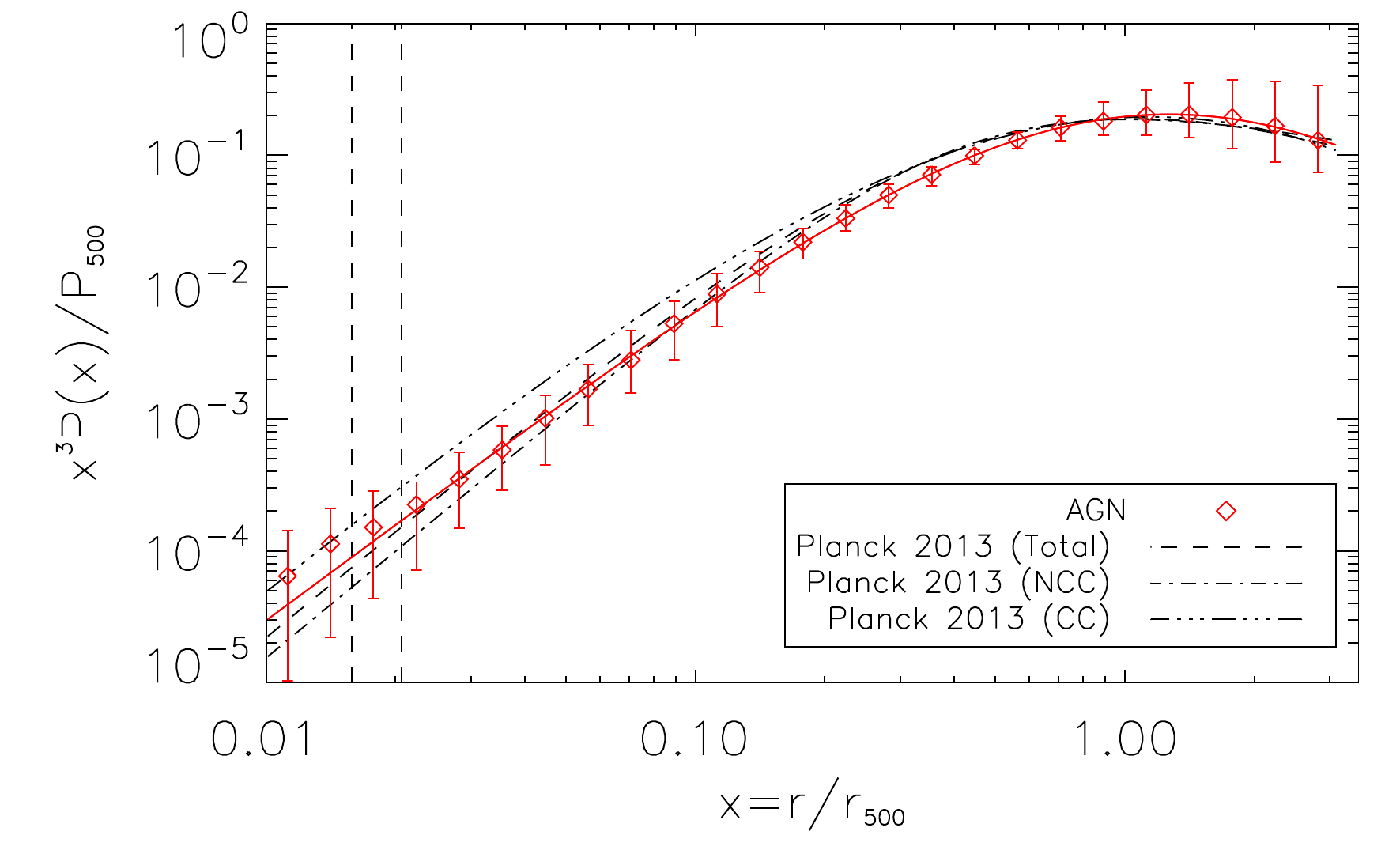}
\caption{Dimensionless pressure profiles for the three radiative models at $z=0$. The red curve is the 
best fitting GNFW model to the median points. Observational fits are also displayed for the cool core, 
non-cool core and total median profiles from \protect\cite{Planck2013}. Other details are as described 
in Fig.~\ref{plot:profiles_tot_frac}.}
\label{plot:profiles_pres}
\end{figure}

It is also useful to study pressure profiles, as the pressure gradient provides hydrostatic support in the cluster.  Furthermore, the pressure 
profile allows us to understand any changes in the Sunyaev-Zel'dovich effect, as the $Y$ parameter can be expressed as the following
integral of the pressure profile for a spherically-symmetric cluster
\begin{equation}
\label{Y_eqn}
D_{\rm A}^2 Y_{\rm SZ} = \frac{\sigma_{\rm{T}} r_{500}^3}{m_{\rm{e}}  c^{2}}  \int_{0}^{1}  P_{\rm e}(x) \, 4 \pi x^{3} \, {\rm dln} x ,
\end{equation}
where $P_{\rm{e}}=n_{\rm{e}}kT_{\rm e}$ is the pressure from free electrons (assumed to be proportional to the hot gas pressure). Plotting
$x^3P(x)$ therefore allows us to assess the contribution to $Y_{\rm SZ}$ from the gas (and therefore its total thermal energy) from each 
logarithmic radial bin.

In Fig.~\ref{plot:profiles_pres}, we show dimensionless pressure profiles (scaled to $P_{500}$; see Appendix~\ref{app:measure}) for
our three radiative models at $z=0$. The median data points are fitted with the GNFW model (e.g. \citealt{Arnaud2010}), defined
as
\begin{equation}
\label{eqn:GNFW}
{P \over P_{\rm 500}}= { P_{0}  \over u^{\gamma} (1+u^{\alpha})^{(\beta-\gamma)/\alpha} },
\end{equation}
where $u=c_{500} \, x$, $c_{500}$ is the concentration parameter, $P_{0}$ is the normalisation and [$\alpha$, $\beta$, $\gamma$] 
are parameters that govern the shape of the profile.  This allows us to make a direct comparison with the
best-fitting GNFW models for the {\it Planck} SZ cluster sample (we show results for their total sample, cool-core clusters and 
non-cool-core clusters; \citealt{Planck2013}). Note that the normalisation of the observed pressure profiles, $P_0$, 
exhibits a weak dependence on mass \citep{Arnaud2010}, which can be summarised as follows
\begin{equation}
P_{0}(M_{500}) = P_{3} \left( \frac{M_{\rm 500}}{3 \times 10^{14} \, {\rm M_{\odot}}}\right)^{0.12},
\label{eqn:pmass}
\end{equation}
where $P_3$ is a free parameter that equals $P_0$ when $M_{500}=3\times 10^{14} \, {\rm M}_{\odot}$. 
For our full sample presented here, we use 
equation~\ref{eqn:pmass} to rescale the profiles for individual clusters before fitting the GNFW model to our median profile 
and comparing with the observed fits for $P_0=P_3$. 

It is immediately apparent that the largest contribution to $Y_{\rm SZ}$ occurs around $r_{\rm{500}}$, far away from the complex physics 
in the cluster core \citep{Kay2012}. At these large radii, there is only a small increase in the pressure 
when going from CSF$\rightarrow$SFB$\rightarrow$AGN, suggesting the $Y_{\rm SZ}$ 
parameter is reasonably insensitive to the physical model used. However, the contribution from regions with $r<r_{\rm{500}}$ cannot be ignored, especially when comparing CSF to SFB/AGN (as we shall see in the next section, this leads to significant differences in the 
$Y_{\rm SZ}-M_{500}$ relation between these models).
The differences in pressure profiles between the runs are largely similar to those seen for the gas density profile; this is because the 
density is a much more sensitive function of radius (varying by orders of magnitude) than the cluster temperature. By design, the 
AGN model provides a good match to the observed data. In detail, it still underestimates the pressure slightly (c.f. the 
{\it Planck} Total profile) except in the very centre (where the gas is unresolved at $r<0.02\,r_{500}$) and at the largest radii ($r>r_{500}$).

\subsection{Spectroscopic-like temperature profiles}

\begin{figure}
\includegraphics[width=80mm]{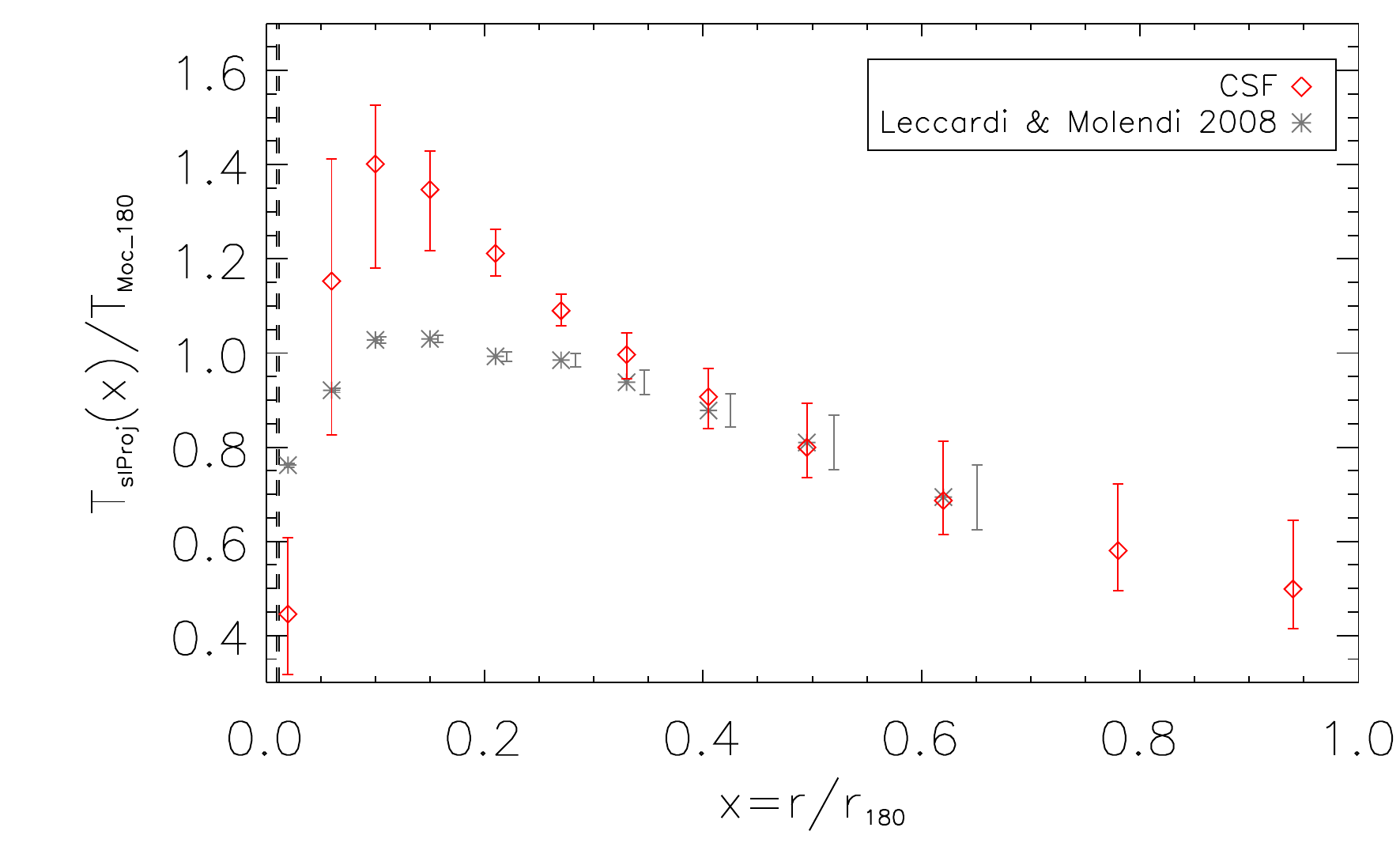}
\includegraphics[width=80mm]{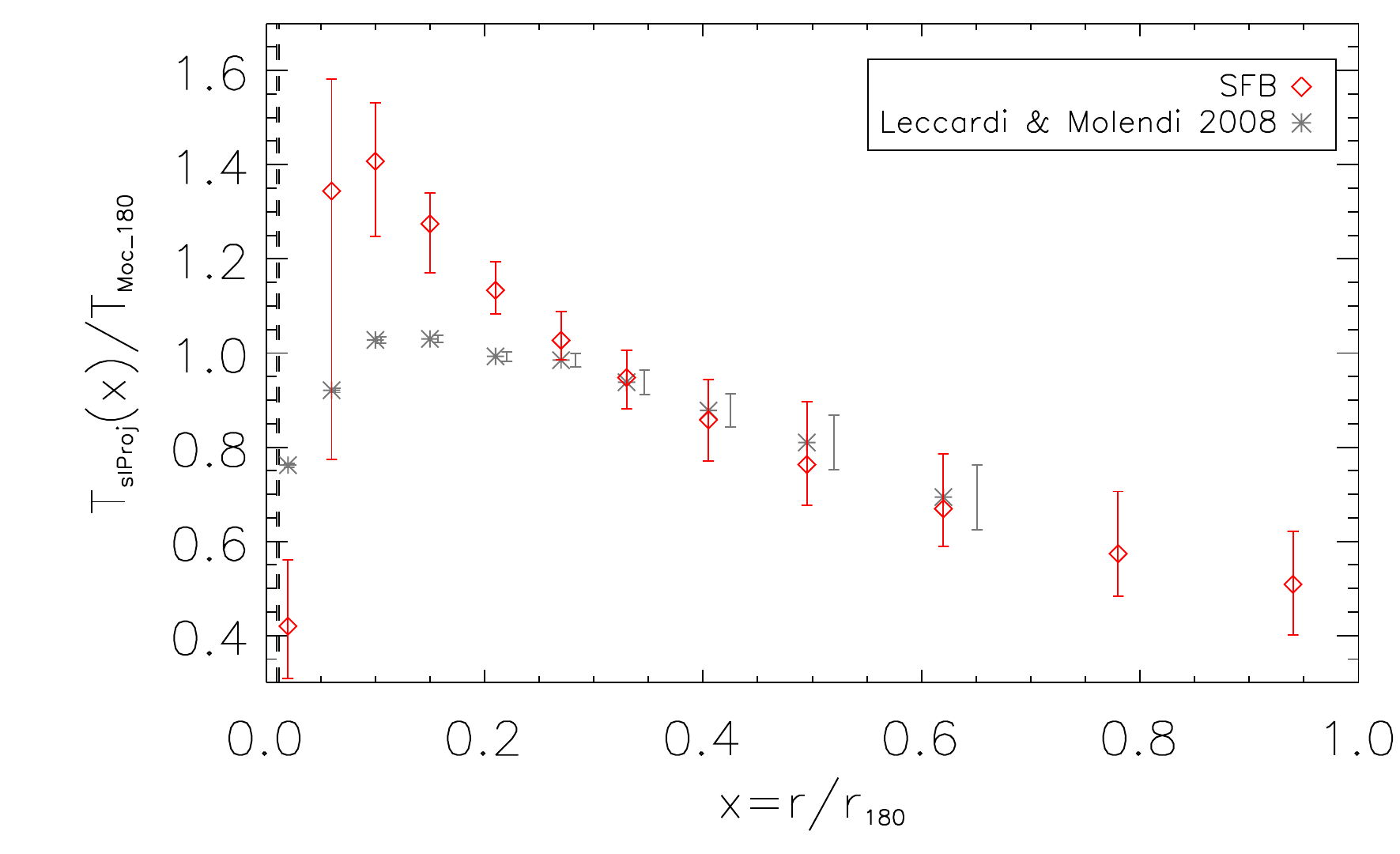}
\includegraphics[width=80mm]{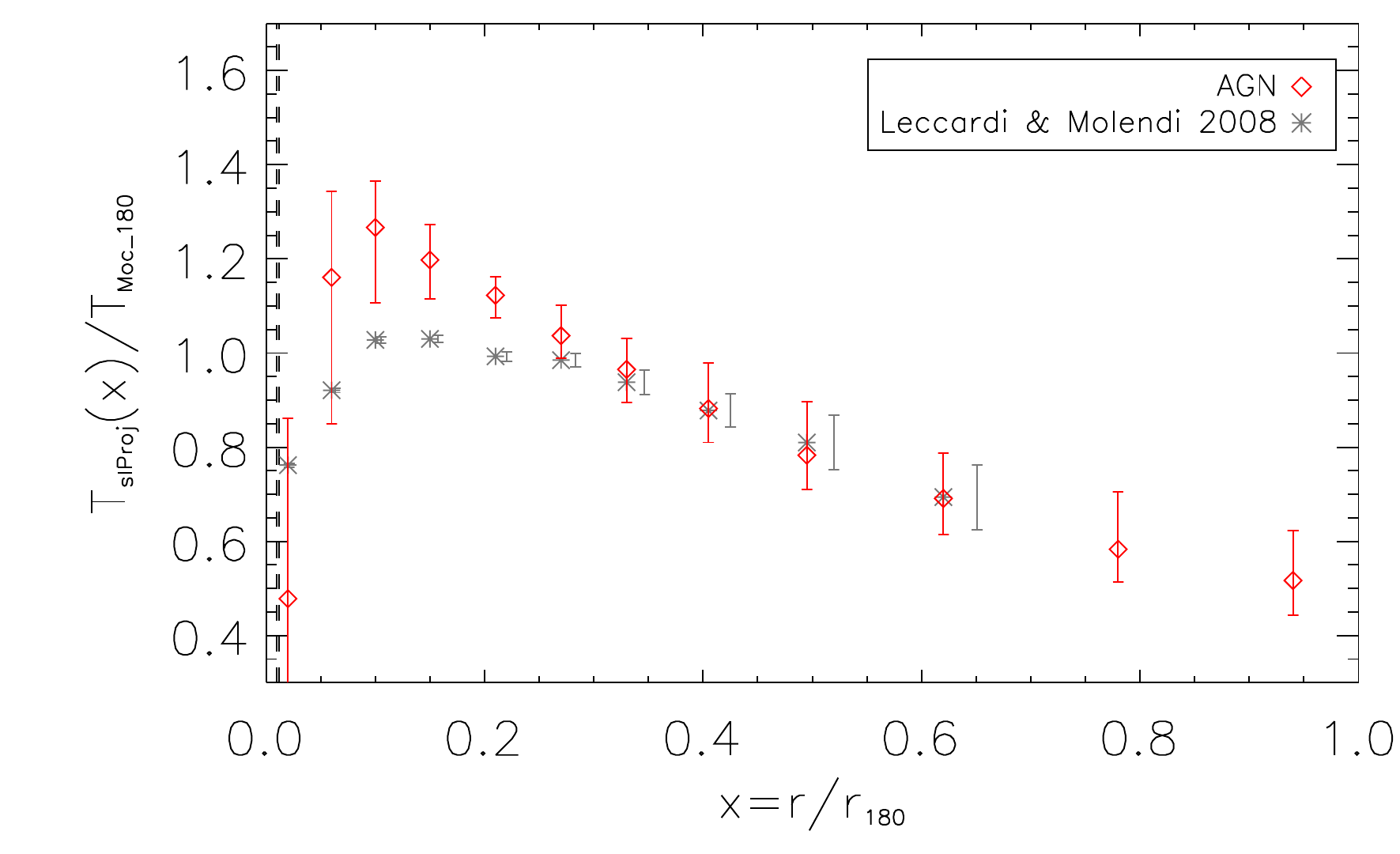}
\caption{Projected spectroscopic-like temperature profiles for the three radiative models at $z=0$. In this case, radii are 
scaled to $r_{180}$ and the temperature in each bin is divided by the average value across the range, $0.1< r/r_{180}<0.7$.  
Grey stars represent the observational data from \protect\cite{Leccardi2008} where the profiles are presented in a similar way.  
Other details are as described in Fig.~\ref{plot:profiles_tot_frac}.}
\label{plot:profiles_spec}
\end{figure}

Projected spectroscopic-like temperature profiles are shown in Fig.~\ref{plot:profiles_spec}, in comparison with 
observational results from \cite{Leccardi2008}. As discussed in 
Section~\ref{sec:method}, we calculated $T_{\rm sl}$ by excluding a very small amount of gas with the highest density 
within each bin \citep{Roncarelli2013}. Failure to do this results in noisier profiles but does not significantly affect
the normalisation (since the profiles are divided by an average temperature). 

All models predict profiles with a qualitatively similar shape, where the temperature declines towards the centre and at large
radii. This shape reflects the underlying gravitational potential (because the gas is approximately in hydrostatic equilibrium).
Comparing the models with the observations in detail, the NR results (not shown) are very similar at all radii, except in the very
centre ($r<0.1r_{180}$) where the observed gas is relatively cooler. The CSF and SFB models predict lower central temperatures
but the profiles have a much higher peak temperature than observed. Again, this is due to cooling: as higher entropy gas flows inwards, it
is adiabatically compressed, as can also be seen from the flattening of the entropy profile (see also \citealt{Tornatore2003,Borgani2004}).
The inclusion of AGN feedback has a more pronounced effect on shape of the inner temperature profile, reducing the peak value and 
the temperature gradient of the gas around it. This result, while a closer match to the observational data, may be due to the feedback 
not acting on enough of the gas in the core (see below).

\subsection{Entropy profiles}

\begin{figure}
  \includegraphics[width=85mm]{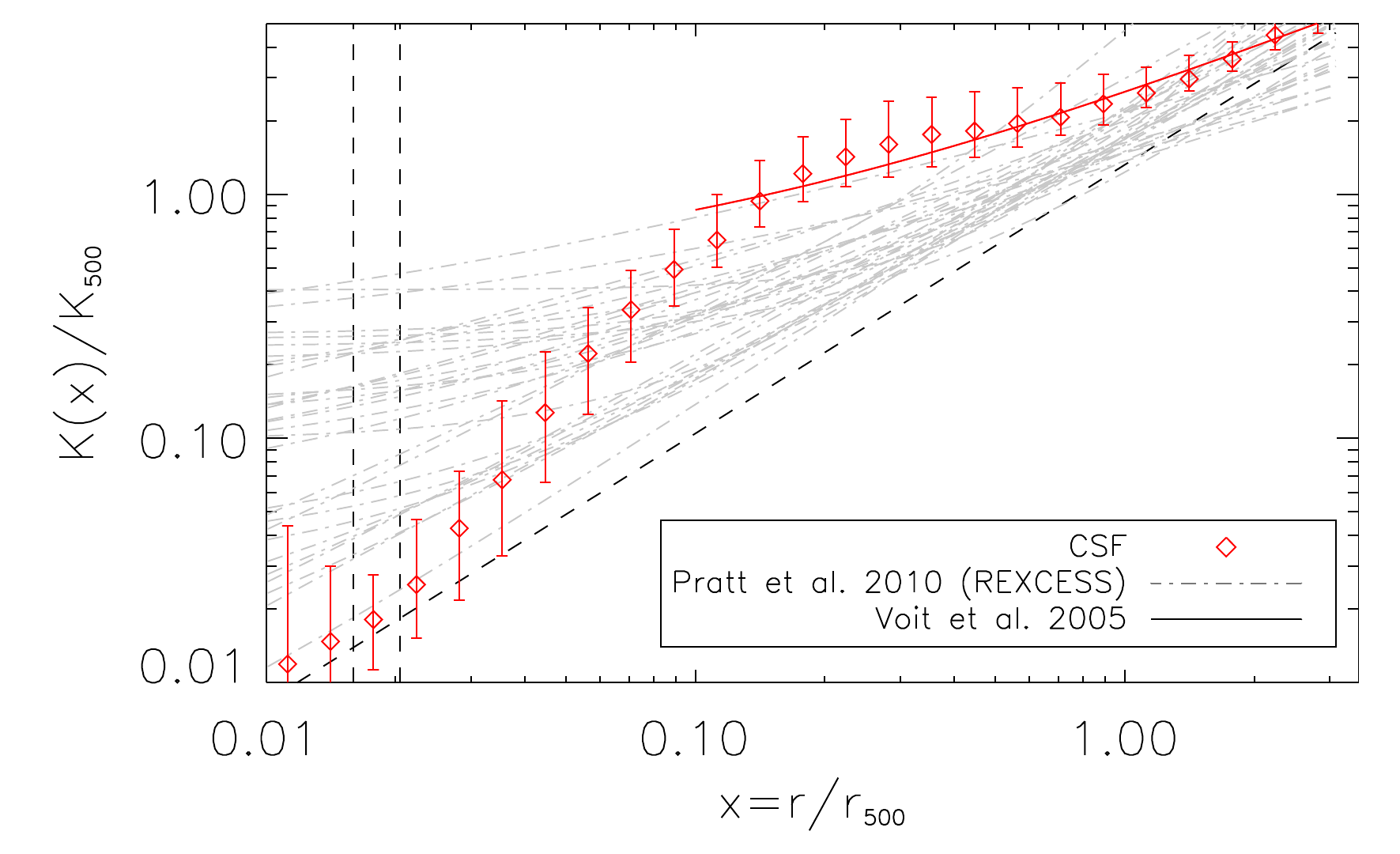}
  \includegraphics[width=85mm]{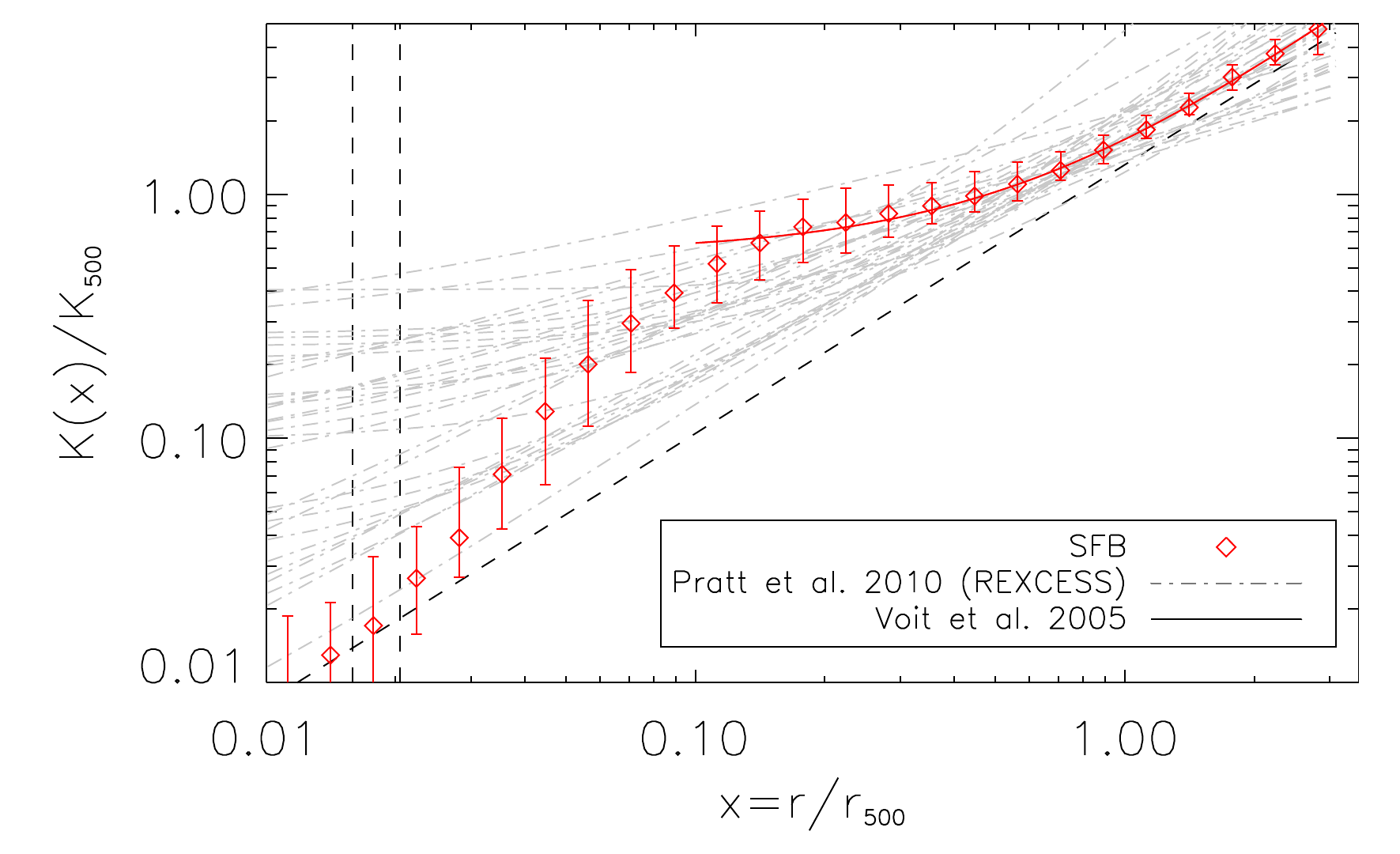}
  \includegraphics[width=85mm]{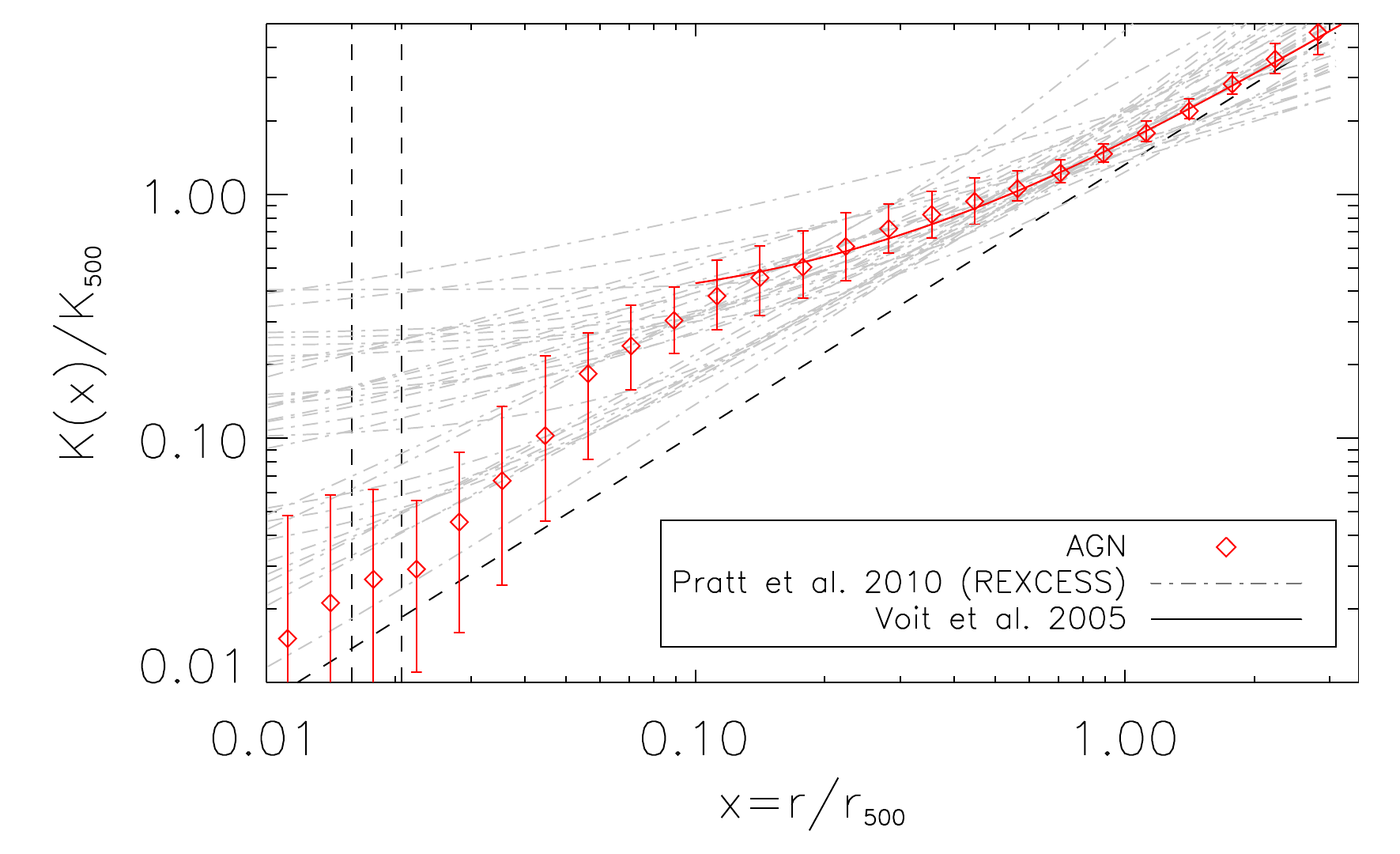}
\caption{Dimensionless entropy profiles for the three radiative models at $z=0$. 
Red curves are model fits to the median data points outside the core region ($x>0.1$; see text for details).
Grey curves are similar fits to observed X-ray clusters from the REXCESS sample, presented in 
\protect\cite{Pratt2010}. The dashed black line is the profile derived from non-radiative
simulations by \protect\cite{Voit2005}, re-scaled for $\Delta=500$. Other details are as described in 
Fig.~\ref{plot:profiles_tot_frac}.}
\label{plot:profiles_ent}
\end{figure}

Entropy profiles show directly the effects of non-adiabatic heating (from feedback, which 
increases the entropy) and radiative cooling (which decreases the entropy). Fig.~\ref{plot:profiles_ent} shows 
dimensionless entropy ($K/K_{500}$) profiles for our simulated clusters. As a guide to the eye, we fit the median 
data points at $x>0.1$ with the function, $K(x) = K_{0} + K_{100}x^{\alpha}$ (shown as the red curve), where 
$x=r/r_{500}$ and $[K_{0}, K_{100}, \alpha]$ are free parameters. We also show similar fits to observational data 
from the REXCESS sample \citep{Pratt2010} and the power-law profile derived from non-radiative simulations by 
\cite{Voit2005}, re-scaled for $\Delta=500$ (assuming a baryon fraction, $f_{\rm b} = 0.15$ and a value,  
$r_{500} /r_{200} = 0.659$, as derived from an NFW profile with concentration, $c_{500} = 3.2$, 
following \protect\citealt{Pratt2010}). 

The NR model (not shown) reproduces the \cite{Voit2005} relation very well, predicting a power-law entropy profile at all resolved radii 
as expected. This result is
below the observational data, owing to the gas density being too high. The entropy profiles in the CSF model show a distinctly
different shape: a sharp rise in entropy with radius until $r=0.1-0.2\, r_{500}$, where it reaches a plateau, before 
rising more gently at larger radius. This shape, at 
odds with the observations, can be understood as follows. As the innermost gas cools and flows towards the centre, higher entropy gas from 
larger distances flows in to replace it, creating the excess in entropy (more than required by the observational data) outside the core. 
At smaller radii, the cooling time becomes sufficiently short (compared to the local dynamical time) that the gas rapidly loses energy, creating a 
steep decline in entropy towards the centre of the cluster. The generation of excess entropy in simulations with cooling has been seen in many 
previous studies (e.g. \citealt{Muanwong2001,Borgani2002,Dave2002}). 

Supernova feedback increases the gas density (and pressure) throughout the cluster, reducing the effects 
of cooling and lowering the entropy profile outside the core, bringing the results into reasonable agreement 
with the observational data. However, the steep decline within the core is still evident as the supernovae are 
unable to provide sufficient energy to offset the cooling that is going on there. The situation is partially improved 
when AGN feedback  is included, where the inner entropy profile is now similar to that of cool-core clusters. 
However, the characteristic break at $r \simeq 0.1\,r_{500}$ is still present.

\begin{figure}
  \includegraphics[width=85mm]{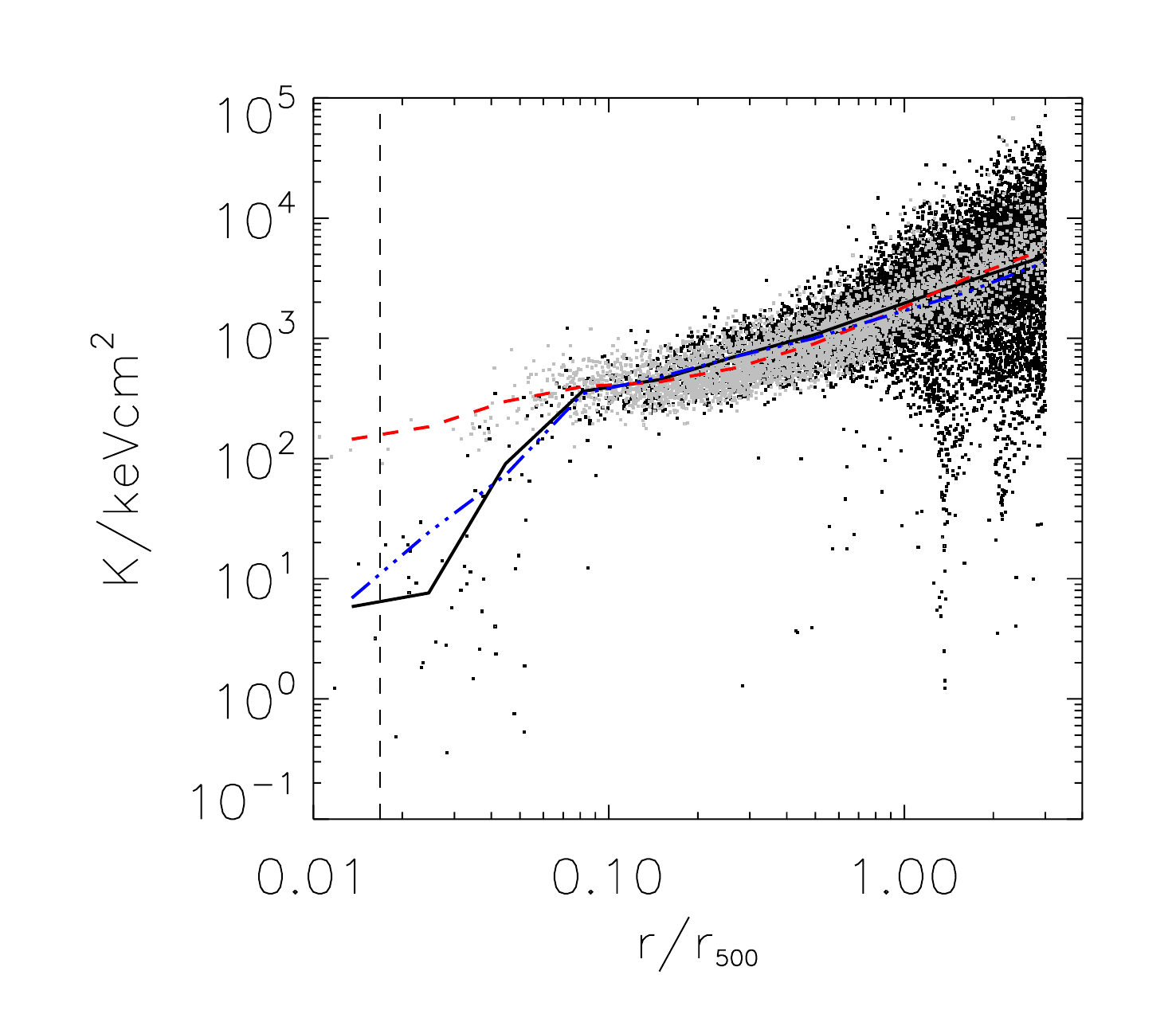}
\caption{Entropy versus radius for a random subset of hot gas particles (black points) in our most massive cluster
at $z=0$, run with the AGN model. The light grey points are a subset of particles directly heated by an AGN.
The solid, triple-dot-dashed and dashed lines are binned 
median profiles for all, SN-heated and AGN-heated particles respectively.}
\label{plot:entprofwind}
\end{figure}

As was discussed in Section~\ref{sec:method}, the AGN heating temperature was tuned to provide approximately 
the correct level of heating across the cluster mass range (as required by matching the pressure profile). However, 
as the simulations do not match the observed entropy profile shape in detail (and the scatter), it is likely that 
there is still something wrong, or incomplete, with our method. To gain some insight into the origin of this discrepancy, 
we show the entropy of a subset of individual gas particles versus radius for our most massive cluster at $z=0$ (black points),
in Fig.~\ref{plot:entprofwind}. As expected, the median profile (solid line) for all gas particles is very
similar to that for the whole cluster sample and the break is clearly present around $r=0.08\,r_{500}$. 
The triple-dot-dashed curve is the profile for the subset of gas particles that were directly heated by supernovae
(these make up around 20 per cent of the gas within $3\,r_{500}$, the maximum radius shown). Clearly, the two
profiles are very similar, as is also the case for AGN-heated particles (light grey points and dashed curve) beyond the break, which make
up only 3 per cent of the gas. This shows that most of the heated particles (from both SNe and AGN) are well mixed with the other gas throughout most of the cluster. Within the central region, however, the AGN-heated gas particles are much hotter 
and thus have much higher entropy ($K \sim 100 \, {\rm keV cm}^{2}$) than the rest of the gas (this is the expected level 
given the typical density, $n_{\rm H}\simeq 0.1 {\rm cm}^{-3}$, of the material, which is being heated to a temperature, 
$T_{\rm AGN}=10^{8.5}$ K). Nevertheless, the average entropy in the core is dominated by the cooler gas and so the break 
persists. We also note that a similar profile shape was found by \cite{McCarthy2010} on group scales (see their Fig.1). This is not 
surprising since we are effectively using the same AGN feedback model as theirs. 

One possible resolution to the problem is to include some degree of entropy mixing in the simulation. It is well known 
that standard SPH algorithms suppress gas mixing e.g. via the Kelvin-Helmholtz instability (see \citealt{Power2013} for recent work). 
Additionally, explicitly including thermal conduction may help \citep{Voit2008}. Discreteness effects from having relatively 
poor numerical resolution may also play a part; as we show in Section~\ref{sec:res}, runs with higher spatial resolution produce 
smoother profiles. Such issues will be investigated in future work.

\begin{figure}
  \includegraphics[width=85mm]{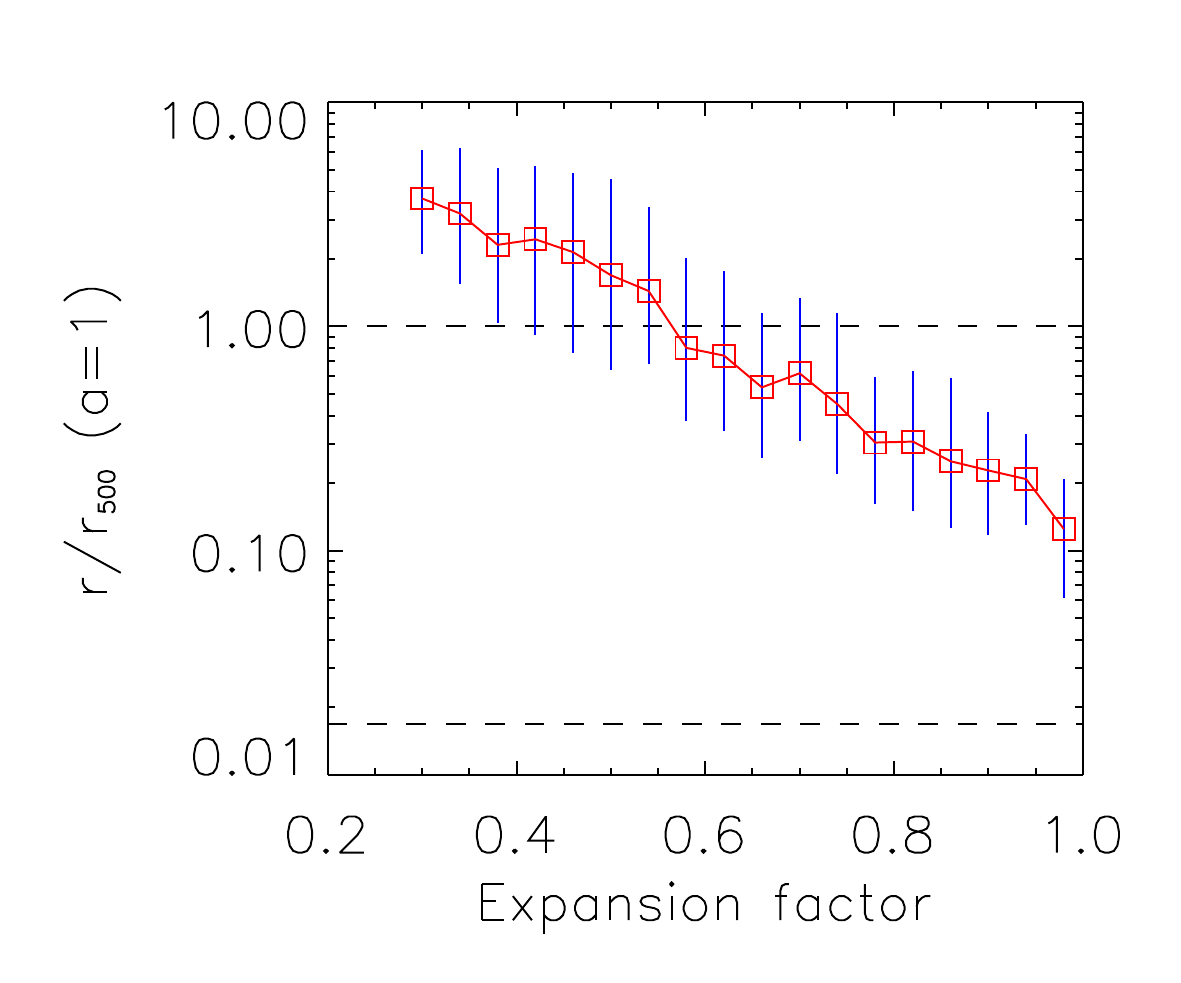}
\caption{Final radius of AGN-heated particles (in $r_{500}$ units) versus the value of $a$ when they were first heated, for
the most massive cluster. Squares are median values while the vertical lines indicate the 10th and 90th percentiles. The horizontal
dashed lines indicate $r_{500}$ and the softening scale.}
\label{plot:windpos}
\end{figure}

We are also interested in {\it when} the feedback happens. In \cite{McCarthy2011}, they argue that 
the AGN feedback largely works in their groups by {\it ejecting} gas from galactic-scale haloes at high redshift ($2<z<4$). Again, 
focussing on our highest mass cluster, we find that nearly all the AGN feedback energy is released at low redshift ($z<1$) 
because that is when most of the black hole growth occurs (see Section~\ref{sec:res}, Fig.~\ref{plot:res_bh}). A plausible explanation
for this difference is that it is harder for a black hole to significantly influence its surrounding environment in a cluster, where the potential
is much deeper, and we have therefore crossed the transition from a feedback-dominated regime to a cooling-dominated regime 
(as argued by \citealt{Stott2012}). The cooling gas in the core (which as we saw above, does not appear to get too disturbed by the
AGN-heated gas) then continues to feed the black hole, leading to significant growth at late times. (This can be true even when the 
accretion rate is small compared with the Eddington rate because the Bondi-Hoyle rate, $\dot{M} \propto M_{\rm BH}^2$.) 
As a result of this late heating, most of the heated gas remains in the cluster by $z=0$, while only gas that is heated earliest ends up 
beyond $r_{500}$. We confirm this in Fig.~\ref{plot:windpos}, where we plot the final radius of the heated particles versus the time when 
they were first heated (we additionally restricted our sample to those particles that were within $0.5 \, r_{500}$ in the NR run, to 
approximately select those particles that were heated by the central black hole). There is a strong negative correlation, with gas heated 
at $a>0.6$ largely remaining inside the cluster. (Heating the gas to a lower temperature reduces the final radius at fixed $a$, as would 
be expected).

\section{Scaling Relations} 
\label{ss:SR}

\begin{table}
\centering
\caption{Fit parameters for the scaling relations at $z=0$. Column 1 lists the relation and model; 
while columns 2-7 give the best-fit values for the normalisation ($A$), slope ($B$) and intrinsic 
scatter ($S$), together with their uncertainties ($\sigma$), estimated using the bootstrap method. 
The quantities $T_{\rm sl,OC}$ and $L_{\rm bol,OC}$ are for when gas from within the core ($r<0.15\,r_{500}$) is omitted.}
\begin{tabular}{lcccccc}
\hline
Run & $A$ & $\sigma_{\rm A}$ & $B$ & $\sigma_{\rm B}$ & $S$ & $\sigma_{\rm S}$ \\
\hline
\multicolumn{7}{l}{$Y_{\rm SZ}-M_{500}$}\\
NR      & -5.588 & 0.005 & 1.66 & 0.01 & 0.027 & 0.004 \\
CSF     & -5.923 & 0.013 & 1.85 & 0.03 & 0.047 & 0.006 \\
SFB     & -5.697 & 0.009 & 1.71 & 0.02 & 0.032 & 0.004 \\
AGN     & -5.653 & 0.008 & 1.70 & 0.02 & 0.034 & 0.004 \\

\hline

\multicolumn{7}{l}{$T_{\rm sl}-M_{500}$}\\
NR  & \phantom{-}0.204 & 0.016 & 0.61 & 0.03 & 0.062 & 0.010 \\
CSF & \phantom{-}0.126 & 0.009 & 0.30 & 0.03 & 0.047 & 0.007 \\
SFB & \phantom{-}0.120 & 0.014 & 0.27 & 0.04 & 0.057 & 0.010 \\
AGN & \phantom{-}0.296 & 0.013 & 0.25 & 0.05 & 0.082 & 0.010 \\

\hline 

\multicolumn{7}{l}{$T_{\rm sl,OC}-M_{500}$}\\
NR & \phantom{-}0.223 & 0.012 & 0.64 & 0.03 & 0.052 & 0.006 \\
CSF & \phantom{-}0.409 & 0.005 & 0.60 & 0.01 & 0.024 & 0.003 \\
SFB & \phantom{-}0.361 & 0.006 & 0.60 & 0.01 & 0.026 & 0.003 \\
AGN & \phantom{-}0.354 & 0.005 & 0.60 & 0.01 & 0.021 & 0.002 \\

\hline 
 
\multicolumn{7}{l}{$L_{\rm bol}-M_{500}$}\\
NR      & \phantom{-}0.517 & 0.048 & 0.97 & 0.09 & 0.179 & 0.024 \\
CSF     & -0.151 & 0.038 & 1.79 & 0.09 & 0.159 & 0.024 \\
SFB     & \phantom{-}0.309 & 0.034 & 1.46 & 0.08 & 0.138 & 0.016 \\
AGN     & -0.158 & 0.027 & 1.54 & 0.06 & 0.123 & 0.019 \\

\hline

\multicolumn{7}{l}{$L_{\rm bol,OC}-M_{500}$}\\
NR & \phantom{-}0.167 & 0.038 & 1.08 & 0.07 & 0.124 & 0.028 \\
CSF & -1.070 & 0.075 & 1.73 & 0.16 & 0.240 & 0.035 \\
SFB & -0.434 & 0.090 & 1.35 & 0.17 & 0.255 & 0.051 \\
AGN & -0.432 & 0.036 & 1.45 & 0.08 & 0.113 & 0.032 \\

\hline

\end{tabular}
\label{tab:scaltab}
\end{table}

While profiles help us to understand the interplay of different physical effects within the clusters, they do not easily 
describe their global properties and how they scale with mass. Scaling relations do this, as well as providing additional, 
important observational tests of the models. Furthermore, observable-mass scaling relations are an important part
of cosmological analyses that use clusters. The primary aim of this section will therefore be to investigate how key observable scaling 
relations ($Y_{\rm SZ}, L_{\rm bol}$ and $T_{\rm sl}$ versus $M_{500}$) vary as we add increasingly realistic physics. 
We also compare our results at $z=0$ with observational determinations, although stress that such a comparison is not 
rigorous as we do not measure the properties in exactly the same way (importantly, we do not investigate the effects of 
hydrostatic mass bias in this paper, which is likely to lead to a small increase in the normalisation of our scaling relations
owing to the hydrostatic masses being lower than the true masses).  

We will also investigate how our 
models differ when the redshift evolution of the scaling relations is considered; for this, we take the most massive progenitor 
of each cluster so our sample contains 30 objects at all redshifts. While our results should be interpreted with some 
caution given that we are not comparing mass-limited samples at each redshift (or indeed, flux-limited 
samples), they are still useful for comparing the relative importance of the different physical processes.

All scaling relations are fit with a power-law model
\begin{equation}
  \label{eqn:scaling}
  E(z)^{\gamma}C_{\rm 500} = 10^{\rm A}(M_{\rm 500}/10^{14}h^{-1}{\rm M}_{\odot})^{\rm B},
\end{equation}
where $E(z)=H(z)/H_0$ and $C_{\rm 500}$ is the observable, that can take the form of $Y_{\rm SZ}$, $L_{\rm bol}$ or $T_{\rm sl}$, with all properties measured within
$r_{500}$. We allow both the normalisation, $A$, and index, $B$, to vary when performing a least-squares fit to the set of data points, 
$(\log_{10}C_{500},\log_{10}M_{500})$. We fix the parameter, $\gamma$, to the self-similar value when studying scaling relations at $z>0$:
for $(Y_{\rm SZ}$, $L_{\rm bol}$,$T_{\rm sl})$, these values are $\gamma=(-2/3,-7/3,-2/3)$ respectively. We also estimate the intrinsic scatter
in each relation using
\begin{equation}
\label{eqn:scatter}
S = \sqrt{ {1 \over N-2} \sum_{i=1}^{N} \, \left[ \log_{10}C_i(M_i) - \log_{10}C_{500}(M_i) \right]^2 },
\end{equation}
where $N=30$ is the number of clusters in our sample, $C_i$ is the value being measured for the $i$th cluster with mass, $M_i$, and
$C_{500}$ is the best-fitting value at the same mass. Uncertainties in $A$, $B$ and $S$ are
estimated using the bootstrap method, re-sampling $10,000$ times and computing the standard deviation of the distribution of best-fit values.
Results for the fits at $z=0$ are summarised in Table~\ref{tab:scaltab} and we discuss each relation in turn (including the evolution of the
parameters with redshift), below.

\subsection{The $Y_{\rm SZ}-M_{500}$ relation}

\begin{figure}
\includegraphics[width=80mm]{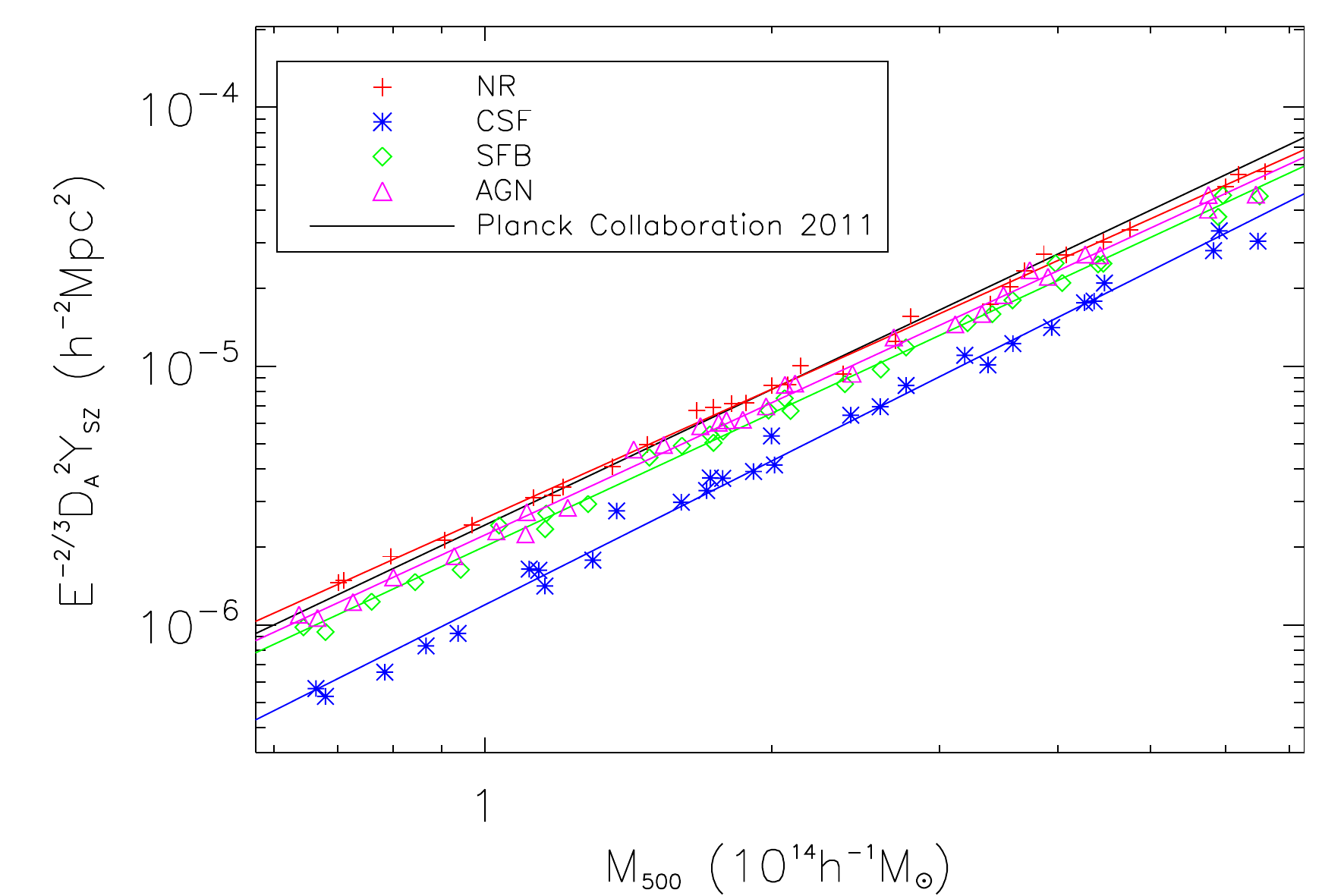}
\includegraphics[width=80mm]{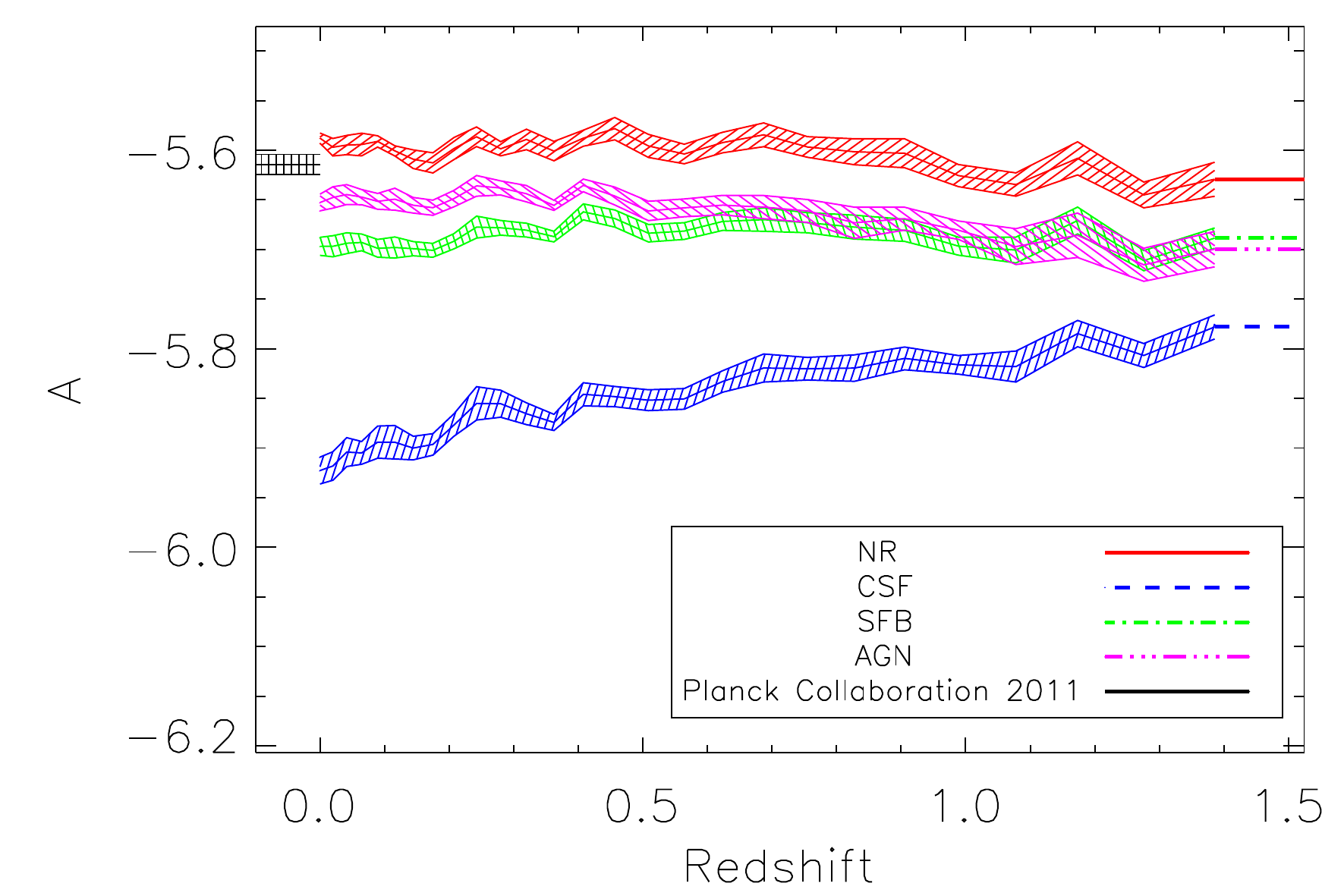}
\includegraphics[width=80mm]{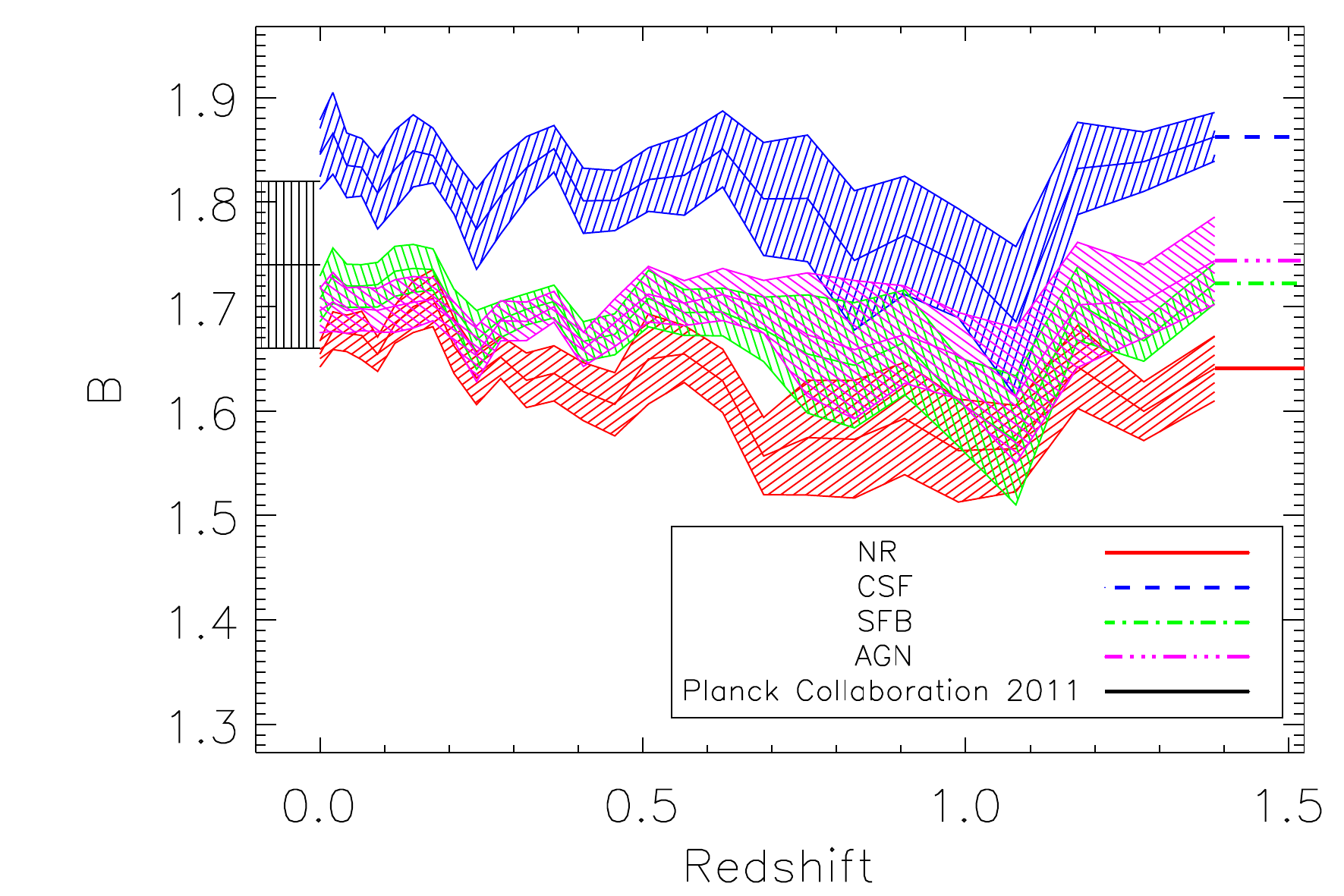}
\caption{$Y_{\rm SZ}-M_{\rm 500}$ relation at $z=0$ (top panel) for the non-radiative (NR; red crosses), 
cooling and star formation (CSF; blue stars), supernova feedback (SFB; green diamonds) and AGN (magenta triangles) 
models respectively. Accompanying solid lines are best-fitting power laws to the data, while the observed relation from the 
{\it Planck} collaboration (\protect\citealt{Planck2011e} ) is also plotted as a black line. The middle and bottom panels show 
the evolution of the normalisation and slope with redshift where the shaded region illustrates the uncertainty in each parameter
(one standard deviation from the best-fitting values of $A$ and $B$). The black line and shaded region represents the best-fit value
and error from observations of low redshift clusters \protect\cite{Planck2011e}.}
\label{plot:ymrel}
\end{figure}

The $Y_{\rm SZ}-M_{500}$ relation is a good basic test, given that $Y_{\rm SZ}$ is proportional to the total thermal energy of the intracluster gas.
Unlike X-ray luminosity, it should be relatively insensitive to non-gravitational physics, a result confirmed with previous simulations
(e.g. \citealt{daSilva2004,Nagai2006,Battaglia2012,Kay2012}).

In the top panel of Fig.~\ref{plot:ymrel} the $Y_{\rm SZ}$-$M_{\rm 500}$ relation at $z=0$ is plotted, where red crosses, blue stars, green diamonds 
and purple triangles are results from the NR, CSF, SFB and AGN runs respectively. We also show the best-fitting relations to each dataset as solid
lines with the same colour as the data points. As an observational comparison, the best-fitting straight line to {\it Planck} and {\it XMM-Newton} data \citep{Planck2011e} is also 
plotted in black. 

As expected, the $Y_{\rm SZ}-M_{500}$ relation is well defined for all the runs with minimal scatter ($S < 0.05$), but it is immediately apparent that the 
CSF relation is a poor match to the observations, whereas the NR, SFB, and AGN runs all do reasonably well (the normalisation agrees to within 10-20 per cent and may be improved once the effect of hydrostatic mass bias is accounted for).
The severe over-cooling present in the CSF run leads to a reduction in $Y_{\rm SZ}$ as the gas cools and provides less pressure support. While the NR model is 
unable to reproduce many other observables, the result here is a good match to the observations, suggesting that the feedback must be strong enough to counteract cooling without increasing the thermal energy significantly (as also seen with the temperature profiles). The similarity between the SFB and AGN runs can be explained by the 
fact that the dominant contribution to $Y_{\rm SZ}$ occurs at $ r \simeq r_{\rm 500}$. In this region the feedback from supernovae is more effective than from AGN, 
but this conclusion may at least in part be affected by our method for incorporating black holes within the simulation. Nevertheless, it emphasises the point that the mitigation
of cooling by supernovae in clusters is an important factor. 

We have also examined the dependence of the fit parameters ($A$ and $B$) on redshift; the lower panels in Fig.~\ref{plot:ymrel} show results for the normalisation, $A$,
and slope, $B$, from each snapshot to $z=1.4$. 
In all models except CSF,  the normalisation evolves in accord with the self-similar scenario (the small amount of drift
at higher redshift is due to changes in the gas temperature, as discussed below). The amount of over-cooling in the CSF runs (which reduces the gas density) 
becomes more severe with time, leading to a normalisation that is around 70 per cent of the observed value at $z=1.4$ and 50 per cent at $z=0$. 
The slope exhibits significantly more scatter between redshifts than the normalisation, but there is still a clear difference between CSF and the other models. 

\subsection{The $T_{\rm sl}-M_{500}$ relation}

\begin{figure*}
\includegraphics[width=80mm]{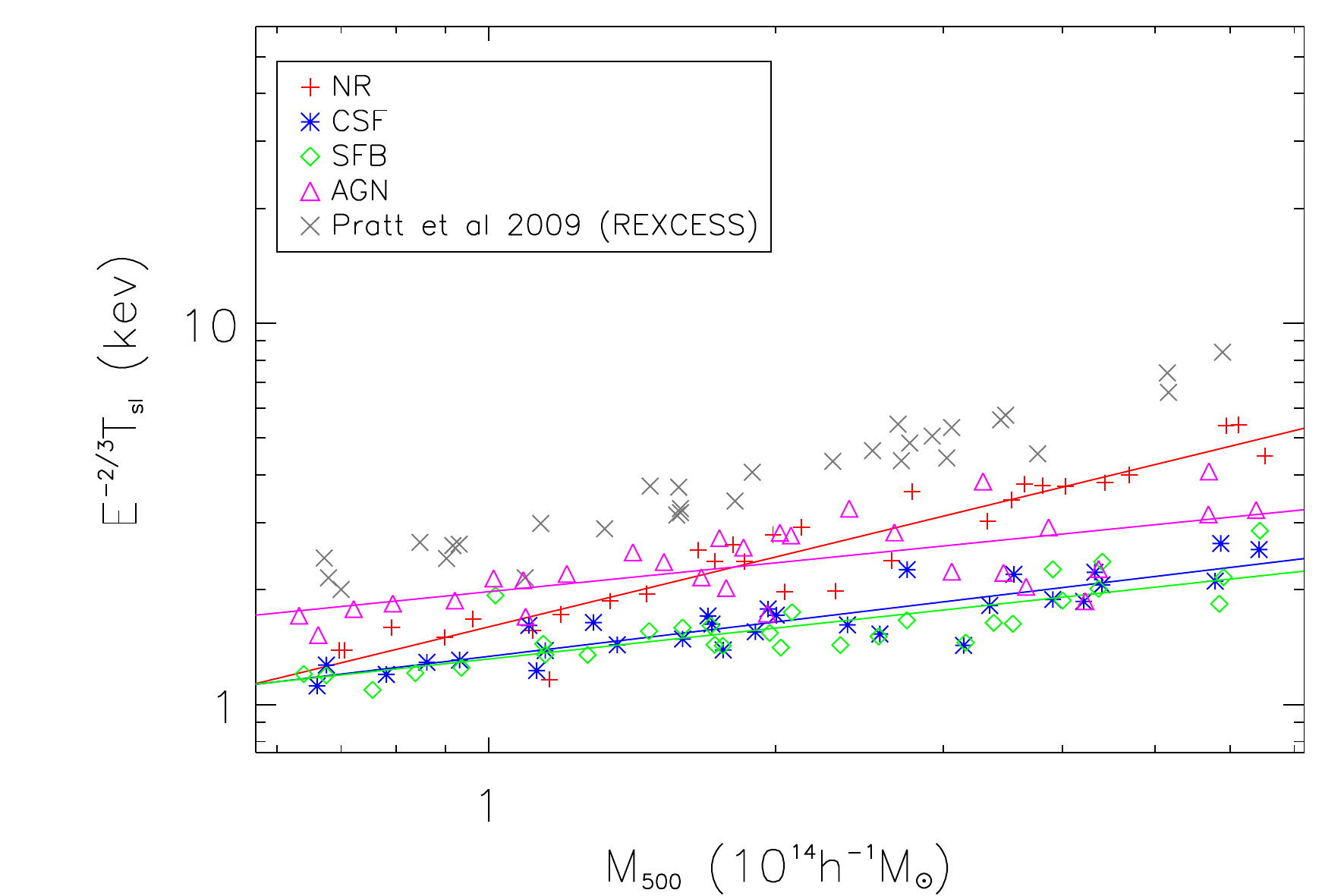}
\includegraphics[width=80mm]{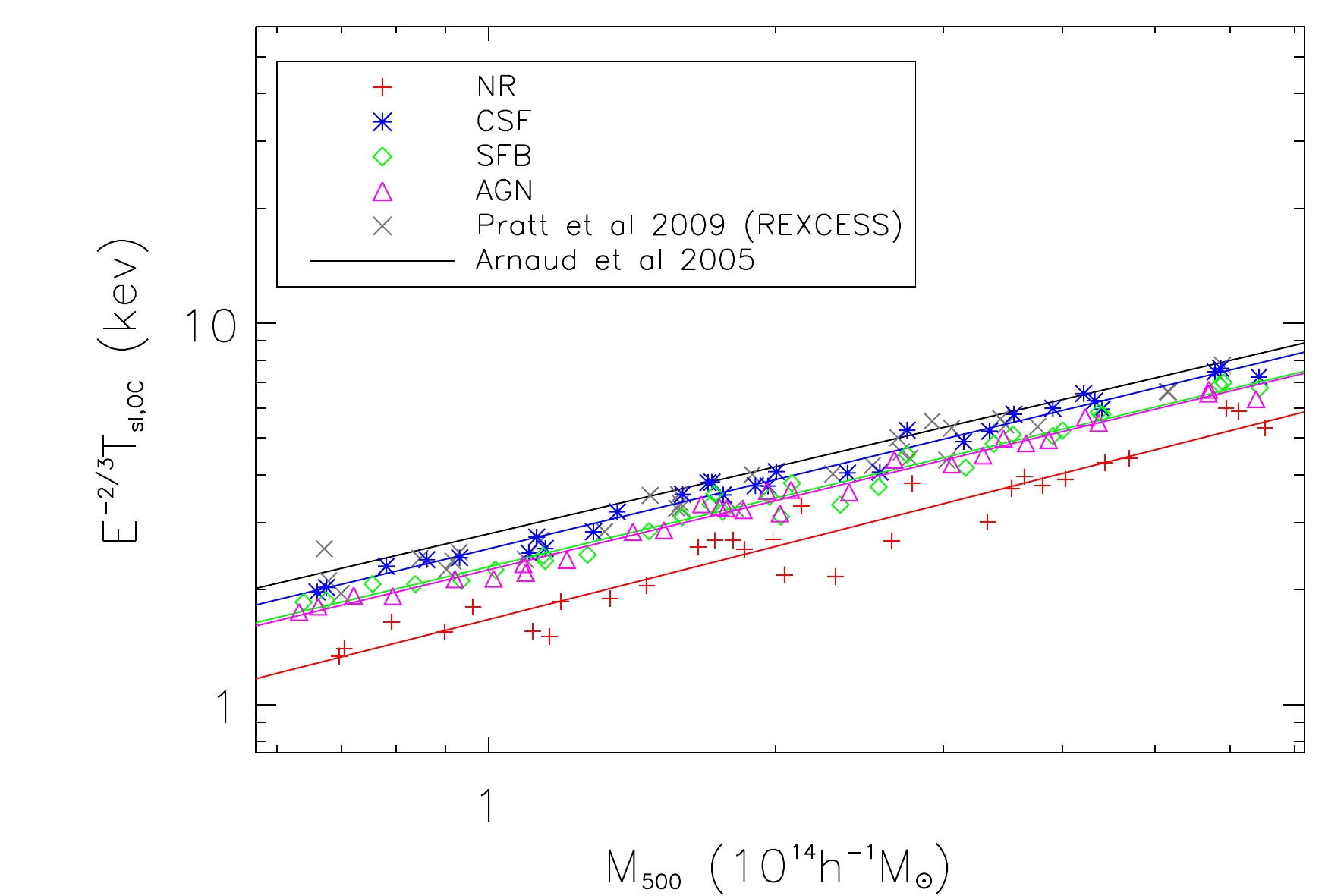}
\includegraphics[width=80mm]{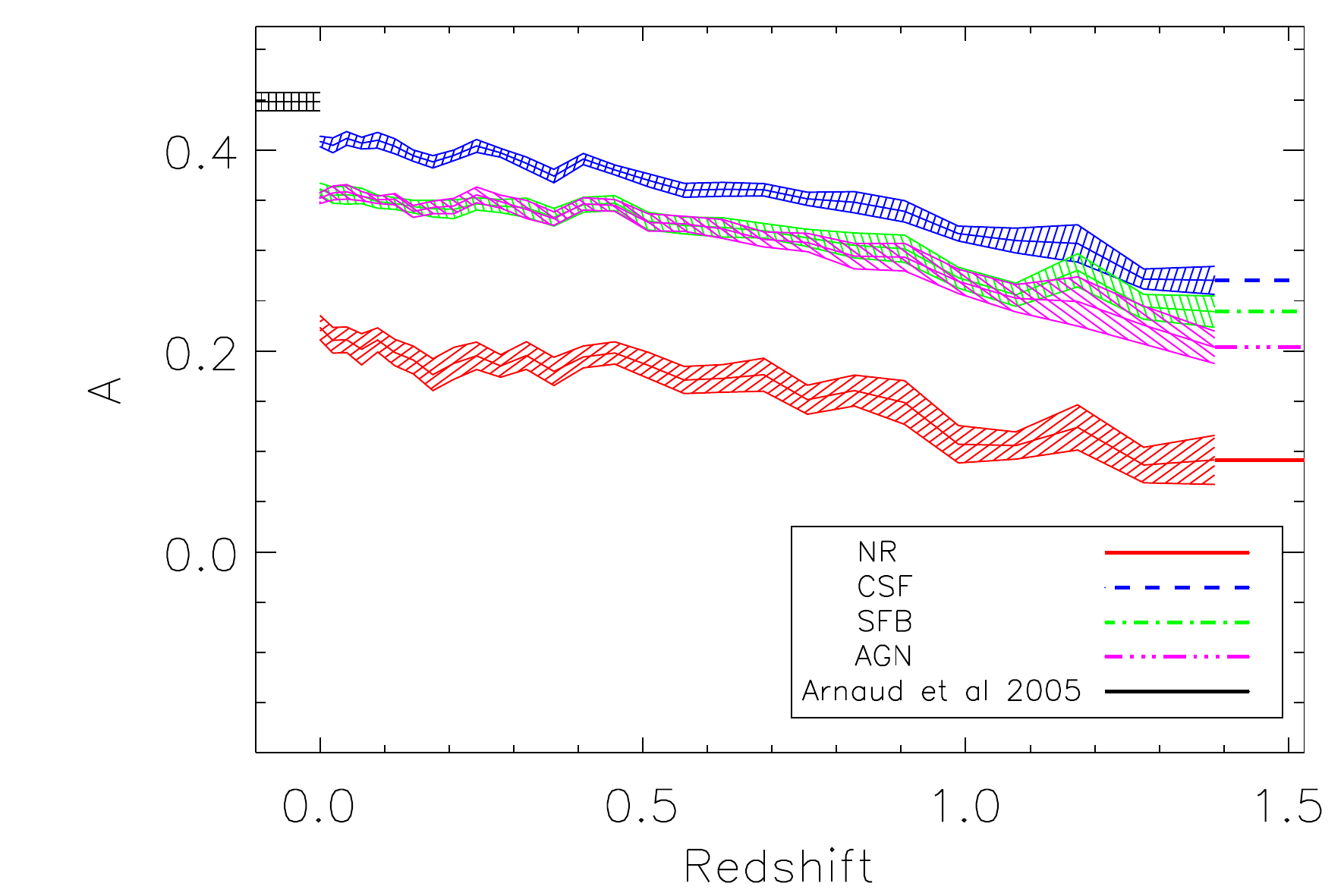}
\includegraphics[width=80mm]{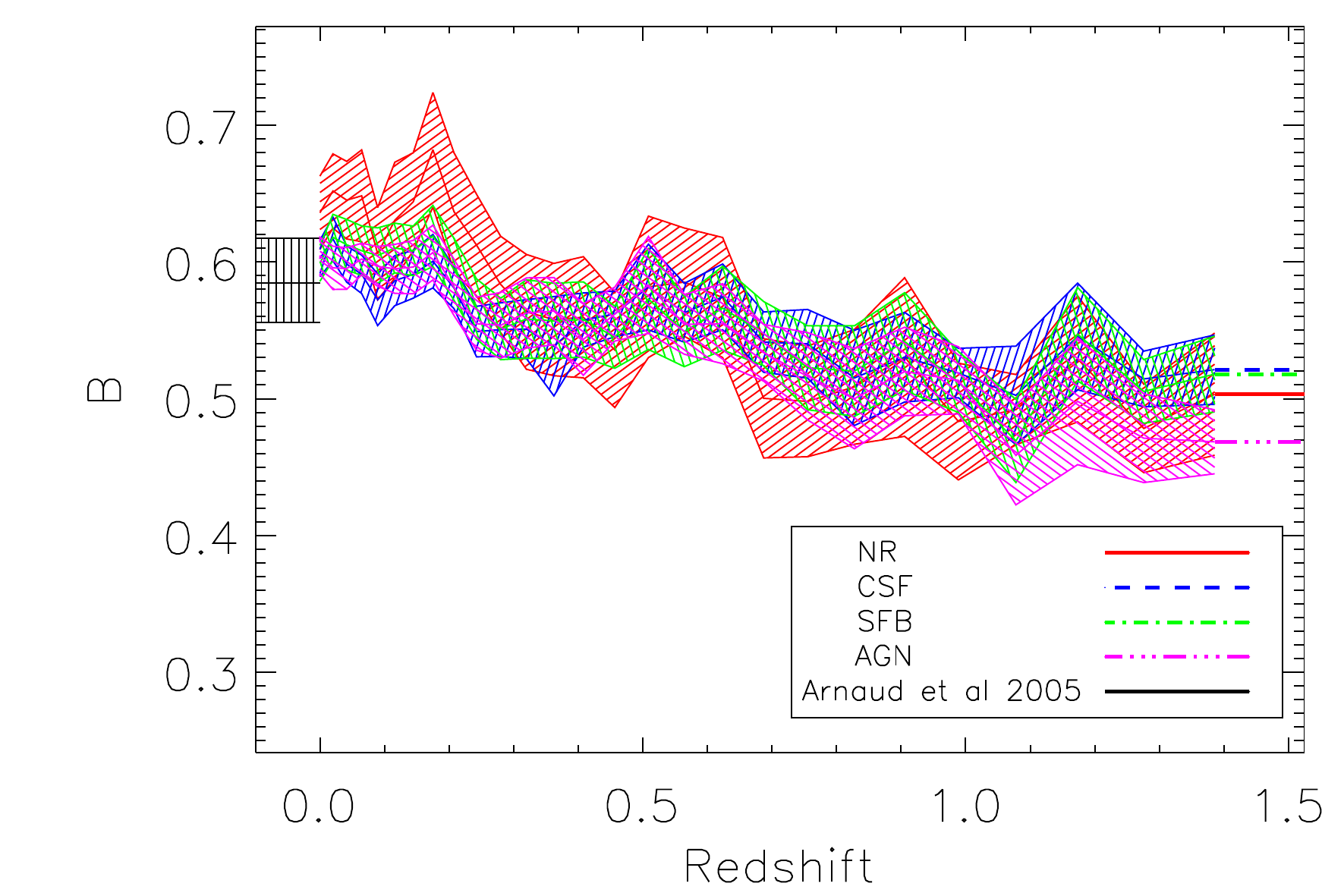}
\caption{$T_{\rm sl}-M_{\rm 500}$ scaling relations for the NR, CSF, SFB and AGN models at $z=0$ (top panels),
when the core is included (top-left) and excluded (top-right) from the temperature calculation. The bottom panels
show the dependence of the normalisation and slope parameters with redshift, when the core region is omitted.
Grey crosses and the black line/shaded region are observational results from \protect\cite{Pratt2009} and 
\protect\cite{Arnaud2005} respectively. All other details are as used in Fig.~\ref{plot:ymrel}.}
\label{plot:Txmrel}
\end{figure*}

Spectroscopic-like temperature versus mass relations are displayed in Fig.~\ref{plot:Txmrel}, where $T_{\rm sl}$
is calculated after the densest gas is removed from each shell \citep{Roncarelli2013}. The top 
panels show results at $z=0$; in the right panel, the core region ($r<0.15\, r_{500}$) was excluded from the temperature
calculation. We also show observational results from the REXCESS sample \citep{Pratt2009}.

None of the models match the observational data when the temperature is measured using all gas within $r_{500}$. The NR 
clusters have temperatures that are around 60 per cent of the observational values, with a slope ($B=0.61$) that is closest to the 
self-similar value ($B=2/3$). Including supernova feedback makes little difference to the temperature; 
only AGN feedback produces a significant increase, with the temperature being around 75 per cent  of the observed temperature at fixed mass.
However, the slope for the AGN model (and the other radiative models) is considerably flatter than the self-similar prediction. This is
because the more massive clusters have significantly higher fractions of cooler gas in the core that is still hot enough ($kT>0.5$ keV) to be included
in the $T_{\rm sl}$ calculation.

When the core is excluded, the NR results change very little at $z=0$ but the runs with cooling all predict temperatures that are closer
to the observational data (80-90 per cent of the observed values at fixed mass).  While the CSF and SFB
runs show the largest change (where feedback is absent and ineffective in the core, for the respective runs), an improved match
is also seen for the AGN model.  Furthermore,
all models have a slope close to the self-similar model at $z=0$, varying from $\sim 0.6$ for the AGN model to 
 $0.7$ for the SFB model; the AGN model is closest to the observational data ($B=0.58$). 

Studying the results at higher redshift, we find that the normalisation evolves negatively with redshift in all models, 
regardless of whether the core is included or not (results for the latter case are shown in the bottom panels of 
Fig.~\ref{plot:Txmrel}). We also checked the mass-weighted temperature-mass relation and a similar result was 
found, suggesting the result may be peculiar to the way in which our clusters were selected (larger, mass-limited 
samples would be required to check this). For the slope, when the core is included the lower values seen in the 
radiative models persist to high redshift, while the non-radiative value decreases slightly. When the core is excluded, 
all models exhibit similar behaviour (again, this is seen when considering the mass-weighted temperature).

\subsection{The $L_{\rm bol}-M_{500}$ relation}

\begin{figure*}
\includegraphics[width=80mm]{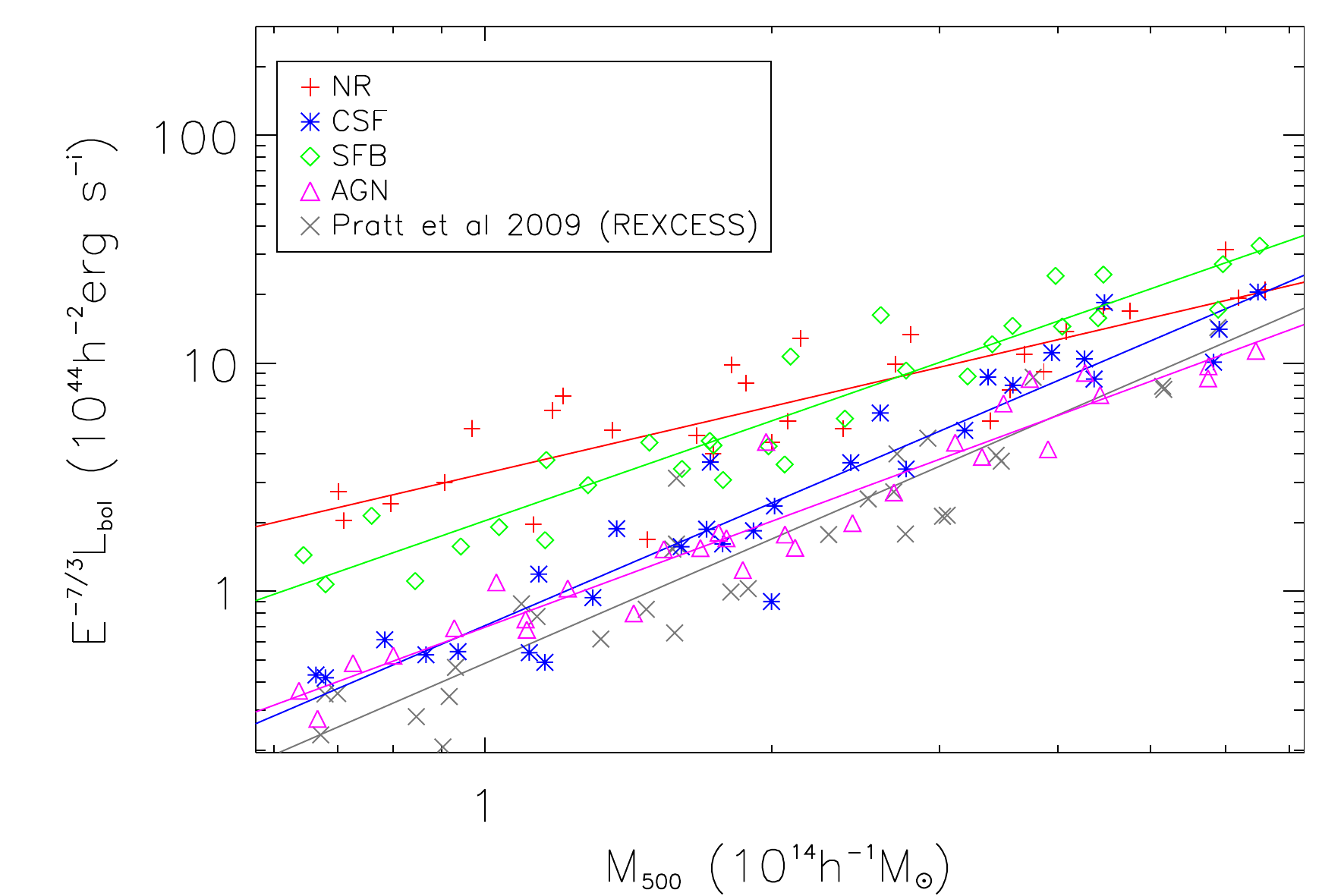}
\includegraphics[width=80mm]{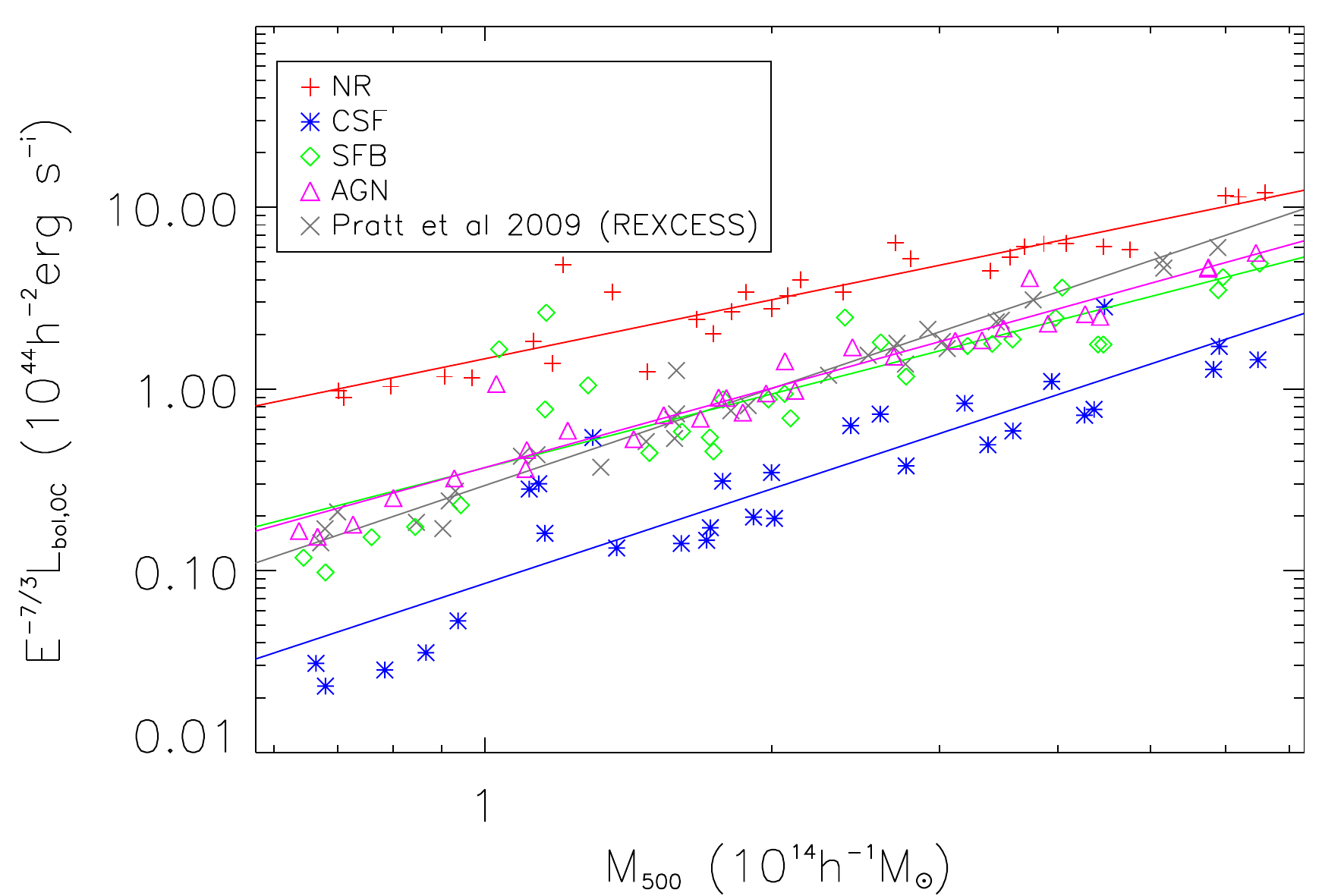}
\includegraphics[width=80mm]{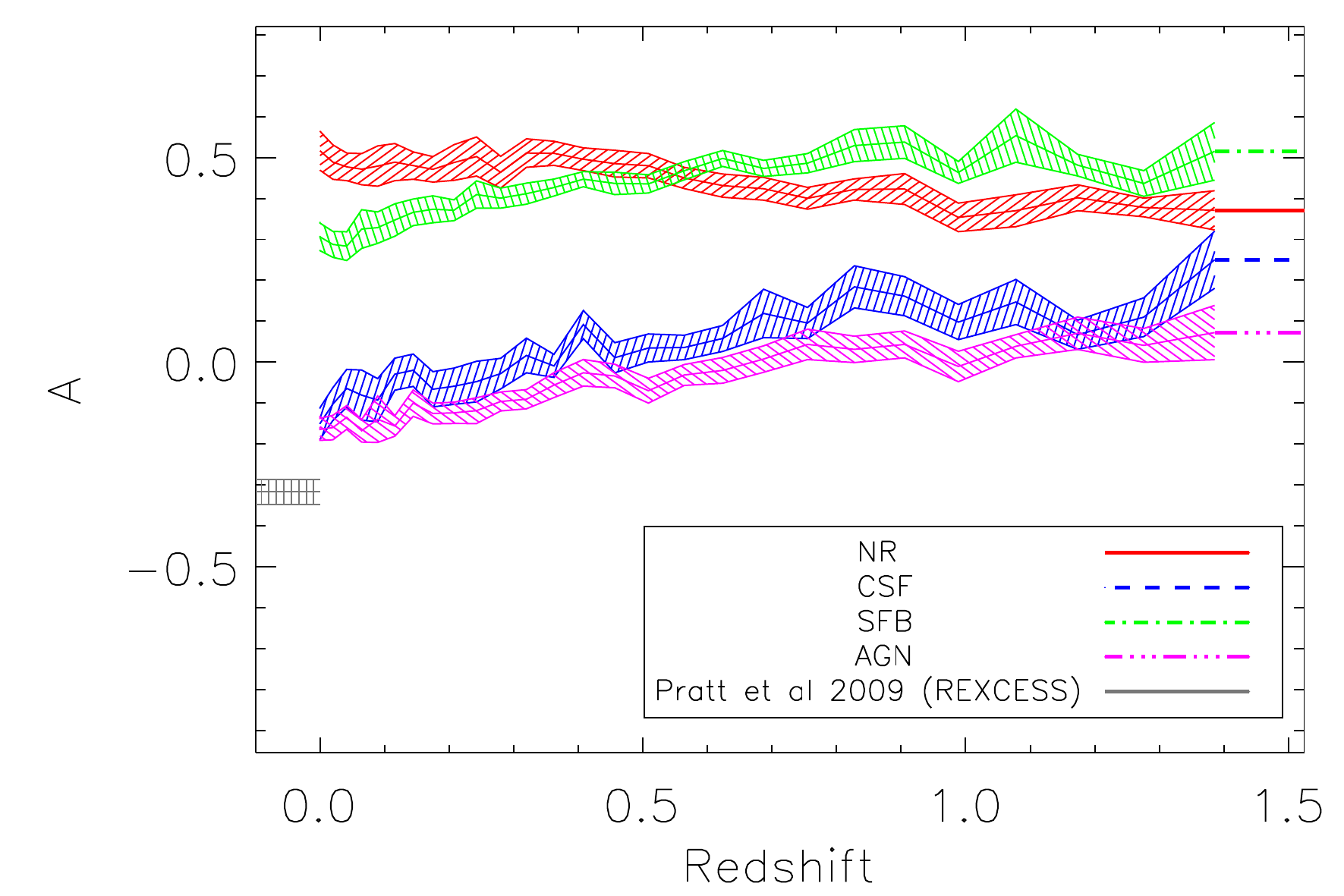}
\includegraphics[width=80mm]{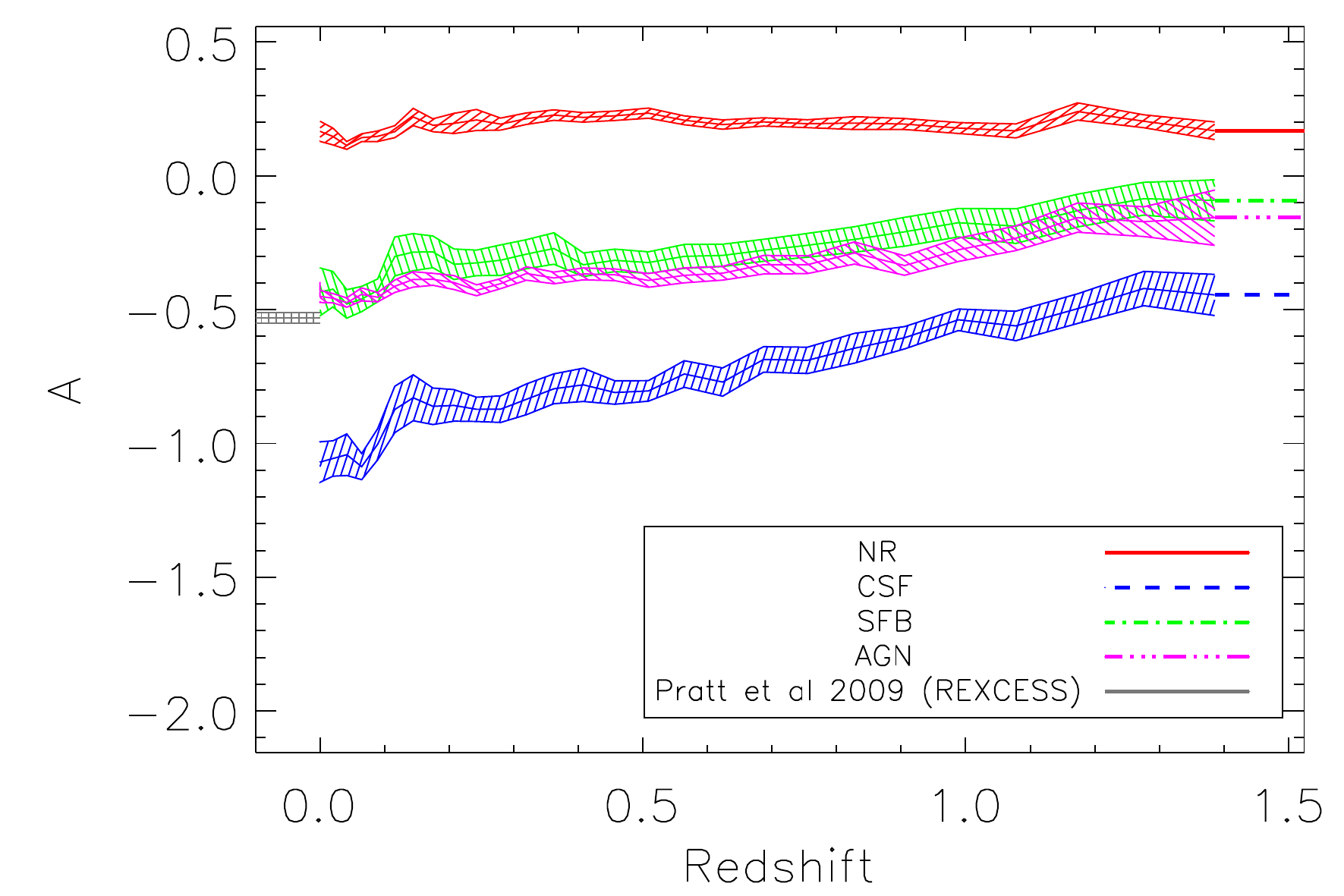}
\caption{$L_{\rm bol}-M_{\rm 500}$ scaling relations for the NR, CSF, SFB and AGN models at $z=0$ (top panels) and
evolution of the normalisation with redshift (bottom panels). Panels on the left (right) show results 
when luminosities are calculated including (excluding) the core region. Grey crosses/line/shaded region are observational 
results from REXCESS \protect\cite{Pratt2009}. All other details are as used in Fig.~\ref{plot:ymrel}.}
\label{plot:Lmrel}
\end{figure*}

Finally, results from X-ray luminosity scaling relations are displayed in Fig.~\ref{plot:Lmrel}. The panels on the left are
for all emission within $r_{500}$ and the right when the core ($0.15 \, r_{500}$) is excluded. Results from
each simulation model are shown as before and we also show observational data points from REXCESS
(\citealt{Pratt2009}; grey crosses). 

As expected, clusters in the NR model are over-luminous, both with and without the core, due to the fact that the 
gas is too dense at all radii. The slope ($B=0.97 \pm 0.09$) is flatter than the self-similar value ($B=4/3$). While
part of this discrepancy could be due to sample selection given the large intrinsic scatter ($S=0.18 \pm 0.02$), 
the main reason is that the lower mass clusters are sufficiently cold ($kT \sim 1-3$ keV) that line emission makes
a significant contribution to the luminosity (the cooling function is approximately constant at these temperatures,
for $Z=0.3\,Z_{\odot}$). 

In the CSF run, cooling causes a significant drop in luminosity, driven primarily by the 
decrease in density as the gas cools below $T=10^5$K and forms stars. It is interesting that the results match the
observations reasonably well when all emission is included, but the CSF clusters are under-luminous when the core
is excluded. Again, the former result is well known (e.g. \citealt{Bryan2000,Muanwong2001}), being due to the effect of
cooling removing the dense, low entropy gas. However, this effect produces density (or entropy) profiles with the
wrong shape: the density is too low beyond the core and too high in the centre. This leads to the core-excluded 
relation being too low.

Supernova feedback increases the density of the gas within $r_{\rm 500}$ due to reduced star formation, resulting 
in a luminosity profile that is higher than for CSF across the whole radial range. This leads to a luminosity that is also too 
high inside the core (due to the supernovae being ineffective at suppressing the cooling there) but matches the observed
luminosities if the core is excluded. Finally when AGN feedback is included the density and therefore luminosity in the 
central region is reduced as gas is expelled, but this has a lesser effect on the outskirts. Thus, the AGN relation provides 
a better match to the observed mean relation in both cases.

The intrinsic scatter in the $L_{\rm bol}-M_{500}$ relation is similar to the REXCESS observations (0.17) for the 
NR and CSF runs, but is too small in the AGN model ($S=0.12 \pm 0.02$). This again points to the fact that, in our
most realistic model, the full range of cool-core and non-cool core clusters is not recovered. When the core is excised,
the scatter decreases in the NR and AGN cases, but actually increases in the CSF and SFB runs. Closer inspection reveals
that this is due to a few objects with unusually high luminosities, caused by the presence of a large substructure outside
the core. The effect of this substructure is diminished in the AGN model, where the extra feedback reduces the amount of cool,
dense gas in the object.

The slope in the NR model does not evolve with redshift and remains $ \sim 60$ per cent of the present observed value. Some
evolution is seen at low redshift for the radiative models, but when the core is excluded there is no evidence for substantial evolution in any
of the models. However, the change in normalisation with redshift is much more interesting and can be seen in the bottom panels
in Fig.~\ref{plot:Lmrel}. All radiative models predict higher luminosities at higher redshift (for a fixed mass) than expected from 
the self-similar model. Importantly, the amount of evolution is similar in the CSF, SFB and AGN models, 
but their normalisation values are offset from one another at a given redshift. In general, the differences we see at $z=0$ are 
largely replicated at the other redshifts. This suggests that the departure from self-similar evolution in the radiative models is 
largely driven by radiative cooling, with both feedback mechanisms largely serving to regulate the gas fraction, with AGN more 
effective in the inner region and supernovae further out. This is consistent with the entropy profile having a similar shape at $z=0$ in
all three radiative models.

\section{Resolution Study}
\label{sec:res}
\begin{table}
\centering
  \caption{Details of the resolution tests performed. Column~1 lists the label given to each run; column~2 the approximate number
  of dark matter within $r_{200}$; column~3 the gas particle mass (in $10^{8} \, h^{-1} \, {\rm M}_{\odot}$) and column~4 the maximum value
  of the (Plummer-equivalent) softening length (in $h^{-1} \, {\rm kpc}$). Labels in bold represent runs with softening lengths chosen
  using the method outlined in \protect\cite{Power2003}.}
  \begin{tabular}{lcrr}
  \hline   
  Label & $N_{200}$ & $m_{\rm gas}$ & $\epsilon_{\rm max}$\\
\hline           
\bf{VLR-LS} & $10^5$ & 12.0 & $19.0$\\
VLR-MS & $10^5$ & 12.0 & $6.0$\\
VLR-SS & $10^5$ & 12.0 & $1.9$\\
\bf{LR-MS} & $10^6$ & 1.3 & $6.0$\\
LR-SS & $10^6$ & 1.3 & $1.9$\\
\bf{HR-SS} & $10^7$ & 0.14 & $1.9$\\
\hline   
\end{tabular}
\label{tab:ResRuns}
\end{table}

An important issue that we have yet to discuss is the effect of numerical resolution. Resolution can be split into two components: 
the spatial resolution which is governed by the gravitational softening length (and minimum SPH smoothing length for the gas), 
and the mass resolution which is governed by the mass of the dark matter, gas and star particles. 

In order to investigate mass resolution effects, new initial conditions were generated for our most massive
cluster ($M_{\rm 200} \simeq 10^{15} \, h^{-1} M_{\odot}$) with ten times fewer ($N_{\rm 200}\simeq 10^5$) and ten times greater 
($N_{\rm 200} \simeq 10^7$) particles than our default value ($N_{200} \simeq 10^6$). When the number of particles was increased,
additional small-scale power was added in the initial conditions allowing smaller mass haloes to be resolved. We shall refer to this 
sequence of runs (going from the smallest to largest particle number) as VLR-LS, LR-MS and HR-SS respectively. 
In all three cases, the softening 
length were computed using the method outlined in \cite{Power2003} and the minimum SPH smoothing length was set equal
to this value. To specifically test the effect of spatial resolution, the LR and VLR clusters were also run with smaller softening lengths. 
Table \ref{tab:ResRuns} summarises the details of the runs. 

We first performed tests for the NR model, which 
allows us to check for the severity of two-body heating effects, expected to occur if the particle mass is too large and
the softening too small. Such heating creates an artificial core in the density profile beyond the softening scale (i.e. for $r>2.8 \, \epsilon$), 
as energy is transferred from the dark matter to the gas. We found evidence for two-body heating in the VLR-MS, VLR-SS and LR-SS runs, 
somewhat vindicating our default choice of softening from \cite{Power2003}.  Two-body heating effects are reduced in the SFB case, 
as the cooling is able to dissipate this additional heat \citep{Steinmetz1997}. This, however, does not mean that two-body heating is no 
longer an issue as it will still affect the evolution of the dark matter, which may in turn affect the gas and stars through changes to the 
gravitational potential. Given this complexity and the limited sample, one must be conservative about any conclusions drawn.
For the remainder of this section, we focus on the AGN model only.

\begin{figure}
  \includegraphics[width=80mm]{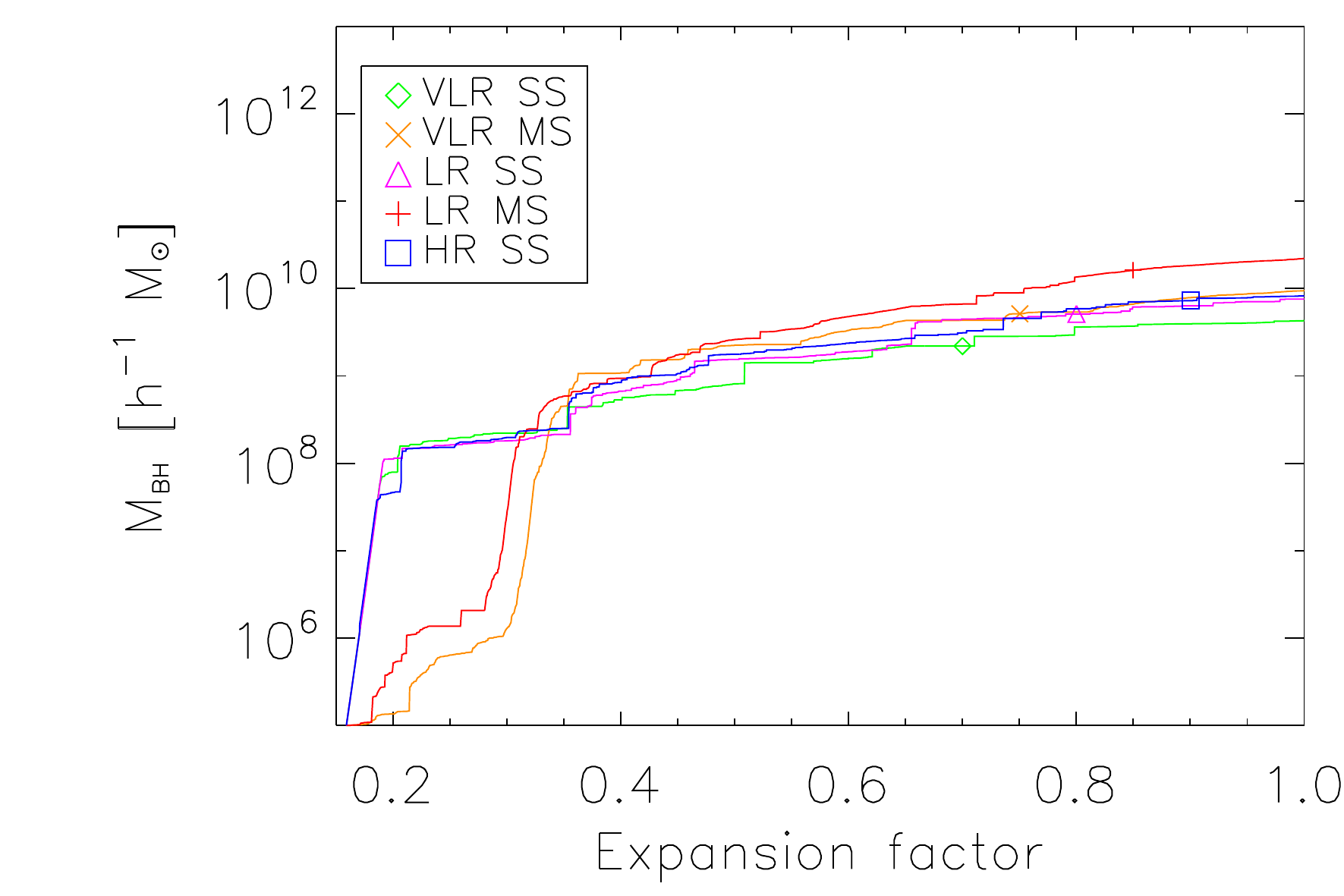}
  \includegraphics[width=80mm]{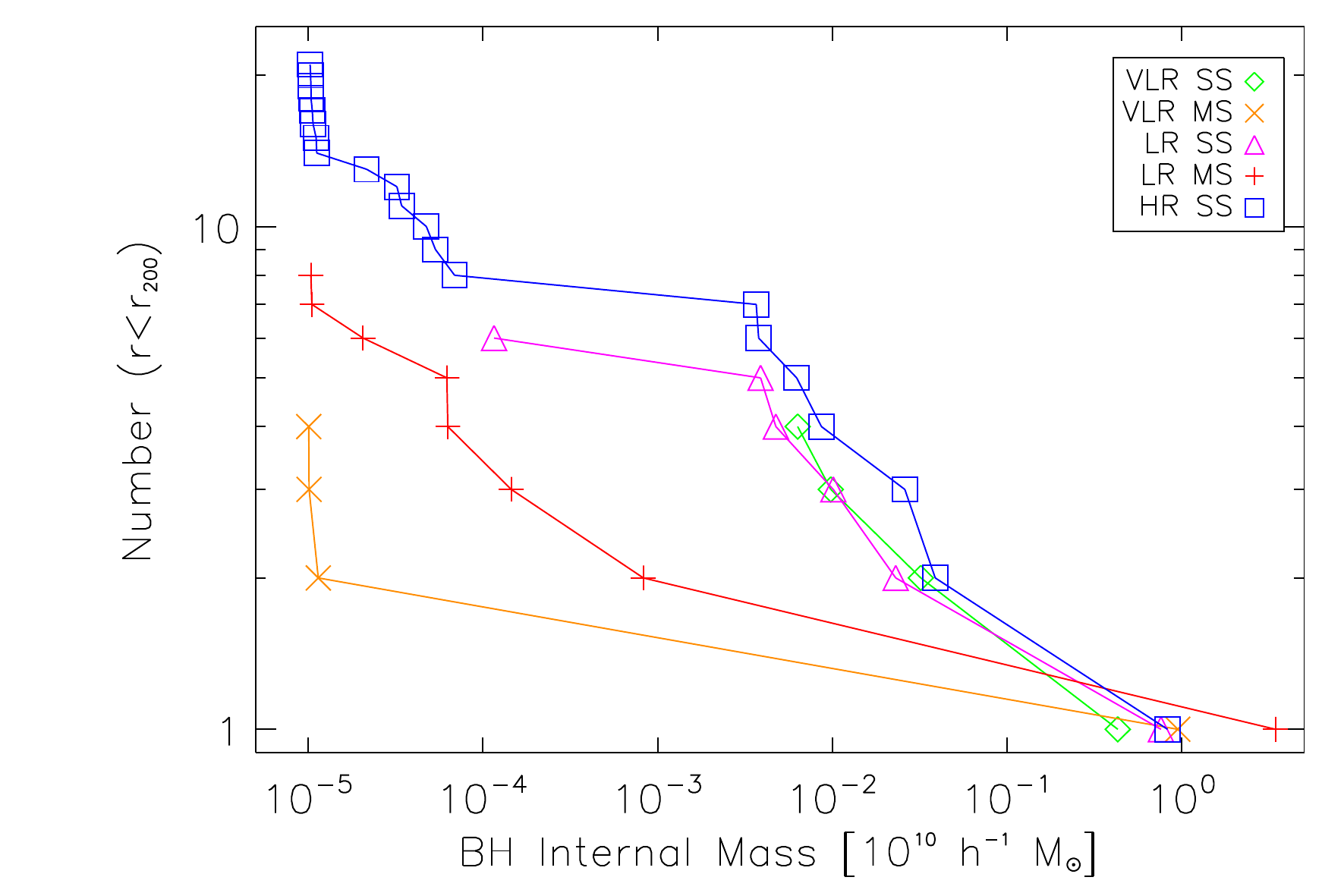}
  \includegraphics[width=80mm]{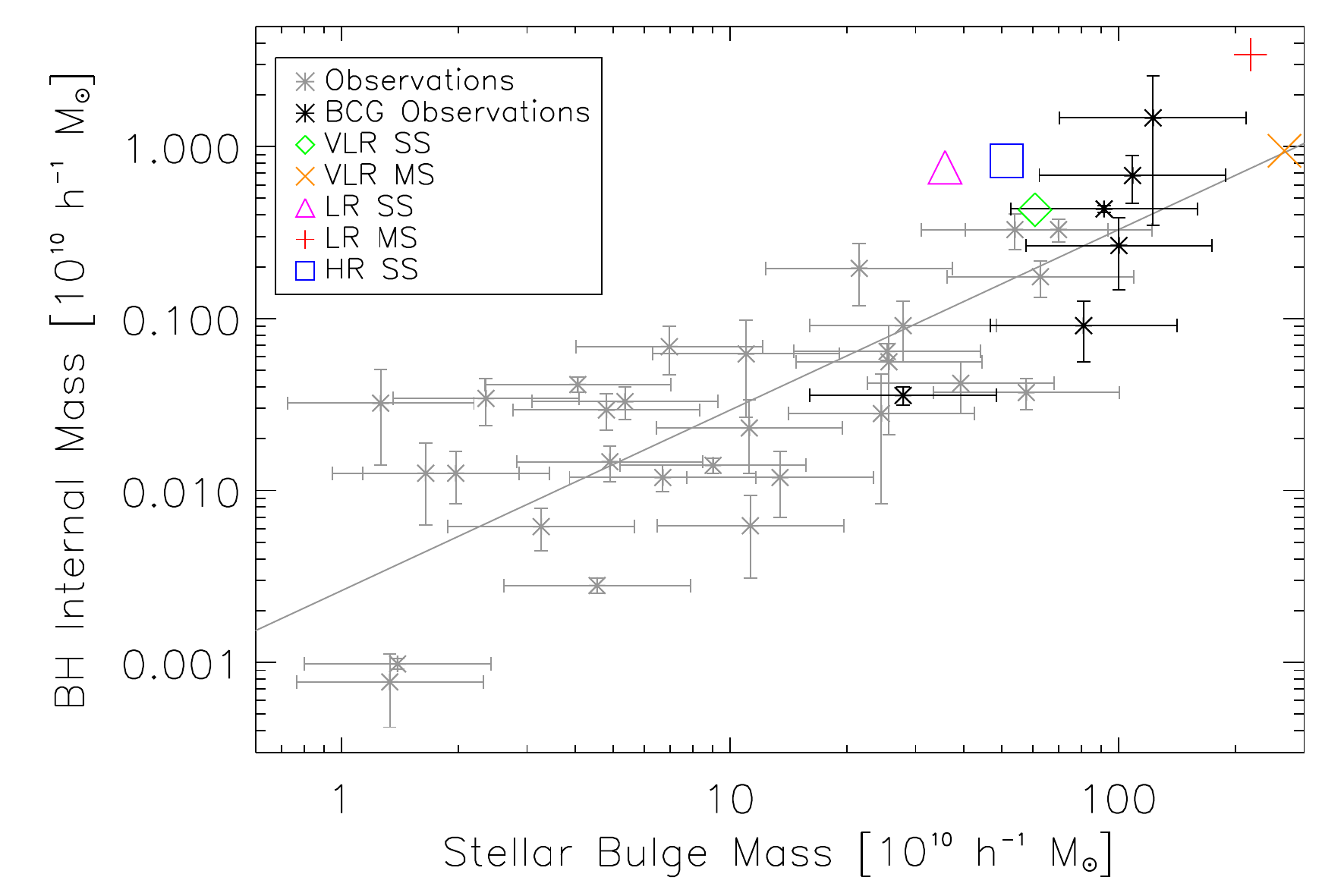}
  \caption{Growth of the largest black hole with expansion factor (top panel), 
  cumulative black hole mass function within $r_{\rm 200}$ at $z=0$ (middle) 
  and central black hole mass versus bulge mass relation (bottom) for our most massive cluster, run with varying resolution and the AGN 
  physics model.}
\label{plot:res_bh}
\end{figure}

\subsection{Black Hole properties}

Fig. \ref{plot:res_bh} shows the growth of the central black hole in the various runs (top panel), the $z=0$ 
cumulative black hole mass function for 
objects within $r_{200}$ (middle) and the $z=0$ central black hole versus stellar bulge mass relation (bottom). 
It is immediately clear that the choice of softening length has a significant effect on the initial growth of the black hole, while
the mass resolution is less important.
The VLR-SS, LR-SS and HR-SS runs, which all have softening lengths of $\epsilon=1.9 \, h^{-1}{\rm kpc}$, exhibit rapid,
Eddington-limited growth until $a \simeq 0.2$. On the other hand, the VLR-MS and LR-MS runs (with $\epsilon=6\,h^{-1}{\rm kpc}$)
do not start growing rapidly until $a \simeq 0.3$. (We found that the black holes in the VLR-LS run were unable to grow at all, so do not
show these here.) The softening length affects the accretion rate in two ways. Firstly, the smaller softening results in a deeper 
gravitational potential around the black hole, allowing a more rapid build-up of mass. Secondly, as the minimum SPH smoothing length
is tied to the softening in our runs, a larger density is estimated for the gas local to the black hole. 

The smaller softening also allows the black holes associated with satellite galaxies to grow more efficiently, as can be seen from the 
black hole mass function. Note that our default choice of resolution (LR-MS) produces the most massive black hole, with the second 
most massive object being more than three orders of magnitude smaller. 
In addition to the effects of the softening on the accretion rate, we also checked whether the black hole mass function 
is affected by over-merging of satellite black holes on to the central object; a smaller softening would make this less likely. However, 
we found this was not important, at least for the most massive objects which are always associated with the same substructures.

Central black hole mass versus stellar bulge mass is displayed in the bottom panel of Fig.~\ref{plot:res_bh}. 
It is clear that, as well as affecting the growth of the largest black hole, resolution also has an effect on the 
final mass of the stellar bulge. Runs with a smaller softening length produce a smaller bulge; efficient early
growth (and therefore AGN feedback) is clearly important for the growth of the central BCG. 

\subsection{Star formation history}

\begin{figure}
\includegraphics[width=80mm]{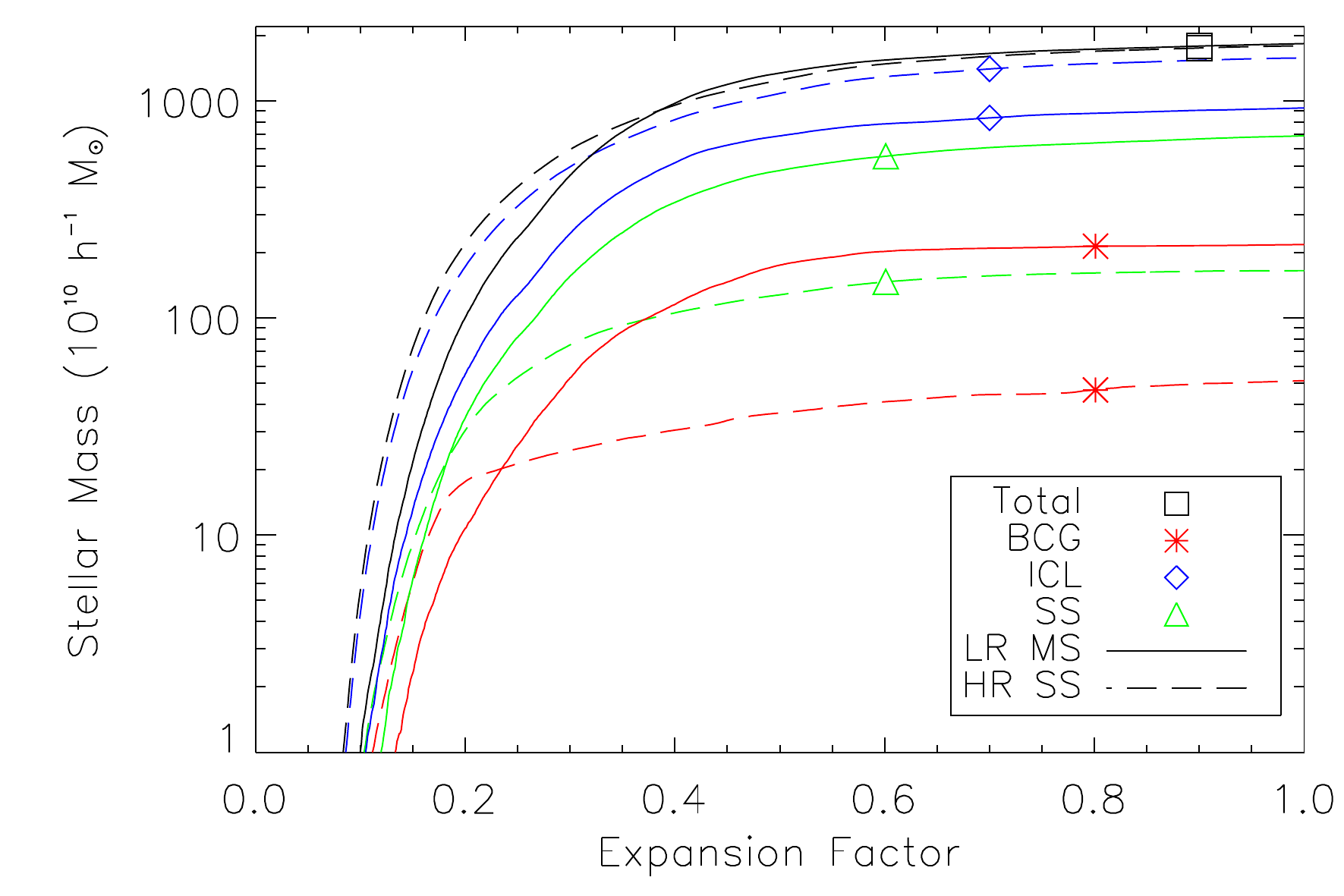}
\includegraphics[width=80mm]{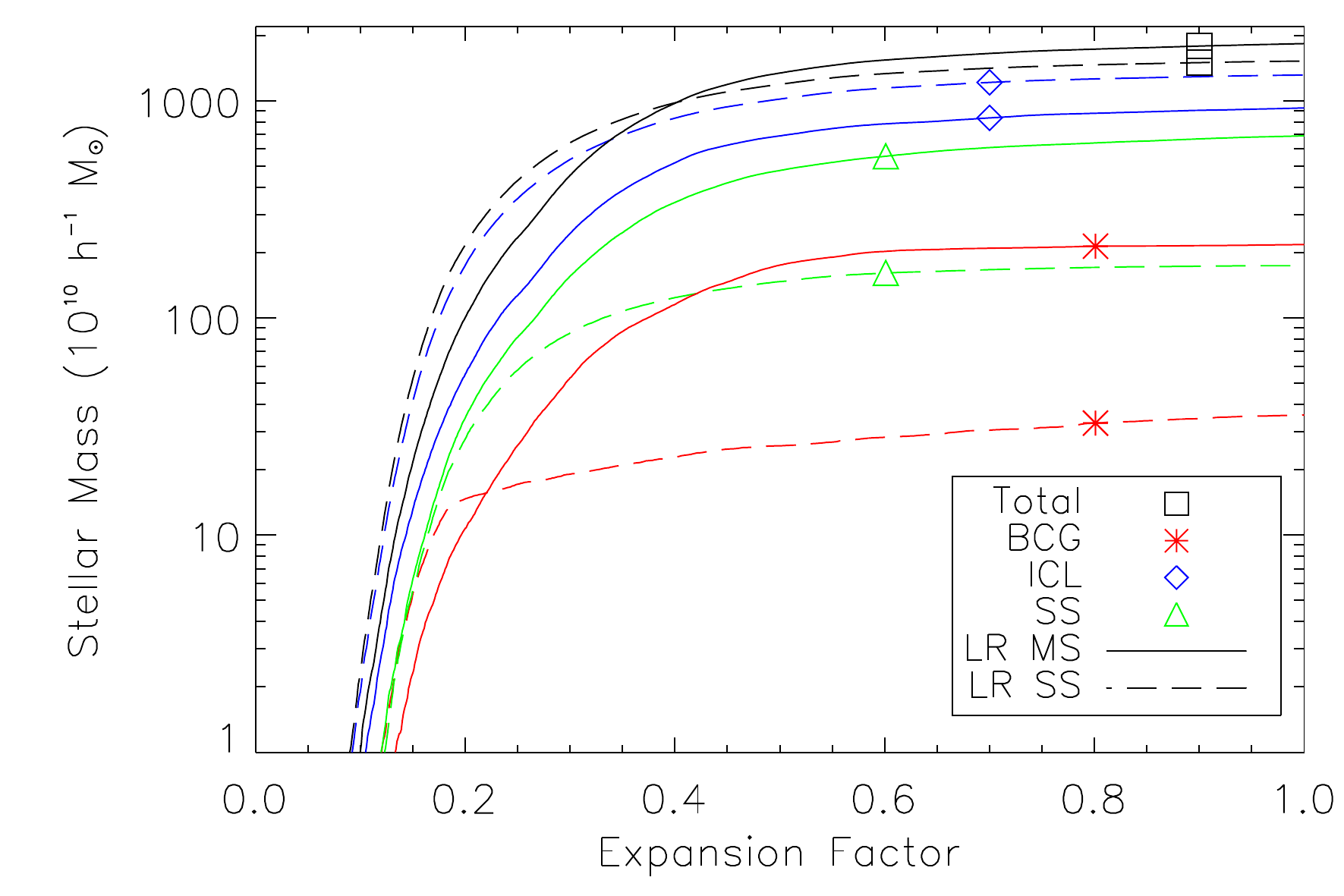}
\caption{Cumulative stellar mass formed at each value of $a$, that ends up within $r_{200}$ at $a=1$. Solid curves are for the
default LR-MS run, while the dashed curves are for the HR-SS (top panel) and LR-SS (bottom panel) runs. The results are also
split into BCG, ICL and SS sub-components (as described in the legend).}
\label{plot:res_sfh}
\end{figure}

To further investigate the effect of resolution on the cluster's star formation history, we show in Fig.~\ref{plot:res_sfh} the 
stellar mass formed by each value of $a$, that ends up within $r_{200}$ at $a=1$. We also split this mass into the various 
sub-components (BCG, ICL and SS) as discussed in Section~\ref{sec:sfh}.

In the top panel, we compare our default-resolution (LR-MS; solid curves) to the higher-resolution simulation (HR-SS; dashed curves), 
while the effect of softening alone can be seen explicitly in the bottom panel (LR-MS with LR-SS; the VLR results are similar). 
Increasing the resolution makes little difference to the final stellar mass in the halo, although more stars form at early times
($a<0.3$). This is expected, given that smaller-mass objects can be resolved in the HR-SS simulation. However, some of the effect 
is also due to the change in softening length (as can be seen from comparing the solid and dashed curves in the
bottom panel). As with the black holes, stars begin to form earlier when the softening length is smaller due to the deeper 
potential and higher gas densities in the halo centres.

As discussed above, the runs with smaller softening lengths also have significantly fewer 
stars in the BCG at $z=0$, but this is also true for the other galaxies (SS). Consequently, the ICL mass has increased, so the runs
with smaller softening lengths appear to have increased amounts of stripping. 
A simple explanation for this is that the stars in the sub-haloes are being puffed up due to two-body heating. While this would be 
expected to be larger in the LR-SS run (due to the smaller softening), it would also be less severe in HR-SS (due to the smaller
dark matter particle mass). It is therefore unlikely that this is the cause, given that a similar increase in ICL mass is seen in both runs.
An alternative explanation is that the stronger feedback at early times leads to cluster galaxies being less bound. Thus, more stars 
are stripped from the SS before they have a chance to merge with the central BCG. 

This also has implications for the evolution of the BCG which, we find, grows much less rapidly at $z<1$ in the runs with smaller softening 
lengths. In the LR-MS run the BCG grows by almost a factor of 30 since $z=1$ (c.f. the sample median value of a factor of 
5, as discussed in Section~\ref{sec:sfh}). However, in the LR-SS run (with the same mass resolution), the BCG has a slightly higher
mass than the LR-MS object at $z=1$ and grows by only a factor of 3 or so by $z=0$. Again, nearly all the growth comes from dry 
mergers but the total mass in these merger events is now much smaller. Thus, in our model, the growth rate of BCGs at $z<1$ 
is also sensitive to the adopted softening length but a slower rate (as desired) comes at the price of a larger ICL component. 
Simulations with higher resolution
will be required to investigate this further, taking also the stellar mass function of galaxies into account.

\subsection{Cluster profiles}

\begin{figure*}
\includegraphics[width=80mm]{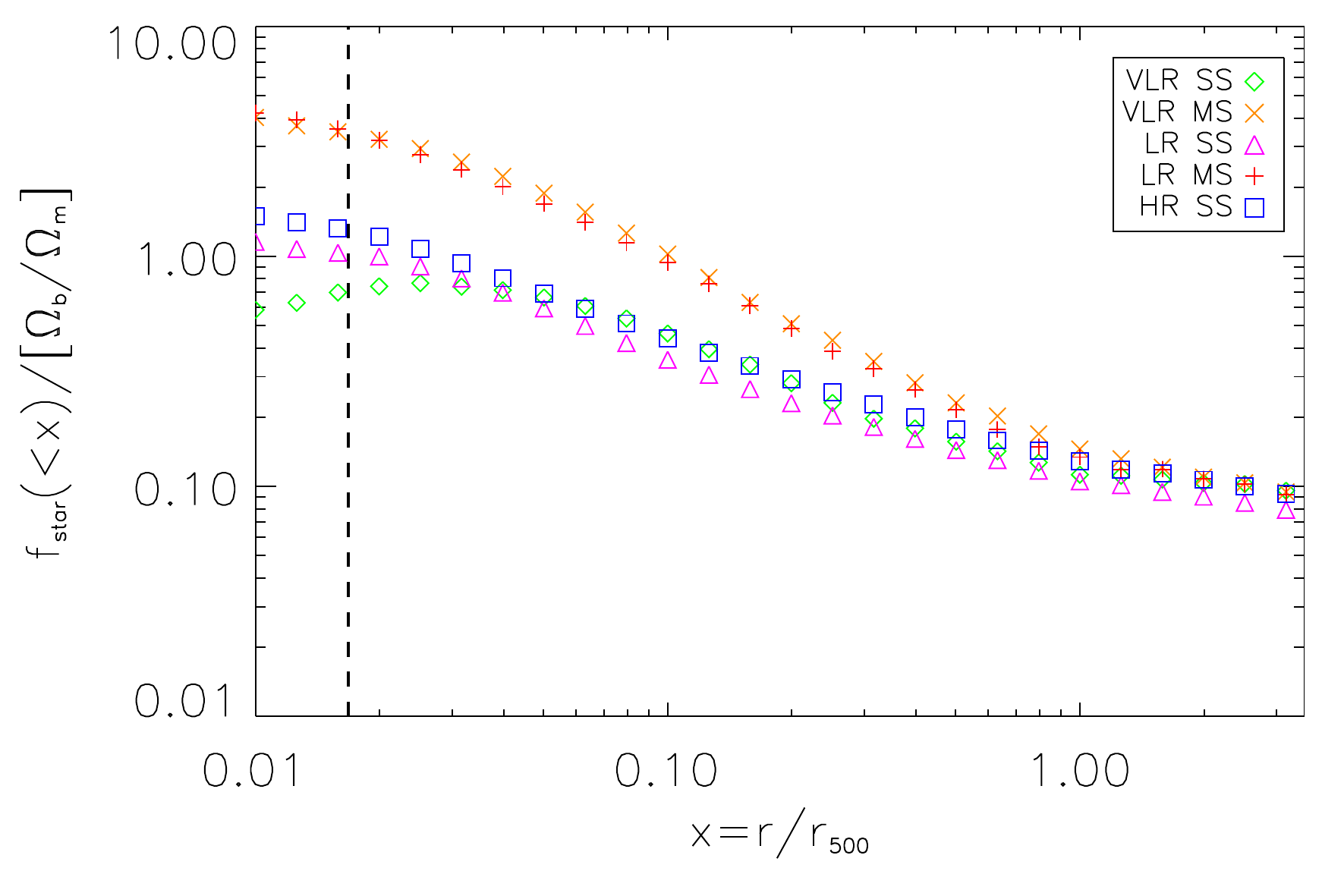}
\includegraphics[width=80mm]{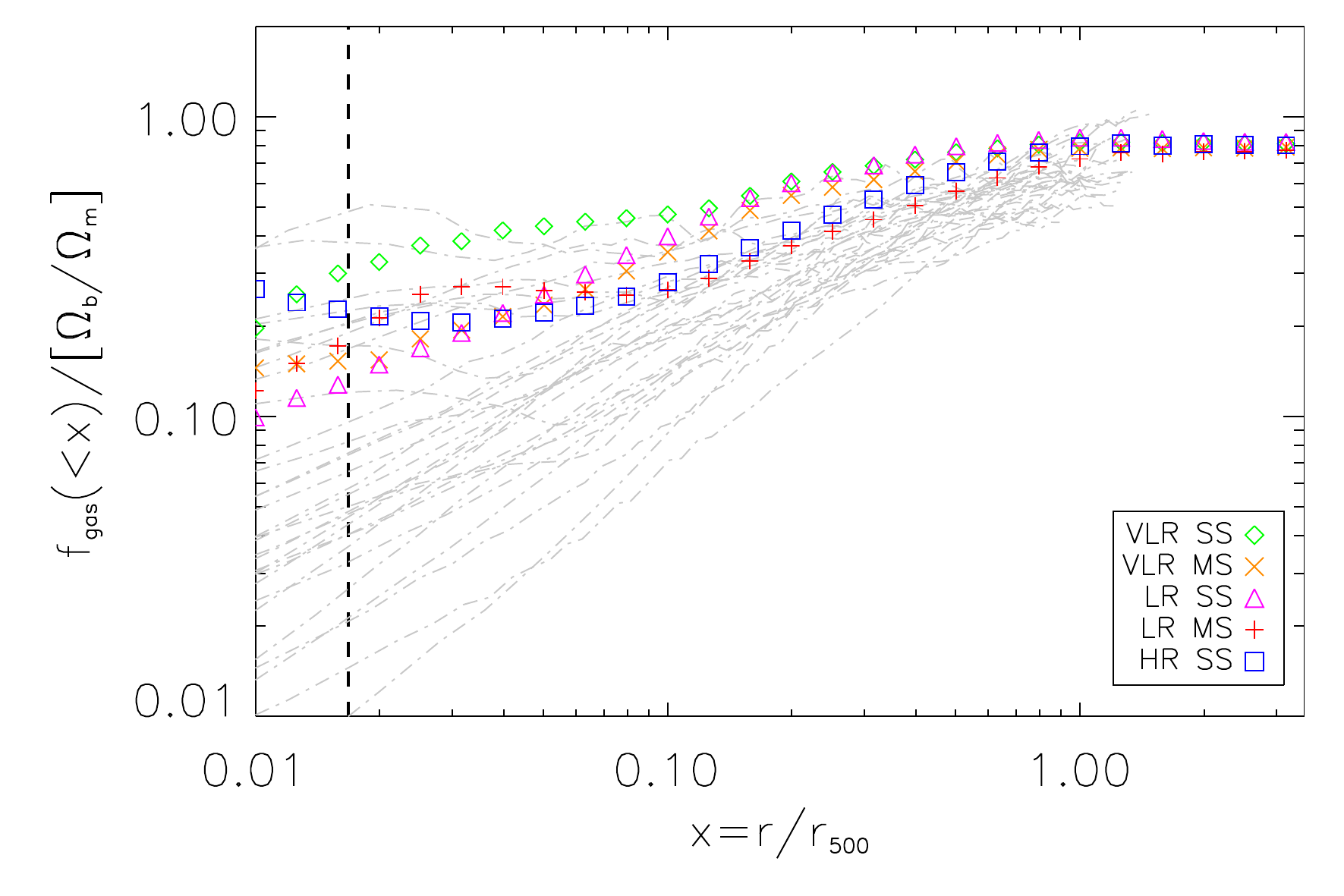}
\includegraphics[width=80mm]{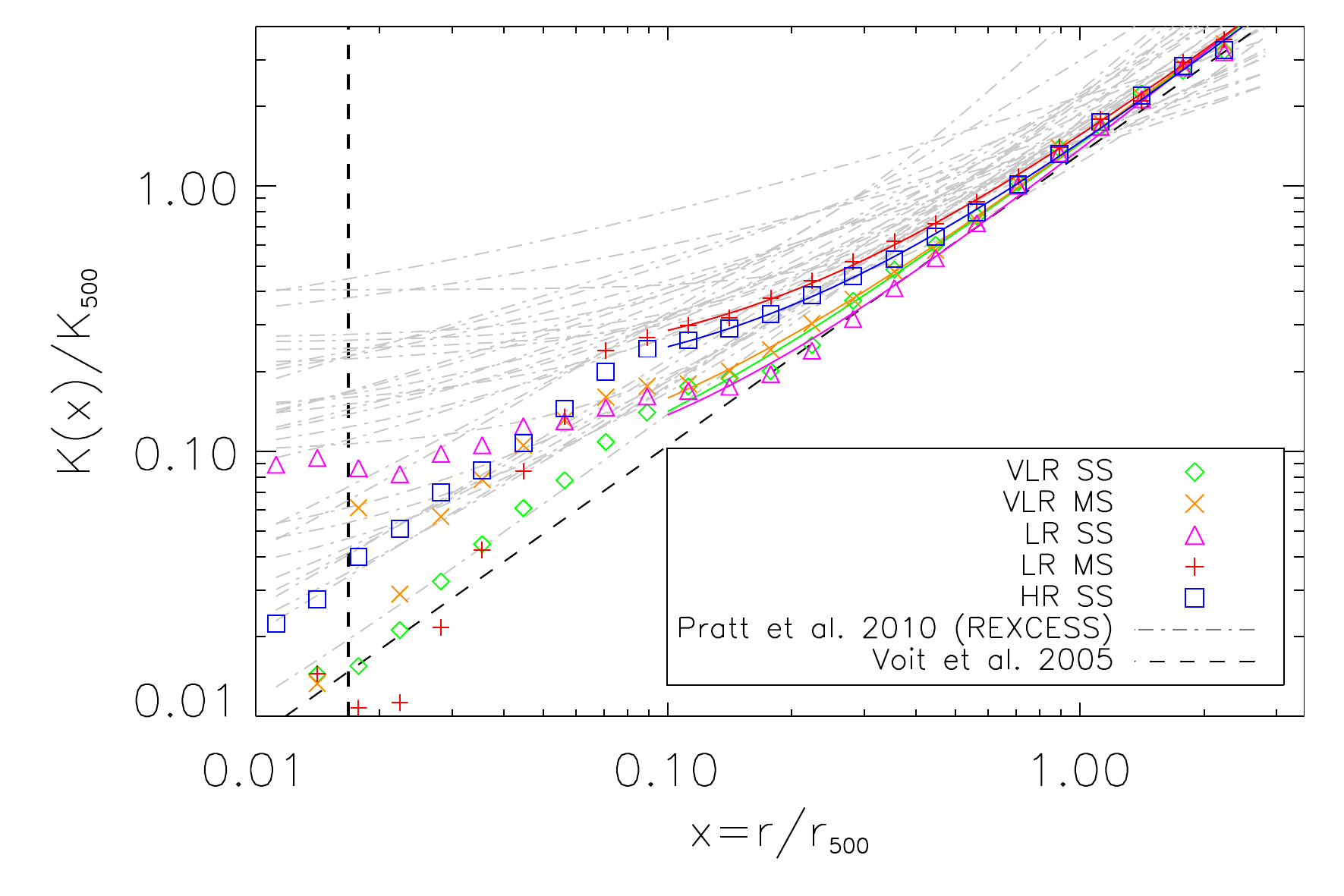}
\includegraphics[width=80mm]{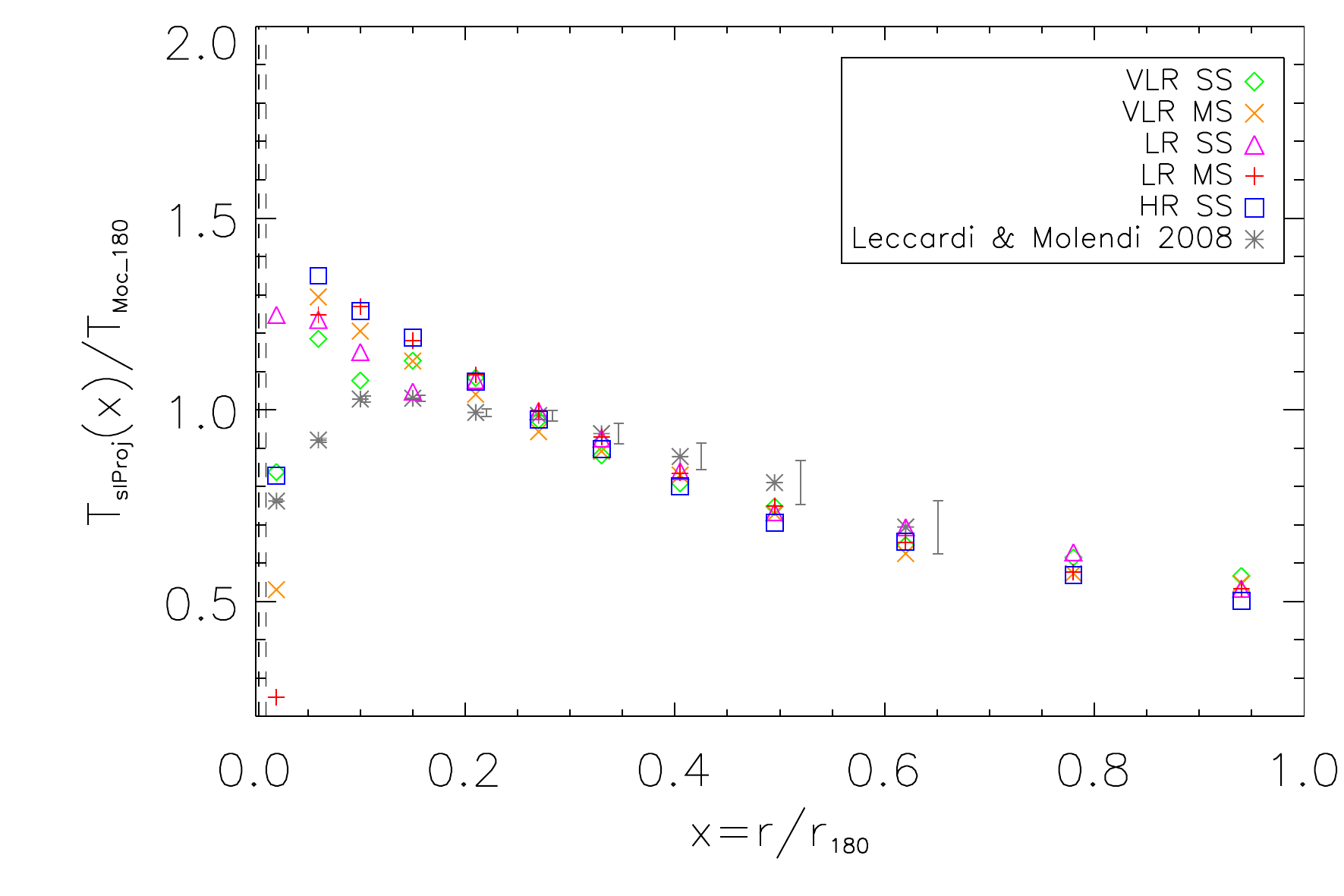}
\caption{Radial gas and star profiles at $z=0$ for the AGN runs with varying resolution. Clockwise from the top-left panel
are the cumulative star fraction, hot gas fraction, projected spectroscopic-like temperature and entropy respectively. The
vertical dashed lines represent the minimum resolved scale for the SS and MS runs respectively.}
\label{plot:res_bary}
\end{figure*}

The top-left panel of Fig. \ref{plot:res_bary} displays the cumulative star fraction profile, allowing us to assess how resolution
affects the final distribution of stars in the cluster. Again, much larger differences are seen when the softening length is varied: the
SS runs have smaller star fractions in the core than the MS runs, due to the smaller BCG that has formed in the former cases. 
The effects of resolution on the cumulative gas fraction are more complex (top-right), exhibiting a dependence on both mass and
spatial resolution. It is interesting to note that our standard set of runs (LR-MS and HR-SS, with a softening length that increases 
with particle mass according to the \citealt{Power2003} formula) agree best outside the core.
For the total baryon fraction profiles (not shown), a similar result to the star fraction profiles is seen in the core as the stars dominate
the baryon budget in this region. However, on large scales ($r>0.5\, r_{500}$) there is good convergence between all runs.

In the lower panels of Fig.~\ref{plot:res_bary}, we show entropy and temperature profiles for the hot gas. We first consider the
entropy profiles; as was the case with the hot gas fraction, the LR-MS and HR-SS runs show the best agreement outside the core. 
At fixed spatial resolution, decreasing the particle mass leads to 
a similar or 
larger entropy at fixed radius (e.g. going from VLR-SS $\rightarrow$ LR-SS $\rightarrow$ HR-SS).
Similarly, increasing the softening length at fixed mass resolution also increases the entropy (e.g. LR-SS $\rightarrow$ LR-MS). These
increases can largely be explained as due to decreases in gas density.
Decreasing the softening length (at fixed mass resolution) produces more feedback at early times, as
expected from the more efficient black hole growth and star formation. This feedback is more effective in keeping the gas from forming
stars in the cluster (hence also the lower star fractions in these runs), leading to a higher gas density (and thus lower entropy). However, 
decreasing the particle mass (at fixed softening length) has a smaller effect on the stellar and black hole masses, suggesting that the
effect on the hot gas is related to how well the outflows are resolved: higher mass resolution appears to lead to more effective outflows
which move gas to larger radii. In summary, when going from LR-MS to HR-SS, the combined effects of less efficient star formation and 
more effective outflows 
approximately
cancel, producing a similar entropy profile outside the core.

Inside the core, the entropy profiles 
show quite a lot of scatter, indicating that the core gas properties are sensitive to the choice of numerical parameters. 
The default LR-MS profile is much steeper than the others, a feature driven by the 
high gas density and low temperature within the central region. However, the highest resolution (HR-SS) run matches the observations
the best across all radii and the entropy within the core does not drop below those of CC clusters. Note that the distinctive feature at 
$\sim 0.06 \, r_{500}$ is still present.

Finally, the spectroscopic-like temperature profiles are displayed in the bottom right of Fig~\ref{plot:res_bary}. Again, we have
applied the method of \cite{Roncarelli2013}, removing the densest gas within each bin. (We found that the removal
of this gas is more important for runs with larger softening lengths, which show significantly more scatter from bin to bin.) In the 
radial range, $0.1 < r/r_{500} < 1$, all runs have similar temperature profiles, however within the central region ($r<0.05 \, r_{500}$) 
the runs with larger softening lengths (VLR-MS and LR-MS) have very low temperatures. It can also be seen that the HR-SS run
has the highest peak temperature and is therefore most discrepant with the observations. While this result is for one object, it 
suggests, as with our entropy profile, that we may be missing important physics in the cluster core.

\section{Summary and Discussion}
\label{sec:discuss}

There is now a wealth of observational data on clusters, offering an important opportunity 
to test how the different physics implemented in simulations affects the evolution of these massive objects. 
In this paper, results for a set of 30 clusters from the Millennium Gas simulation with a mass range of 
$10^{14} h^{-1} {\rm M}_{\odot} < {\rm M}_{\rm 200} < 10^{15} h^{-1} {\rm M}_{\odot}$ were presented, with 
increasingly realistic sub-grid physics, in order to ascertain the effects of each physical process. The models 
implemented were a non-radiative (NR) model, which only included gravity and hydrodynamics; a model that also
included radiative cooling and star formation (CSF); a model including supernova feedback (SFB), where powerful 
gas heating occurs (to $10^7$K) when a star is formed; and an AGN feedback model where a fixed fraction of energy 
from the accretion of mass on to a super-massive black hole was used to heat the gas to high temperature (scaling
with the virial temperature of the cluster). Each component is important in order to make progress towards the simulation 
of realistic clusters, which we summarise as follows:
\begin{itemize}
 \item
Radiative cooling permits gas to lose energy and become more dense as it flows to the centre of the cluster. The gas is then 
likely to undergo star formation, but without any form of feedback, too many stars are created, leading to the classic 
{\it over-cooling} problem. However, cooling also causes higher entropy gas to flow inwards and this process is behind the
similarity-breaking seen in all our radiative models (as is evident from the entropy profile shapes and the redshift evolution 
of the X-ray luminosity-mass relation).
\item
Supernova feedback provides energy to heat the gas surrounding stars and thus reduce the star formation rate. Although this 
energy is distributed throughout the cluster halo (in galaxies), it is only effective at larger radii: in the central region the energy released in 
supernovae is insufficient to prevent dense, cold gas cores that again result in forming too many stars.
\item
Super-massive black holes are distributed throughout the cluster halo in galaxies, however most of the energy released is from 
the largest black hole in the centre of the cluster. As a result, AGN feedback has a dramatic effect on the core gas, reducing the size
of the central, brightest cluster galaxy. 
\end{itemize}

In order to ascertain the ideal properties for supernova and AGN feedback, different parameters where tested on both a low and 
high-mass halo, and compared to observations when appropriate. From this study, we conclude that:
\begin{itemize}
\item
While AGN feedback has an important effect on the star formation rate in the central region of the cluster, powerful supernova 
feedback is also required in order to reproduce the observed gas and star fractions within $r_{\rm 500}$. 
\item
The only AGN feedback parameter which has a large effect on the evolution of the cluster is the heating temperature. If a temperature 
is chosen that is too low, the heated gas is unable to escape the deep potential and instead creates a core of warm, dense gas. If the 
heating temperature is higher, gas then escapes out of the central region, reducing the gas density, but also taking most of the thermal energy with it.
\item
A better match to the observed pressure profiles is obtained when the AGN heating temperature is tuned to scale with the final virial 
temperature of the halo. As such, the current model suffers from a fine-tuning problem.
\item
While the AGN feedback efficiency has little effect on the cluster's evolution, it does allow the mass of the black hole to be tuned; a lower
efficiency allows the black hole to become more massive and vice-versa.
\end{itemize}
 
With the feedback parameters chosen as detailed above, our simulations with the AGN model are capable of reproducing a 
range of observable properties of clusters, including baryon, gas and star fraction within $r_{\rm 500}$; gas density and pressure profiles; 
and the $Y_{\rm SZ}-M_{\rm 500} $ and $L_{\rm bol}-M_{\rm 500} $ scaling relations. However, the simulations failed to resolve a 
number of issues. 
Firstly, observables that are more sensitive to the temperature than the density of the gas are simulated with some success. A 
small amount of low-entropy gas particles, both inside and outside substructures, serve to make the spectroscopic-like 
temperature much noisier than the mass-weighted temperature. When this gas was removed, the match to observations 
(especially outside the core) improves.  A second issue is that while the AGN feedback significantly reduces the stellar mass fractions within 
 BCGs, they are still around a factor of 3 larger than observed. A large part of this problem appears to be due to the fact that the AGN are not 
efficient enough at high redshift; a similar problem was seen by \cite{Ragone2013}, who also performed cosmological simulations of 
clusters with (a different model of) AGN feedback. A third issue is that the entropy profiles in the AGN model do not match the observations 
inside the core region ($r < 0.15 \, r_{500}$). A characteristic break is seen at this point, inside which the entropy declines rapidly 
to the centre. Further investigation revealed that the AGN heats and ejects gas from the central region, largely without disturbing the 
surrounding, cooling material. Creating the extreme profile of a non-cool-core cluster would require the core gas to be mixed much more 
efficiently than what is seen in our simulations, while cool-core clusters may be approximated with some additional variation in the heating temperature. 

Finally, we considered the effect of varying the spatial and mass resolution for one of our clusters. We found that reducing the 
gravitational softening length (and thus also the minimum SPH smoothing length for the gas) had the largest effect on the black holes
and stars; a smaller softening leads to earlier central black hole growth and larger black holes in satellite galaxies. A smaller 
softening also affects the star formation history of the cluster, producing a smaller BCG, but more stars are found in the diffuse 
component. However, the resolution appears to affect the hot gas in a more complex way: outside the core, 
a higher spatial resolution and mass resolution increase and decrease the gas density respectively, leading to broadly similar results. 
The highest-resolution run produces a higher core entropy that is more consistent with cool-core clusters though the break in the
profile remains. The problem with the X-ray temperatures also diminishes, but is still present, in the high-resolution runs.

\subsection{Comparison with recent work}

While this paper was being written, we became aware of two other new studies that are qualitatively similar to ours (i.e.
designed to study the effects of AGN feedback on the galaxy cluster population). Firstly, \cite{Planelles2014} performed
simulations of 29 Lagrangian regions, producing 160 objects down to group scales (see also \citealt{Planelles2013}). 
We subsequently refer to this work as P14. 
Secondly, \cite{LeBrun2013} analysed a set of large-volume ($400 \, h^{-1}{\rm Mpc}$) cosmological simulations with 
AGN feedback (cosmo-OWLS). Their analysis also goes down to group scales and state that there are approximately
$14,000$ objects in their non-radiative run at $z=0$. We refer to this work as LeB14. 

All three studies differ in the cosmological model adopted, the resolution
of the simulations and the implementation of the sub-grid physics. Regarding cosmological parameters, LeB14
adopt values derived from {\it Planck} data for their main results; the main relevant change is the baryon fraction, which 
decreases from $\sim 0.17$ for our study (and that of P14) to $\sim 0.15$. A lower baryon fraction will
reduce the efficiency of radiative cooling (for a fixed metallicity and temperature) and therefore not require as much feedback
to reproduce the observed ratio of gas to stars. 

For numerical resolution, our default dark matter particle mass varies such that the approximate number of particles 
within $r_{200}$, $N_{200} \simeq 10^6$. In P14 and LeB14, the particle mass is kept constant so $N_{200}$ varies with
halo mass. For the same range of masses as our sample ($M_{200}=1-10\times 10^{14} \, h^{-1}{\rm M}_{\odot}$), 
$N_{200}\simeq 10^{5}-10^{6}$ for P14 (thus matching our resolution for the most massive objects) and 
$N_{200}\simeq 2\times(10^{4}-10^{5})$ for LeB14 (i.e. at least a factor of five smaller than ours). Regarding the gravitational
softening length, P14 and LeB14 use similar values to ours (5 and 4 $h^{-1}{\rm kpc}$ respectively), however both studies allow
the SPH smoothing length to decrease below this value (by a factor of 2 and 5 respectively). As we discussed in 
Section~\ref{sec:res}, a smaller softening length (and SPH smoothing length) has a significant impact on the growth of the black 
holes and star formation rate. While this may be necessary in order to grow black holes when the resolution is low, 
one also has to be cautious given that a smaller value can also lead to spurious two-body heating effects.

For the gas physics, both P14 and LeB14 include metal-dependent cooling whereas our study assumed a metal-free gas. Including
metals would likely require us to re-tune the feedback parameters due to the increase in cooling efficiency; this will be especially 
true at high redshift, where the gas density is higher and temperatures lower. As in our case, P14 and LeB14 also include 
supernovae-driven winds in their main simulations, but add the energy in kinetic form whereas we adopt the thermal feedback
approach. Such models ought to produce a similar outcome when the cooling time of the gas is sufficiently long, but the limited resolution 
of the simulations will likely lead to some differences. Finally, regarding the AGN feedback, our study and LeB14 both 
used the method of \cite{Booth2009}, whereby energy is stored until there is enough to heat one particle to a fixed temperature, 
$T_{\rm AGN}$, whereas P14 use the kernel-weighted feedback implementation of \cite{SpringelDMH2005}, where the energy
is shared immediately between nearby gas particles. A key difference between the AGN model in this paper and those in 
LeB14 is that we use a value of $T_{\rm AGN}$ that scales with the final virial temperature of the halo, whereas they find 
$T_{\rm AGN}=10^8\,{\rm K}$ gives the best results over their whole mass range.

As with the work presented in this paper, both P14 and LeB14 find that their AGN feedback models are in good agreement
with observational data for many global properties (e.g. gas and star fractions, X-ray and SZ scaling relations). One property
that our models do not predict as well as the others is the X-ray temperature of the cluster gas; in our case, we must remove
the densest gas otherwise it substantially down-weights the spectroscopic-like temperature. The reason for this is unclear, however
we first note that LeB14 estimate temperatures by directly fitting plasma models to simulated X-ray spectra, as is done for the
observations (they also adopt hydrostatic mass estimates in the scaling relations and thus factor in the effect of hydrostatic mass
bias, which we ignore in this paper due to the problems encountered with temperature measurements). It may be that the
spectroscopic-like formula is incorrectly tuned to the radiative simulations presented here (\citealt{Mazzotta2004} used non-radiative
simulations in their study). However, P14 also use $T_{\rm sl}$ in their analysis, suggesting the cold, dense gas in their simulations 
(which are similar resolution to ours for high-mass objects) is less of a problem. It may be that metal enrichment plays a part (allowing
more rapid cooling out of the hot phase). Alternatively, the smaller adopted minimum SPH smoothing length could affect the results, allowing
gas to reach higher density and cool more efficiently. As discussed in Section~\ref{sec:res}, this can have a noticeable effect on the core
temperatures.

Gas density and entropy profiles in LeB14 match those in our own work over the radial ranges displayed, which is unsurprising considering 
the similarity of the feedback models. However, our simulations have higher resolution, allowing us to probe smaller radii where the entropy 
and temperature are too low, whilst the gas density too high. It should also be noted that LeB14 achieved their best match to observations
using a fixed AGN heating temperature. However, as they allude to in their conclusions, the lower baryon fraction in the {\it Planck}  
cosmology appears to play an important role in this difference (I. McCarthy, private communication). Finally, we note that the pressure
profile of P14 matches the observations (and therefore, by design, our own results).  However, although our entropy profiles agree 
with theirs outside the core, there are noticeable differences within this region (P14 overestimate the observed core entropy). Interestingly, 
unlike the simulations presented in this paper (or those in LeB14), the gas profiles presented in P14 seem very similar in their runs
with and without AGN feedback on cluster scales, with only the stellar fractions in being affected by the AGN. It thus seems that their 
implementation of AGN feedback, whilst key in regulating the star formation, does not significantly effect gas profiles.
\vspace{0.5cm}

In conclusion, our study reinforces those of P14 and LeB14 that simulations incorporating radiative cooling and simple models for the
feedback of energy from supernovae and AGN, are able to successfully reproduce many key observational properties of clusters. However,
a detailed match to the spatial distribution of gas and stars is still wanting, especially in the cluster cores. An important step forward will
be to compare many of the AGN models run on the same initial conditions, so we can remove the uncertainty from cosmological parameters
and numerical resolution. This, combined with progress in modelling cluster physics and performing higher resolution simulations 
(an important but reachable goal will be to resolve the Jeans length of warm interstellar gas) should allow us to produce even more
realistic cluster simulations and understand more about their formation and evolutionary history. Such progress will be crucial for improving
the use of clusters as cosmological probes and our understanding of galaxy formation in extreme environments.

\section*{Acknowledgments}
The simulations used in this paper were performed on the 
ICC Cosmology Machine, which is part of the DiRAC Facility jointly
funded by STFC, Large Facilities Capital Fund of BIS and
Durham University. SRP and RDAN were both supported by STFC 
studentships while this work was being done. 
RDAN also acknowledges the support received from the Jim Buckee Fellowship.
STK, PAT and ARJ acknowledge support from STFC, grant numbers ST/L000768/1, 
ST/J000652/1 and ST/I001166/1 respectively.

\bibliographystyle{mn2eFix}
\bibliography{manuscript}

\appendix
\section{Measuring Simulated Cluster Properties}
\label{app:measure}

In this appendix, we summarise how various cluster properties are estimated from the simulated data. For
each cluster, we start with a list of all gas, star and dark matter (DM) particles that are located within a radial 
distance $r<r_{\Delta}$ from the position of the most bound particle (as found by SUBFIND). The outer radius,
$r_{\Delta}$ is defined in the usual way
\begin{equation}
M_{\Delta} = {4 \over 3} \, \pi \, \Delta \, \rho_{\rm cr}(z) \, r_{\Delta}^3,
\end{equation}
where $M_{\Delta}$ is the total enclosed mass and $\rho_{\rm cr}(z)$ is the critical density.
By default, we set $\Delta=500$ but occasionally use other values where appropriate. 
Hot gas is defined as those gas particles with 
temperatures $T>10^{6}\, {\rm K}$ except when X-ray temperatures are estimated (see below). 
When estimating cluster profiles, we sub-divide the cluster volume into spherical shells, equally spaced in 
$\log_{10}(r/r_{500})$. 

Baryon, hot gas and star fractions are calculated via 
\begin{equation}
  \label{eqn:mass_del}
  f_{\rm type}=\frac{\sum_{i=1}^{N_{\rm type}} m_{i} }{ \sum_{i=1}^{N} m_{i} },
\end{equation}
where {\it type} refers to gas, stars or both (with total number $N_{\rm type}$); $m_{i}$ is the the mass of the $i$th particle and $N$ is the 
total number of particles (including DM) in the region being summed. The mass density of each species is similarly calculated as
\begin{equation}
\rho_{\rm type} = {1 \over V} \, \sum_{i=1}^{N_{\rm type}} \, m_i,
\end{equation}
where the $V$ is the volume of the region where the density is being estimated. For example, the density profile
is estimated using concentric spherical shells, each with volume $V=(4\pi/3)(r_{\rm out}^3-r_{\rm in}^3)$, where
$r_{\rm in}$ and $r_{\rm out}$ are the inner and outer shell radii respectively.

For the electron pressure of the hot gas ($P_{e} = n_{e} kT$) we use
\begin{equation}
\label{eqn:Pest}
P_{e} = {1 \over V} \frac{k}{\mu_{e} m_{\rm H}} \sum^{N_{\rm hot}}_{i=1}m_{i}T_{i} ,
\end{equation}
where $T_i$ is the temperature of the $i$th hot gas particle and $\mu_e=1.14$ is the 
assumed mean atomic weight per free electron. Similarly, we estimate the 
entropy ($K \equiv kT/n_e^{2/3}$) using
\begin{equation}
\label{eqn:Kest}
K  = V^{2/3} k (\mu_e m_{\rm H})^{2/3} \, \frac{\sum_{i=1}^{N_{\rm hot}}m_{i} T_{i}}{\left( \sum_{i=1}^{N_{\rm hot}}m_{i} \right)^{5/3}}.
\end{equation}
When presenting pressure and entropy profiles, we follow convention and express these quantities in
units of a characteristic scale, appropriate for self-similar, isothermal systems (e.g. \citealt{Nagai2007}).
Starting from a characteristic density 
\begin{equation}
\rho_{500}=500 \, (\Omega_{\rm b}/\Omega_{\rm m}) \, \rho_{\rm cr}(z) = 
\left[ 1500 \over 8\pi G \right] \, (\Omega_{\rm b}/\Omega_{\rm m}) \, H(z)^2,
\end{equation}
and temperature 
\begin{equation}
kT_{\rm 500} = {GM_{500} \mu m_{\rm H} \over 2r_{500}} = 5\mu m_{\rm H} \, \left[ {GH(z)M_{500} \over 2} \right]^{2/3},
\end{equation}
the entropy scale can be written as
\begin{equation}
K_{500} = \left[ {4 \pi^2 G^4 \mu^3 \mu_e^2 m_{\rm H}^5 \over 4500 (\Omega_{\rm b}/\Omega_{\rm m})^2 }\right]^{1/3} \, H(z)^{-2/3} \, 
 M_{500}^{2/3}, 
\end{equation}
where $\rho_{\rm cr}=3H^2/(8\pi G)$,  
$H(z)=H_0E(z)=H_0 [\Omega_{\rm m}(1+z)^3 + \Omega_{\Lambda}]^{1/2}$ is the Hubble parameter (assuming
a flat universe) and $\mu=0.59$ is the mean atomic weight for a fully-ionised, primordial gas. 
Similarly, for the pressure scale 
\begin{equation}
P_{500} = {3 (\Omega_{\rm b}/\Omega_{\rm m})(\mu/\mu_{e}) \over 8\pi} \, 
\left[ {500 \over 2 G^{1/4}} \right]^{4/3} \, H(z)^{8/3} \, M_{500}^{2/3}.
\end{equation}

When X-ray temperatures are presented, we use the approximation suggested by \cite{Mazzotta2004}
\begin{equation}
\label{eqn:Test}
T_{\rm sl}=\frac{\sum_{i=1}^{N_{\rm X}}\rho_{i} T_{i}^{1/4}}{\sum_{i=1}^{N_{\rm X}}\rho_{i} T_{i}^{-3/4}},
\end{equation}
where $N_{\rm X}$ is the number of particles with $kT>0.5$ keV. For hot clusters ($kT>2$ keV), this 
spectroscopic-like temperature was shown to be a better estimate of the X-ray temperature than a simple mass-weighted
temperature and preferentially weights cooler, denser gas. (Note, however, that we have not checked the weighting is
optimal for the models presented in this paper; \citealt{Mazzotta2004} tuned it to non-radiative simulations.)
As discussed in Section~\ref{sec:method}, we found that this estimate is significantly affected by the presence
of a small number of dense gas particles that may be spurious due to the lack of entropy mixing in standard SPH. To
reduce this effect, we adopt the method discussed in \cite{Roncarelli2013}, which starts by ranking all gas particles
in each radial shell (used to calculate profiles) by volume, $V_i = m_i / \rho_i$. Particles with the largest $V_i$ that make
up 99 per cent of the shell volume are retained and the rest (which are the densest particles by construction) discarded.

Finally, we also consider two integrated properties, both observable: the bolometric X-ray luminosity and the SZ $Y$
parameter. The luminosity is estimated as
\begin{equation}
\label{eqn:Lest}
L_{\rm bol}=\frac{m_{\rm gas}}{(\mu m_{H})^2} \sum_{i=1}^{N_{hot}} \, \rho_{i} \, \Lambda(T_i,Z),
\end{equation}
where $\rho_{i}$ is the SPH density of the $i$th hot gas particle and $\Lambda(T,Z)$ is the same cooling function
used in the simulation. Although we assume $Z=0$ for our main radiative runs, we adopt $Z=0.3\,Z_{\odot}$ when
calculating luminosities as this is the typical metallicity of the ICM.

Cluster cores are the hardest part to simulate and 
both $L_{\rm bol}$ and $T_{\rm sl}$ are dominated by the central region. Results excluding gas from the inner
region ($r<0.15 r_{500}$) will therefore also be considered for these quantities and written as $L_{\rm bol,OC}$ and 
$T_{\rm sl,OC}$ for the luminosity and temperature respectively.

For the SZ $Y$ parameter we absorb the angular diameter dependence
\begin{equation}
\label{eqn:y_del}
D_{\rm A}^2 Y_{\rm SZ}= \frac{\sigma_{\rm T} m_{gas} k}{\mu_{e}m_{\rm H}m_{e}c^2} \sum^{N_{hot}}_{i=1} T_{i}.
\end{equation}
This quantity is proportional to the total thermal energy of the hot gas and ought to be less sensitive to cooling and
feedback processes than the X-ray luminosity.

\section{Observational data}
\label{app:obs}

We use a number of results from observational datasets to compare with our simulations in this paper. Firstly,
we compare our  baryon, hot gas and star fractions with the observational constraints from \cite{Giodini2009}.  
They analysed 91 groups and poor clusters at redshift, $0.1 < z < 1$, selected from the COSMOS 2 deg$^2$ 
survey, and 27 nearby clusters with robust total and stellar masses inside $r_{\rm 500}$. For hot gas fractions, 
we also make use of the {\it XMM-Newton} results of \cite{Arnaud2007} for a sample of 10 relaxed clusters, 
and the larger REXCESS sample \citep{Bohringer2007}, with data taken from \cite{Croston2008}. The REXCESS 
sample is a representative sample of low redshift X-ray clusters and contains 33 objects over a mass range 
$10^{14}<M_{500}/{\rm M}_{\odot}<10^{15}$. We also use this sample when comparing our gas density 
\citep{Croston2008} and entropy \citep{Pratt2010} profiles, as well as our X-ray scaling relations between luminosity, 
temperature and mass \citep{Pratt2009}. 

For the star fractions, we additionally compare our results with the best-fit relation (between star fraction and halo mass) 
presented in \cite{Budzynski2014}. In that study, 20,171 large groups and clusters with a mass 
$M_{\rm 500}>10^{13.7}\, {\rm M}_{\odot}$ were optically selected from Sloan Digital Sky Survey data 
at $0.15 < z < 0.4$. The objects were then separated into 4 mass bins and stacked in order to calculate more robust stellar 
fractions (including the contribution from a low surface brightness component). 

When presenting temperature profiles, we compare our results with those from \cite{Leccardi2008}, where 
50 objects were selected with $M_{\rm 500}>10^{14}{\rm M_{\odot}}$ and observed with 
{\it XMM-Newton}. Some of these objects are also in the REXCESS sample. For the pressure profiles, we compare
against fits to the total, cool-core and non-cool-core samples by \cite{Planck2013}. They analysed {\it Planck} SZ + 
{\it XMM-Newton} X-ray data for 62 nearby massive clusters with a mass range of 
$2 \times 10^{14}<M_{500}/{\rm M}_{\odot}<2 \times 10^{15}$.
We also compare our simulated $Y_{\rm SZ}-M_{500}$ relations against the results from an earlier study by the Planck
Collaboration using the same data \citep{Planck2011e}.

\label{lastpage}

\end{document}